\documentclass[acmsmall,screen]{acmart}

\usepackage{algorithm}
\usepackage{algcompatible}
\usepackage{balance}
\usepackage{makecell}
\AtBeginDocument{%
  
}

\setcopyright{acmlicensed}
\acmJournal{TODS}
\acmYear{2025} \acmVolume{1} \acmNumber{1} \acmArticle{1} \acmMonth{1}\acmDOI{10.1145/3722102}

\def\Uber{Uber, Inc.}
\author{Ahmed Metwally}
\email{ametwally@uber.com}
\orcid{0000-0002-1084-8078}
\affiliation{
  \institution{\Uber{}}
  \streetaddress{190 Mathilda Pl}
  \city{Sunnyvale}
  \state{CA}
  \country{USA}
  \postcode{94086}
}
\renewcommand{\shortauthors}{A. Metwally}

\begin{document}

\title{Scaling and Load-Balancing Equi-Joins}

\def\Section{\S}
\def\Equation{Eqn.}
\def\Inequality{Ineq.}
\def\Relation{Rel.}
\def\Figure{Fig.}
\def\Table{Table.}
\def\Algorithm{Alg.}
\def\naive{na\"\i ve}
\def\Naive{Na\"\i ve}

\newcommand{\ceilX}[1]{\left \lceil #1 \right \rceil }
\newcommand{\floorX}[1]{\left \lfloor #1 \right \rfloor }

\newenvironment{packed_enum}{
\begin{enumerate}
  \setlength{\itemsep}{0pt}
  \setlength{\parskip}{0pt}
  \setlength{\parsep}{0pt}
}{\end{enumerate}}

\newlength \smallfigwidth
\newlength \bigfigwidth
\makeatletter
\if@twocolumn
  \setlength \smallfigwidth {0.9 \columnwidth}
\else
  \setlength \smallfigwidth {0.45 \textwidth}
\fi
\setlength \bigfigwidth {\textwidth}
\makeatother

\def\FrameworkFullName{Adaptive-Multistage-Join}
\def\FrameworkName{AM-Join}
\def\FrameworkCodeName{amJoin}

\def\MultistageJoinFullName{Tree-Join}
\def\MultistageJoinCodeName{treeJoin}

\def\SmallLargeJoinFullName{Index-Broadcast-Join}
\def\SmallLargeJoinName{IB-Join}
\def\SmallLargeJoinCodeName{indexBroadcastJoin}
\def\SmallLargeLeftOuterJoinFullName{Index-Broadcast-Left-Outer-Join}
\def\SmallLargeLeftOuterJoinName{IB-LO-Join}
\def\SmallLargeLeftOuterJoinCodeName{indexBroadcastLeftOuterJoin}
\def\SmallLargeFullOuterJoinFullName{Index-Broadcast-Full-Outer-Join}
\def\SmallLargeFullOuterJoinName{IB-FO-Join}
\def\SmallLargeFullOuterJoinCodeName{indexBroadcastFullOuterJoin}

\algnewcommand\ASSUMES{\item[\textbf{Assumes:}]}
\algnewcommand\LOCAL{\item[\textbf{Local:}]}
\algnewcommand\INDENTLOCAL{\item[ \quad \quad \; ]}
\algnewcommand\INPUT{\item[\textbf{Input:}]}
\algnewcommand\INDENTINPUT{\item[ \quad \quad \; ]}
\algnewcommand\OUTPUT{\item[\textbf{Output:}]}
\algnewcommand\INDENTOUTPUT{\item[ \quad \quad \; \; \ ]}
\algnewcommand\CONSTANT{\item[\textbf{Constant:}]}
\algnewcommand\INDENTCONSTANT{\item[ \quad \quad \quad \; \; \ ]}
\algnewcommand\RETURN{\item[\textbf{\quad \ \ \ Return}]}
\algnewcommand\INDENT{\item[ \quad \quad \quad ]}
\algnewcommand\CONTINUELINE{\item[ \newline \quad \quad \quad ]} 
\algnewcommand\CONTINUELINEINDENTED{\item[ \newline \quad \quad \quad \; \ ]}

\begin{abstract}

The task of joining two tables is fundamental for querying databases. In this paper, we focus on the equi-join problem, where a pair of records from the two joined tables are part of the join results if equality holds between their values in the join column(s). While this is a tractable problem when the number of records in the joined tables is relatively small, it becomes very challenging as the table sizes increase, especially if hot keys (join column values with a large number of records) exist in both joined tables.

This paper, an extended version of \cite{metwally2022scaling}, proposes \FrameworkFullName{} (\FrameworkName{}) for scalable and fast equi-joins in distributed shared-nothing architectures. \FrameworkName{} utilizes (a) \MultistageJoinFullName{}, a proposed novel algorithm that scales well when the joined tables share hot keys, and (b) Broadcast-Join, the fastest known algorithm when joining keys that are hot in only one table.

Unlike the state-of-the-art algorithms, \FrameworkName{} (a) holistically solves the join-skew problem by achieving load balancing throughout the join execution, and (b) supports all outer-join variants without record deduplication or custom table partitioning. For the fastest \FrameworkName{} outer-join performance, we propose the \SmallLargeJoinFullName{} (\SmallLargeJoinName{}) family of algorithms for Small-Large joins, where one table fits in memory and the other can be up to orders of magnitude larger. The outer-join variants of \SmallLargeJoinName{} improves on the state-of-the-art Small-Large outer-join algorithms.

The proposed algorithms can be adopted in any shared-nothing architecture. We implemented a MapReduce version using Spark. Our evaluation shows the proposed algorithms execute significantly faster and scale to more skewed and orders-of-magnitude bigger tables when compared to the state-of-the-art algorithms.

\end{abstract}


\maketitle

\section{Introduction}

Retrieval of information from two database tables is critical for data processing, and impacts the computational cost and the response time of queries. In the most general case, this operation entails carrying out a cross join of the two tables. The more common case is computing an equi-join, where two records in the two tables are joined if and only if equality holds between their \emph{keys} (values in the join column(s)). The algorithms for equi-joins have been optimized regularly since the inception of the database community \cite{codd1969derivabilityj,codd1970relational,selinger1979access,bernstein1981query,cheung1982method,date1983outer,dewitt1987single,schneider1989performance,lakshmi1990effectiveness,shasha1991optimizing,dewitt1992practical}. 

Significant research has been done to enhance the sequential equi-join algorithms on multi-core processors \cite{kim2009sort,blanas2011design,albutiu2012massively,balkesen2013main,balkesen2013multi,schuh2016experimental,barthels2015rack,bandle2021partition} and on GPUs \cite{he2008relational,kaldewey2012gpu,chen2016accelerating,guo2019distributed,sioulas2019hardware,paul2020revisiting,rui2020efficient,paul2021mg}. However, the proliferation of data collection and analysis poses a challenge to sequential join algorithms that are limited by the number of threads supported by the processing units. Scaling equi-joins had to progress through distributed architectures, which is the direction adopted in this work.

This work is motivated by equi-joining industry-scale skewed datasets in a novel way. We tackle \emph{natural self-joins} \cite{codd1969derivabilityj,codd1970relational} at the intersection of equi-joins \cite{Codd72relationalcompleteness}, inner-joins, and self-joins \cite{Hursch1989relational}. This join semantic is an integral operation in the similarity-based join algorithms (e.g., \cite{metwally2012v, metwally2024similarity}) used for fraud detection. While natural self-joins can be more efficient to perform than the general equi-joins, the state-of-the-art equi-join algorithms failed to scale to our large and skewed datasets. This motivated us to develop the fast, efficient, and scalable \FrameworkFullName{} (\FrameworkName{}) that scales, not only to our specialized use case, but also to the most challenging equi-joins.

In this paper, an extended version of \cite{metwally2022scaling}, we first propose \MultistageJoinFullName{}, a novel algorithm that scales well by distributing the load of joining a key that is \emph{hot} (i.e., high-frequency or shared by a large number of records) in both tables to multiple executors. Such keys are the scalability bottleneck of most of the state-of-the-art distributed algorithms. We give special attention to balancing the load among the executors throughout the \MultistageJoinFullName{}  execution.

We then tackle Small-Large joins, where one table fits in memory and the other can be up to orders of magnitude larger. We devise the \SmallLargeJoinFullName{} (\SmallLargeJoinName{}) family for Small-Large joins, and show analytically their outer-join variants improve on the state-of-the-art Small-Large outer-join algorithms \cite{xu2010new,cheng2017design}.

The \MultistageJoinFullName{}, and the Broadcast-Join algorithms are the building blocks of \FrameworkName{}. \FrameworkName{} achieves (a) high scalability by utilizing \MultistageJoinFullName{} that distributes the load of joining a key that is hot in both relations to multiple executors, and (b) fast execution by utilizing the Broadcast-Join algorithms that reduce the network load when joining keys that are hot in only one relation. \FrameworkName{} extends to all outer-joins elegantly without record deduplication or custom table partitioning, unlike the state-of-the-art industry-scale algorithms \cite{bruno2014advanced}. The outer-join variants of \FrameworkName{} achieves fast execution by utilizing the \SmallLargeJoinName{} family of algorithms.

All the proposed algorithms use the basic MapReduce primitives only, and hence can be adopted on any shared-nothing architecture. We implemented a MapReduce version using Spark \cite{zaharia2010spark}. Our evaluation highlights the improved performance and scalability of \FrameworkName{} when applied to the general equi-joins. When compared to the sate-of-the-art algorithms \cite{bruno2014advanced,gavagsaz2019load}, \FrameworkName{} executed comparably fast on weakly-skewed synthetic tables and can join more-skewed or orders-of-magnitude bigger tables, including our real-data tables. These advantages are even more pronounced when applying the join algorithms to natural self-joins. The proposed \SmallLargeJoinName{} outer-join algorithm executed much faster than the algorithms in \cite{xu2010new,cheng2017design}.

The rest of the paper is organized as follows. We formalize the problem and its variations and lay out the necessary background in \Section~\ref{sec:formalization}. We review the related work in \Section~\ref{sec:related_work}. We propose \MultistageJoinFullName{} in \Section~\ref{sec:tree_join}. We then discuss Small-Large joins and propose \SmallLargeJoinFullName{} in \Section~\ref{sec:broadcast}. The \FrameworkFullName{} is described in \Section~\ref{sec:main_algorithm}. In \Section~\ref{sec:finding_hot_keys}, we discuss identifying hot keys, which is an integral part of \FrameworkFullName{}. We report our evaluation results in \Section~\ref{sec:results}, and summarize our contributions in \Section~\ref{sec:conclusion}.

\section{Formalization}
\label{sec:formalization}

We now introduce the concepts used. As the readers progress through the paper, they are referred to \Table~\ref{table:symbols} for the symbols used.

\begin{table}
  \caption{The symbols used in the paper.}
  \label{table:symbols}
  \scriptsize
  \centering
  \begin{tabular}{ c | l }
  \toprule

\textbf{Symbol} & \textbf{Meaning} \\

  \midrule

$\mathcal{R}$, $\mathcal{S}$        & The joined relations/tables. \\
$\mathcal{Q}$                                & The relation/table storing the join results. \\
$id_{rel}$                                       & The relation identifier ($0$ for $\mathcal{R}$ and $1$ for $\mathcal{S}$). \\
$Key$.                                           & The join attribute(s)/column(s). \\
$Attrib_{\mathcal{R}}$, $Attrib_{\mathcal{S}}$     & The remaining attribute(s) in $\mathcal{R}$ and $\mathcal{S}$. \\
$key_{rec}$.                                  & The value of $Key$ in record $rec$. \\
$attrib_{\mathcal{R}_{r}}$              & The value of $Attrib_{\mathcal{R}}$ in record $r$. \\
$attrib_{\mathcal{S}_{s}}$              & The value of $Attrib_{\mathcal{S}}$ in record $s$. \\
$\ell$, $\ell_{\mathcal{R}}$, $\ell_{\mathcal{S}}$  & The number of records, a.k.a., frequency, \\
                                                                                       & with a specific key in general, in $\mathcal{R}$ and in $\mathcal{S}$. \\
$\ell_{max}$                                   & The maximum value of $\ell$. \\
$\lambda$                                      & The relative cost/runtime of  \\
                                                       & sending data over the network vs.\\
                                                       & its IO from/to a local disk. $\lambda \geq 0$. \\
$\ell'$                                              & For a given key, the number of records a \\
                                                       & subsequent \MultistageJoinFullName{} executor receives, \\
                                                       & where $\ell' = \ell^p$, for some $p$, s.t. $0 \leq p \leq 1$. \\
$\delta(\ell)$                                   & The number of sub-lists produced by the \\
                                                       & splitter in \MultistageJoinFullName{} for a list of length $\ell$. \\
$t$                                                  & The number of iterations of \MultistageJoinFullName{}. \\
$\kappa_{\mathcal{R}}$, $\kappa_{\mathcal{S}}$        & The hot keys in $\mathcal{R}$ and $\mathcal{S}$. \\
$|\kappa_{\mathcal{R}}|_{max}$, $|\kappa_{\mathcal{S}}|_{max}$.   & The maximum numbers of hot keys \\
                                                       & collected for $\mathcal{R}$ and $\mathcal{S}$. \\
$M$                                                & The available memory per executor. \\
$m_{id}$                                         & The size of a record identifier in bytes. \\
$m_{Key}$                                     & The average size of $Key$ in bytes. \\
$m_\mathcal{R}$, $m_\mathcal{S}$                            & The average size of records in $\mathcal{R}$ and $\mathcal{S}$ in bytes. \\
$\Delta_{operation}$                      & The expected runtime of \emph{operation}.\\
$n$                                                 & The number of executors.\\
$e_i$                                              & A specific executor.\\
$\mathcal{R}_i$, $\mathcal{S}_i$. & The partitions of $\mathcal{R}$ and $\mathcal{S}$ on $e_i$.\\

  \bottomrule

\end{tabular}
\end{table}

\subsection{Equi-Joins, Inner-Joins and Self-Joins}

\begin{figure*}[!ht]
\centering
\includegraphics[width = \bigfigwidth]{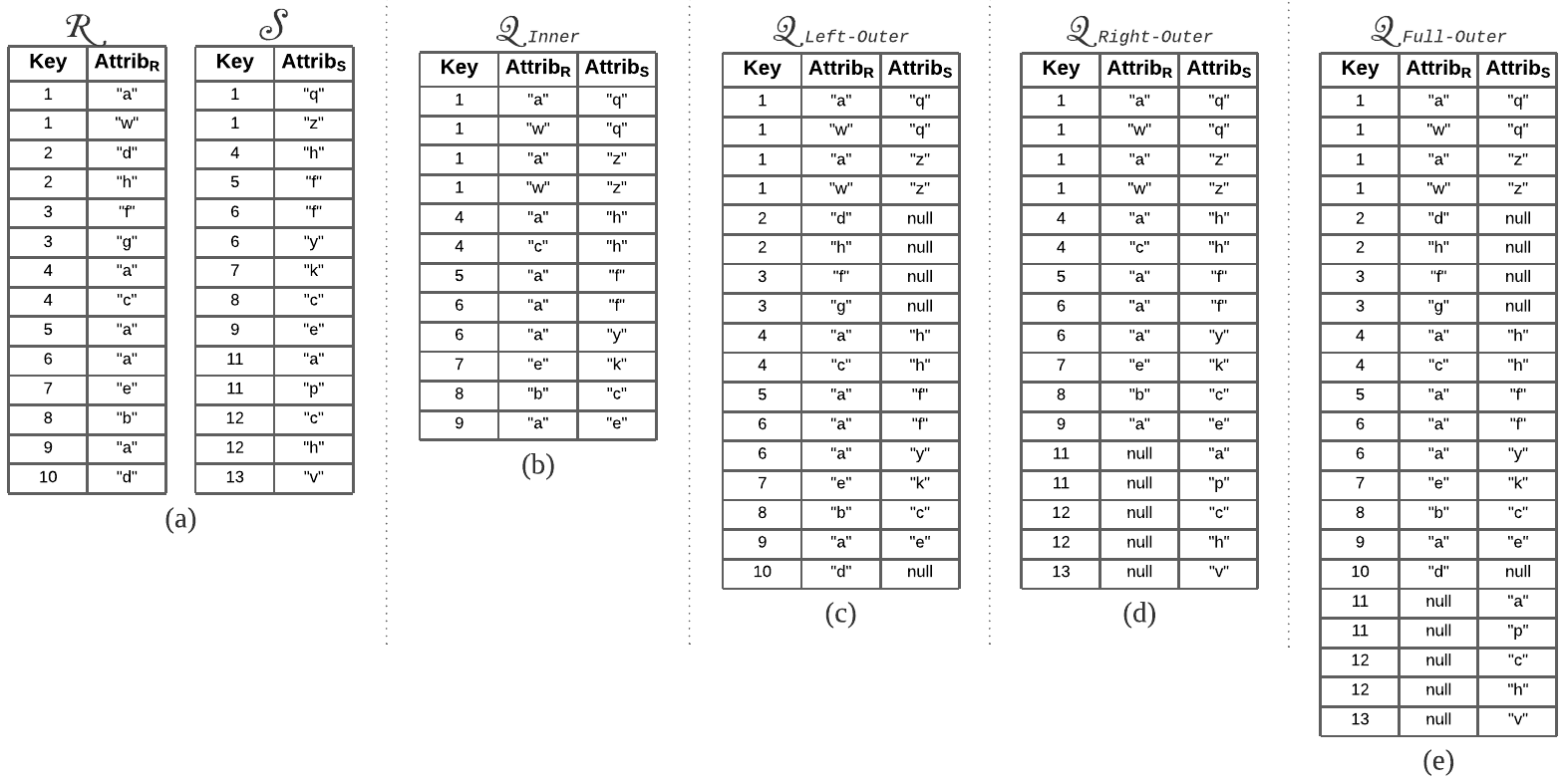}
\caption{
An example for joining two relations. (a) shows the input relations. (b) through (e) show the results of the inner, left-outer, right-outer, and full-outer-joins, respectively. The tables are sorted for readability, but sorting is not guaranteed in practice.
}
\label{fig:main_join_example_inner_and_outer}
\end{figure*}

The equi-join operation combines columns from two tables, a.k.a. relations in the relational database community, based on the equality of the column(s), a.k.a., join attribute(s). We focus on the case where two relations are joined. Other work, e.g., \cite{afrati2010optimizing,nica2012analyzing,chu2015theory,afrati2018sharesskew}, focused on extending equi-joins to multiple relations.

Given two relations $\mathcal{R}$ and $\mathcal{S}$, where $\mathcal{R} = (Key, Attrib_\mathcal{R})$ and $\mathcal{S} = (Key, Attrib_\mathcal{S})$, the equi-join results in a new relation $Q = (Key, Attrib_\mathcal{R}, Attrib_\mathcal{S})$, where $Key$ is the join attribute(s), and $Attrib_\mathcal{R}$ and $Attrib_\mathcal{S}$ are the remaining attribute(s) from $\mathcal{R}$ and $\mathcal{S}$, respectively, and are collectively referred to as $Attrib$. Any pair of records, $r$ from $\mathcal{R}$ and $s$ from $\mathcal{S}$ whose join attribute(s) have equal values, i.e., $key_r=key_s$, is output to $Q$. This is captured by the notation $\mathcal{Q} = \mathcal{R} \Join_{key_r = key_s} \mathcal{S}$ or by the shorter notation $\mathcal{Q} = \mathcal{R} \Join_{Key} \mathcal{S}$.

In case a record in $\mathcal{R}$ or $\mathcal{S}$ is output to $Q$ if and only if its join attribute(s) do have a match on the other side of the join, this is called an \emph{inner-join}. If a record in $\mathcal{R}$ ($\mathcal{S}$) is output to $Q$ whether or not it has a match on its join attribute(s) in $\mathcal{S}$ ($\mathcal{R}$), this is called a \emph{left-outer-join} (\emph{right-outer-join}). If any record in $\mathcal{R}$ or $\mathcal{S}$ is output to $Q$ even if its join attribute(s) do not have a match on the other side of the join, this is called a \emph{full-outer-join}. \Figure~\ref{fig:main_join_example_inner_and_outer} shows an example of joining the two relations, $\mathcal{R}$ and $\mathcal{S}$, shown in \Figure~\ref{fig:main_join_example_inner_and_outer}(a). The results of the inner, left-outer, right-outer, and full-outer-joins are shown in \Figure~\ref{fig:main_join_example_inner_and_outer}(b) through \Figure~\ref{fig:main_join_example_inner_and_outer}(e), respectively.

In case $\mathcal{R}$ and $\mathcal{S}$ are the same table, this is called \emph{self-join} \cite{Hursch1989relational}. A table can be self-joined using different join attributes, $Key_1$ and $Key_2$ on the two sides of the join, captured by the notation $\mathcal{R} \Join_{key_{1_{r_1}} \theta key_{2_{r_2}}} \mathcal{R}$, where $\theta$ is the function that has to hold between the $Key_1$ and $Key_2$ attribute(s) of records $r_1$ and $r_2$ respectively for the pair $r_1$-$r_2$ to belong to $\mathcal{Q}$. If the same join attributes are used on both sides of the self-join, i.e., $Key1 = Key2 = Key$, and $\theta$ is the equality function, i.e., the join condition is $key_{r_1} = key_{r_2}$, then this is an inner-join by definition, and is captured by the notation $\mathcal{R} \Join_{key_{r_1} = key_{r_2}} \mathcal{R}$, or $\mathcal{R} \Join_{Key} \mathcal{R}$ for short. We call this special case \emph{natural self-join}. The $\mathcal{Q}$ produced by a natural self-join contains duplicate pairs of records, but in reverse order. To eliminate the redundancy in $\mathcal{Q}$, we drop the join completeness: (a) a pair $r-r$ should be in $\mathcal{Q}$ exactly once, and (b) a pair $r_2$-$r_1$ should not be in $\mathcal{Q}$, if $r_1 \neq r_2$ and $r_1$-$r_2$ is in $\mathcal{Q}$.

\subsection{Popular Distributed-Processing Frameworks}
\label{sec:mapreduce}

MapReduce~\cite{dean2008mapreduce} is a popular framework with built-in fault tolerance. It allows developers, with minimal effort, to scale data-processing algorithms to shared-nothing clusters, as long as the algorithms can be expressed in terms of the functional programing primitives \textit{mapRec} and \textit{reduceRecs}.

\begin{flalign*}
& mapRec: \langle key_1, value_1 \rangle \rightarrow \langle key_2, value_2 \rangle^* &\\
& reduceRecs : \langle key_2, value_2^* \rangle \rightarrow  value_3^* &
\end{flalign*}

The input dataset is processed using \emph{executors} that are orchestrated by the \emph{driver} machine. Each record in the dataset is represented as a tuple, $\langle key_1, value_1 \rangle$. Initially, the dataset is partitioned among the \textit{mappers} that execute the map operation. Each mapper applies \textit{mapRec} to each input record to produce a potentially-empty list of the form $\langle key_2, value_2 \rangle^*$. Then, the \textit{shufflers} group the output of the mappers by the key. Next, each \textit{reducer} is fed a tuple of the form $\langle key_2, value_2^* \rangle$, where $value_2^*$, the reduce\_value\_list, contains all the $value_2$'s that were output by any mapper with the $key_2$. Each reducer applies \textit{reduceRecs} on the $\langle key_2, value_2^* \rangle$ tuple to produce a potentially-empty list of the form $value_3^*$. Any MapReduce job can be expressed as the lambda expression below.

\begin{flalign*}
& MapReduce(dataset): & \\
& \quad dataset.map(mapRec).groupByKey.reduce(reduceRecs)
\end{flalign*}

In addition to $key_2$, the mapper can optionally output tuples by a secondary key. The reduce\_value\_list would also be sorted by the secondary key in that case\footnote{Secondary keys are not supported by the public version of MapReduce, Hadoop \cite{Hadoop}. Two ways to support secondary keys were proposed in \cite{lin2010data}. The first entails loading the entire reduce\_value\_list in the reducer memory, and the second entails rewriting the shuffler. The second solution is more scalable but incurs higher engineering cost.}. Partial reducing can happen at the mappers, which is known as \emph{combining} to reduce the network load. For more flexibility, the MapReduce framework allows for loading external data when mapping or reducing. However, to preserve the determinism and the purity of the \textit{mapRec} and \textit{reduceRecs} functions, loading is restricted to the beginning of each operation. 
The Spark framework \cite{zaharia2010spark} is currently the \textit{de facto} industry standard for distributing data processing on shared-nothing clusters. Spark offers the functionality of MapReduce\footnote{While Spark uses the same nomenclature of MapReduce, the MapReduce map function is called \emph{flatMap} in Spark. We use the MapReduce notation, introduced in \cite{dean2008mapreduce}.}, as well as convenience utilities that are not part of MapReduce, but can be built using MapReduce. One example is performing tree-like aggregation of all the records in a dataset \cite{lammel2008google}. The \emph{treeAggregate} operation can be implemented using a series of MapReduce stages, where each stage aggregates the records in a set of data partitions, and the aggregates are then aggregated further in the next stage, and so on, until the final aggregate is collected at the driver.

Spark supports other functionalities that cannot be implemented using the map and reduce operations. One example is performing hash-like lookups on record keys over the network. These Spark-specific functionalities perform well under the assumption the dataset partitions fit in the memory of the executors.

Spark is designed as an in-memory framework. It relies on the executor memory, rather than their local disks, to store and process data. This in-memory processing is a key feature that gives Spark its speed and efficiency, but constrains the size of the records (including the reduce\_value\_lists) that can be input or output by any executor at any time. Even more, while Spark can persist dataset partitions on local disks or checkpoint them on a distributed file system, when Spark processes a data partition, it loads the entire partition in memory\footnote{The reader is referred to the Spark programming guide, \url{https://spark.apache.org/docs/latest/rdd-programming-guide.html\#rdd-persistence}.}.

\section{Related Work}
\label{sec:related_work}

We review the distributed algorithms that are generally applicable to equi-joins, and those targeting Small-Large outer-joins. We only review equi-joins in homogenous shared-nothing systems, where the processing powers of the executors are comparable. We neither review the algorithms devised for heterogeneous architectures (e.g., \cite{tian2016building}) nor in streaming systems (e.g., \cite{das2003approximate,gulisano2016scalejoin,lin2015scalable}), nor the approximate equi-join algorithms (e.g., \cite{quoc2018approxjoin}).

\subsection{General Distributed Equi-Joins}

The work most relevant to ours is that of distributing equi-joins on MapReduce \cite{dean2008mapreduce} and on general shared-nothing \cite{stonebraker1986case} architectures.

\subsubsection{MapReduce Equi-Joins}

The seminal work in \cite{blanas2010comparison} explained \emph{Map-Side} join (a.k.a., \emph{Broadcast} or \emph{Duplication} join in the distributed-architecture community), which is discussed in more detail in \Section~\ref{sec:broadcast}. This work also extended the Sort-Merge join \cite{blasgen1977storage} to the MapReduce framework. It also proposed using secondary keys to alleviate the bottleneck of loading the records from both joined relations in memory. This Sort-Merge join is considered a \emph{Reduce-Side} join (a.k.a., \emph{Shuffle} join in the distributed-architecture community). However, the work in \cite{blanas2010comparison} overlooks the skew in key popularity.

The SAND Join \cite{atta2011sand}, a rediscovery of \cite{wolf1993parallel}, extends the Reduce-Side joins by partitioning data across reducers based on a sample. The work in \cite{okcan2011processing} improves this key-range partitioning using quantile computation. It extends key-range partitioning to combinations of keys in the join results, and maps combinations of keys to specific reducers to achieve load-balancing. The algorithms in \cite{vitorovic2016load,gavagsaz2019load} use sampling to build the cross--relation histogram, and balance the load between the executors based on the sizes of the relations and the estimated size of the join results.

There are major drawbacks with key-range-division approaches \cite{wolf1993parallel,atta2011sand,okcan2011processing,vitorovic2016load,gavagsaz2019load}. In the pre-join step, the key-range computation and communication over the network to the mappers incurs significant computational \cite{chen2020survey} and communication cost \cite{yi2013optimal}. The computational and network bottlenecks are more pronounced when computing key-ranges for combinations of attributes \cite{okcan2011processing}\footnote{This problem was the focus of \cite{alway2016constructing} in the context of estimating join results size \cite{sun1993instant,wang2003containment,vengerov2015join,chen2017two}.}. These approaches implicitly assume that the keys are distributed evenly within each key range, which is rarely true for highly-diverse key spaces, highly-skewed data, or practical key-ranges. 

The multi-executor-per-key Shuffle-Join algorithms in \cite{afrati2010optimizing,beame2014skew} splits the processing of each hot key among multiple executors. The executors that process each key are organized in a grid whose dimensions, length and width, are proportional to the frequencies of the key on $\mathcal{R}$ and $\mathcal{S}$, respectively. Each record from $\mathcal{R}$ is sent to all the executors of a random grid row, and each record from $\mathcal{S}$ is sent to all the executors of a random grid column. Each executor then outputs the cartesian product of all pairs it receives.

\begin{figure*}[!ht]
\centering
\includegraphics[width = \bigfigwidth]{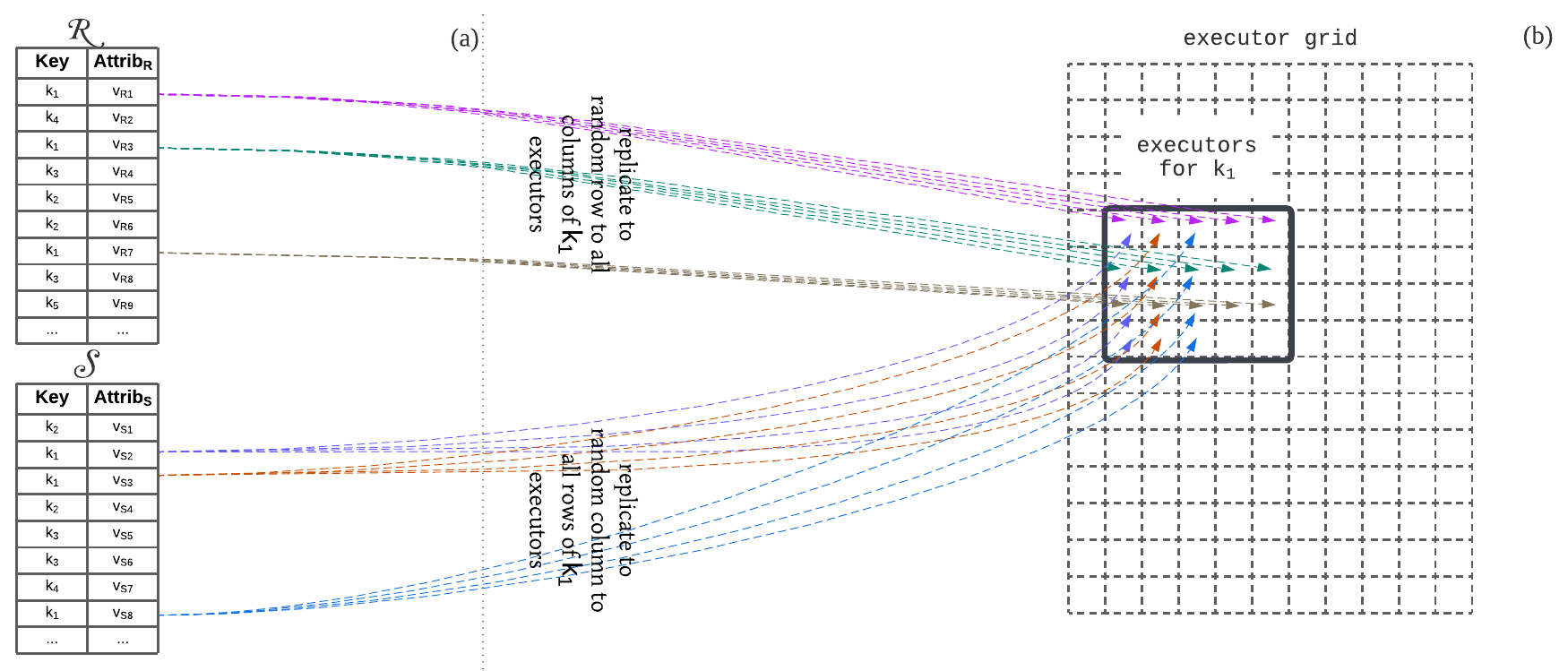}
\caption{A schematic example for joining a specific hot key, $k_1$, using the multi-executor-per-key Shuffle-Join algorithms \protect\cite{afrati2010optimizing,okcan2011processing,beame2014skew,li2018submodularity}. (a) The $\mathcal{R}$ and $\mathcal{S}$ relations to be joined. (b) The records with key $k_1$ replicated among the rows and columns of the grid of executors assigned to $k_1$.
}
\label{fig:grid_schematic}
\end{figure*}

This is illustrated in \Figure~\ref{fig:grid_schematic} for an example key, $k_1$. \Figure~\ref{fig:grid_schematic}(a) shows the $\mathcal{R}$ and $\mathcal{S}$ relations to be joined. Figure~\ref{fig:grid_schematic}(b) shows each $k_1$ $\mathcal{R}$ record assigned to a specific random grid row, and replicated to all grid columns. On the other hand, each $k_1$ $\mathcal{S}$ record is assigned a random grid column, and replicated to all grid rows. This ensures each pair of $k_1$ records, one from $\mathcal{R}$ and another from $\mathcal{S}$, meet at exactly one grid executor. Each $k_1$ grid executor outputs each assigned record from $\mathcal{R}$ with all assigned records from $\mathcal{S}$. The aggregate output of all the $k_1$ grid executors comprise the join results of $\mathcal{R}$ and $\mathcal{S}$ on key $k_1$.

This gird approach was modified into a triangle for natural self-joins in the context of record deduplication \cite{chu2017distributed} to reduce roughly half the processing cost. If \Figure~\ref{fig:grid_schematic} was to illustrate a self-join, both $\mathcal{R}$ and $\mathcal{S}$ in \Figure~\ref{fig:basic_multistage_join_tree}(a) would be the same relation. The executor grid in \Figure~\ref{fig:grid_schematic}(b) would be an upper triangular matrix, and the lower triangle cells would not be assigned to executors.

ExpVar-Join \cite{li2018submodularity} goes a step further by trying to find the optimal grid dimensions (or triangle side length for natural self-joins) for each hot key that would balance the load on all executors. The algorithm executes a loop, and in each iteration it computes the expected runtime and variance in the executor load, and hence the name of the algorithm. The search loop is executed until the expected runtime in the last $n$ iterations stabilizes (varies by less than $1\%$), where $n$ is the number of executors. In each iteration, the algorithm finds the hot key dimension that would reduce the load variance the most if assigned an extra executor. The estimation is done using a greedy approach that has near-optimality guarantees assuming (a) a custom shuffle is implemented, and (b) each grid dimension for each key can grow until it is within a factor of $O(\lambda)$ of the key's frequency, where $\lambda$ is the relative speed of the network vs. a local disk. The runtime is estimated by building a linear regression model that needs to be executed in a pre-processing phase on a sample of the data.

Building a linear regression model on a sample of join runs to estimate the runtime has some drawbacks. The model training time is significant, and may be slower than the join itself, given that several sample joins have to be computed to build the model. The model used by ExpVar-Join is a piecewise linear regression model. This results in training multiple models. The boundaries between the models are hand-picked, which may result in inaccurate estimates of the runtime. A better approach would have been to use a multivariate adaptive regression spline (MARS) model that automatically determines the segments based on the data.

\subsubsection{Shared-Nothing Equi-Joins} 

The algorithm in \cite{hassan2009efficient} is similar to that in \cite{okcan2011processing} discussed above. It allows for grouping the records into evenly distributed groups that are executed on individual executors. However, it suffers from the same hefty overhead.

Map-Reduce-Merge \cite{yang2007map} extends MapReduce with a \emph{merge} function to facilitate expressing the join operation. Map-Join-Reduce \cite{jiang2010map} adds a join phase between the map and the reduce phases. These extensions cannot leverage the MapReduce built-in fault tolerance, and are implemented using complicated custom logic.

The Partial Redistribution \& Partial Duplication (PRPD) algorithm in \cite{xu2008handling} is a hybrid between hash-based \cite{kitsuregawa1983application,dewitt1990bricker} (as opposed to Sort-Merge \cite{graefe1994sort, balkesen2013main, barthels2017distributed}) and duplication-based join algorithms \cite{dewitt1992practical}. The PRPD algorithm collects $\kappa_{\mathcal{R}}$, and $\kappa_{\mathcal{S}}$, the hot (a.k.a. high-frequency) keys in $\mathcal{R}$ and $\mathcal{S}$, respectively. For the correctness of the algorithm, a key that is hot in both $\mathcal{R}$ and $\mathcal{S}$ is only assigned to either $\kappa_{\mathcal{R}}$ or to $\kappa_{\mathcal{S}}$. PRPD splits $\mathcal{S}$ into (a) $\mathcal{S}_{high}$ with keys in $\kappa_{\mathcal{S}}$, whose records are not distributed to the executors, (b) $\mathcal{S}_{dup}$ with keys in $\kappa_{\mathcal{R}}$, whose records are broadcasted to all the executors containing $\mathcal{R}_{high}$ records, and (c) $\mathcal{S}_{hash}$ the remaining records that are distributed among executors using hashing. $\mathcal{R}$ is split similarly into $\mathcal{R}_{high}$, $\mathcal{R}_{dup}$, and $\mathcal{R}_{hash}$. $\mathcal{S}_{hash}$ is joined with $\mathcal{R}_{hash}$ on the executors they were hashed into, and $\mathcal{S}_{high}$ ($\mathcal{R}_{high}$) is joined with $\mathcal{R}_{dup}$ ($\mathcal{S}_{dup}$) on the executors containing $\mathcal{S}_{high}$ ($\mathcal{R}_{high}$). The union of the three joins comprise the final join results\footnote{The unioning of datasets in a distributed system is typically executed by unioning the sets of files containing them, where each file is specified by both the executor it resides on, and the path to the file on that matchine.}.

Track-Join \cite{polychroniou2014track} aims at minimizing the network load using a framework similar to PRPD. For any given key, Track-Join migrates its records from one relation to a few executors \cite{polychroniou2018distributed}, and \emph{selectively broadcasts}, i.e., multicasts, the records from the other relation to these executors. This record migration process is expensive. Moreover, identifying the executors on which the join takes place is done in a preprocessing \emph{track} phase, which is a separate distributed join that is sensitive to skew in key popularity.

The PRPD algorithm is developed further into the SkewJoin algorithm in \cite{bruno2014advanced}. Three flavors of SkewJoin are proposed. The Broadcast-SkewJoin (BSJ) is the same as PRPD. On the other end of the spectrum, Full-SkewJoin (FSJ) distributes $\mathcal{S}_{high}$ and $\mathcal{R}_{high}$ in a round-robin fashion on the executors using a specialized partitioner. This offers better load balancing at the cost of higher distribution overhead. The Hybrid-SkewJoin (HSJ) is a middle ground that retains the original data partitioning on the non-skewed side.

The PRPD algorithm is extended by \cite{cheng2014robust} to use distributed lookup servers\footnote{Utilization of lookup servers has been proposed before by the same research group \cite{cheng2013qbdj}.}. However, the algorithm in \cite{cheng2014robust} splits $\mathcal{S}$ into $\mathcal{S}_{hash}$ and $\mathcal{S}_{high}$. $\mathcal{S}_{hash}$, along with the distinct keys of $\mathcal{S}_{high}$, are hashed into the executors. All $\mathcal{R}$ is hashed into the executors. On each executor, $e_{i}$, $\mathcal{R}_i$, the partition of $\mathcal{R}$ on $e_{i}$, is joined with $\mathcal{S}_{hash_i}$, and with $\mathcal{S}_{high_i}$. The results of joining $\mathcal{R}_i$ with $\mathcal{S}_{hash_i}$ are reported as part of the final results. The results of joining $\mathcal{R}_i$ with $\mathcal{S}_{high_i}$ are augmented with the right $\mathcal{S}$ records using the distributed lookup servers.

Flow-Join \cite{rodiger2016flow} is similar to PRPD, but it detects the hot keys while performing the cross-network join. Flow-Join utilizes (a) \emph{Remote Direct Memory Access} (RDMA) \cite{li2016accelerating} over high-speed network \cite{rodiger2015high, galakatos2016end, salama2017rethinking} and (b) work-stealing across cores and Non-Uniform Memory Access (NUMA) \cite{rodiger2015high} regions for local processing on rack-scale systems.

\subsubsection{Comments on the General Distributed Equi-Joins}

MapReduce is more restrictive than Spark that supports distributed lookups, for example. Spark is more restrictive than the general shared-nothing architecture that supports multicasting, for example. An algorithm proposed for a more restrictive environment can be adopted for a less restrictive one, but the opposite is not true.

All the above algorithms that mix Broadcast-Joins and Shuffle-Joins \cite{xu2008handling, polychroniou2014track, polychroniou2018distributed, bruno2014advanced, cheng2014robust, rodiger2016flow} do not extend smoothly to the outer-join variants. Mixing these two join techniques in the PRPD fashion results in having \emph{dangling tuples} (tuples that do not join on the other side) of the same key distributed across the network. Deduplicating and counting dangling tuples across the network is a non-trivial operation, and was only discussed in \cite{bruno2014advanced, rodiger2016flow}.
 
The \emph{single-executor-per-key} Shuffle-Join (Hash or Sort-Merge) algorithms that assume the Cartesian product of the records of any given key is computed using one executor fail to handle highly skewed data, where one key can bottleneck an entire executor. Consequently, all the other executors remain under-utilized while waiting for the bottleneck executor to finish \cite{suri2011counting}.

On the other hand, the \emph{multi-executor-per-key} Shuffle-Join algorithms that distributes the work of one hot key on multiple executors in \cite{afrati2010optimizing,okcan2011processing,beame2014skew,li2018submodularity}  have two major drawbacks. First, computing a decent assignment of the executors to the grids of the hot keys is fairly costly. The most optimized work in \cite{li2018submodularity} balances the load between the executors in a near-optimal way as described above and under the assumptions stated above. Estimating the variance is $\Theta(n \log(n))$ for increasing a grid dimension by one executor, where $n$ is the number of executors. This is done for all hot keys to find the best reduction in variance, which is $\Theta(\min(|\kappa_{\mathcal{R}}|_{max}$, $|\kappa_{\mathcal{S}}|_{max}))$, where $|\kappa_{\mathcal{R}}|_{max}$ and $|\kappa_{\mathcal{S}}|_{max}$ are the maximum numbers of hot keys collected from $\mathcal{R}$ and $\mathcal{S}$, respectively. Finally, there are $\Omega(n)$ iterations executed to stabilize the expected run time. Hence, the work done by the driver before assigning hot keys to executors is $\Omega(n^2 \log(n)) \times \min(|\kappa_{\mathcal{R}}|_{max}$, $|\kappa_{\mathcal{S}}|_{max})$.

The second drawback of the multi-executor-per-key approaches is that they duplicate the records across the rows and columns of a grid all at once, and they produce the output of each grid executor all at once. This puts tremendous pressure on the memory of the executors, especially in modern memory-processing frameworks, e.g., Spark. The input and output records of any given key on any grid executor has to be in memory all at once to produce the join from that executor.

The strongest drawback of all the algorithms that use Broadcast-Join is low scalability when there exist keys that are hot in both joined relations. In such cases, the broadcasted keys are hot, and their records may not fit in memory. Even if the records of the hot keys may fit in memory, the entire broadcasted relation may not fit in memory. The discussion assumes the basic Broadcast-Join implementation. The Broadcast-Join algorithm can be made more scalable by pre-splitting $\mathcal{S}$ into multiple sub-relations, $\mathcal{S}_1, \dots, \mathcal{S}_i, \dots \mathcal{S}_m$, for some $m$, such that each sub-relation, $\mathcal{S}_i$ fits in memory, running a Broadcast-Join between $\mathcal{R}$ and each sub-relation $\mathcal{S}_i$, and unioning the results of all the joins. However, the standard implementations of Broadcast-Join in major DBMSs are not known to provide this \emph{partition and broadcast} \cite{li2018submodularity} optimization by default.

\subsection{Small-Large Outer-Joins}

We now discuss the outer-joins when one relation can fit in memory and the other is orders of magnitude larger. Broadcast-Joins are the fastest-known for these \emph{Small-Large} join cases. A solution to this problem is essential for extending \FrameworkName{} to outer-joins.

For a left-outer-join, the Duplication and Efficient Redistribution (DER) algorithm \cite{xu2010new} broadcasts the small relation, $\mathcal{S}$, to all $n$ executors. On each executor, $e_{i}$, $\mathcal{S}$ is inner joined with $\mathcal{R}_i$. The joined records are output, and the ids of the \emph{unjoined} records of $\mathcal{S}$ are distributed to the $n$ executors based on their hash. If an executor receives $n$ copies of a record id, then this $\mathcal{S}$ id is \emph{unjoinable} with $\mathcal{R}$ (since it was unjoined on the $n$ executors). The unjoinable ids are inner Hash-Joined with $\mathcal{S}$ and $null$-padded. The left-outer-join result is the union of the two inner join results.

The Duplication and Direct Redistribution (DDR) algorithm \cite{cheng2017design} is similar to DER, but for the records failing the first inner join, each executor hashes the entire unjoined record, instead of only its id, over the network. Because entire records, instead of their ids, are hash-distributed, (a) the first join can be executed as an out-of-the-box left-outer-join, and (b) the second inner join is not needed. Extending DER and DDR to full-outer-joins is achieved by changing the first join to output the unjoined records on $\mathcal{R}_i$.

\section{The \MultistageJoinFullName{} Algorithm}
\label{sec:tree_join}

As a first step towards \FrameworkName{}, we first propose \MultistageJoinFullName{}, a novel algorithm that can execute over multiple stages, each corresponding to a MapReduce Job. \MultistageJoinFullName{} scales well even in the existence of keys that are hot in both joined relations. If a single-executor-per-key Shuffle-Join (e.g., Hash-Join) is executed, a bottleneck happens because a single executor has to process and output a disproportionately large number of pairs of records that have the same hot key. \MultistageJoinFullName{}, on the other hand, alleviates this bottleneck by utilizing the idle executors without doing any duplicate work, while adding minimal overhead. If a multi-executor-per-key Shuffle-Join (e.g., ExpVar-Join) is executed, a bottleneck happens because all the input and output of each grid executor have to fit in memory all at once. \MultistageJoinFullName{}, on the other hand, alleviates this bottleneck by forming the join results gradually through multiple stages. 

We start by describing the basic algorithm in \Section~\ref{sec:tree_join_basic}, we then take the first step towards load-balancing in \Section~\ref{sec:chunking}, and describe the final load-balanced algorithm in \Section~\ref{sec:load_balancing_first_iteration}. We discuss handling natural self-joins in \Section~\ref{sec:tree_self_join}. We analyze some of the algorithm parameters in \Section~\ref{sec:what_is_hot}, and establish the execution stages are very limited in number in \Section~\ref{sec:num_iterations}. 

\subsection{The Basic \MultistageJoinFullName{} Algorithm}
\label{sec:tree_join_basic}

We start by describing a basic version of the inner variant of \MultistageJoinFullName{} (\Algorithm~\ref{algo:tree_join_basic}). The outer-join variants are straightforward extensions. The algorithm starts by building a distributed \emph{joined index} (\Algorithm~\ref{algo:build_joined_index}). This is done by hashing the records of the two relations based on their keys using the same hash function into the different partitions of the joined index. Hence, the records of each key from both relations end up in the same partition.

\begin{figure}[H]
\begin{center}
\begin{minipage}[t]{\smallfigwidth}
    \centering
\begin{algorithm}[H]
\scriptsize
\caption{\small \mbox{\textit{\MultistageJoinCodeName{}Basic}}($\mathcal{R}$, $\mathcal{S}$)}
\label{algo:tree_join_basic}
\begin{algorithmic}[1]
\INPUT Two relations to be joined.
\OUTPUT The join results.

 \STATE $joined\_index$ = $buildJoinedIndex$($\mathcal{R}$, $\mathcal{S}$)
 \STATE $\mathcal{Q}$ = empty Dataset
 \WHILE{$joined\_index$.nonEmpty}
  \STATE $\langle partial\_results, new\_index \rangle$ = 
  \CONTINUELINEINDENTED $\MultistageJoinCodeName{}Iteration$($joined\_index$)
  \STATE $\mathcal{Q}$ = $\mathcal{Q} \cup partial\_results$
  \STATE $joined\_index$ = $new\_index$.randomShuffle
 \ENDWHILE
 \RETURN $\mathcal{Q}$
\end{algorithmic}
\end{algorithm}
\end{minipage}
\end{center}
\end{figure}

Each key in the joined index has two \emph{joined lists} of records, one list coming from each relation. The joined index is built by invoking the \emph{buildJoinedIndex} algorithm (\Algorithm~\ref{algo:build_joined_index}). The \emph{buildJoinedIndex} algorithm (a) invokes \mbox{\textit{map$_{buildJoinedIndex}$}} (\Algorithm~\ref{algo:map_build_joined_index}) on each record in $\mathcal{R}$ and $\mathcal{S}$ (Lines $1$ and $2$ in \Algorithm~\ref{algo:build_joined_index}), (b) unions the mapped records (Lines $3$ and $4$ in \Algorithm~\ref{algo:build_joined_index}), (c) groups by key (Line $5$ in \Algorithm~\ref{algo:build_joined_index}), and (d) reduces the results using \mbox{\textit{reduce$_{buildJoinedIndex}$}} (\Algorithm~\ref{algo:reduce_build_joined_index}) (Line $6$ in \Algorithm~\ref{algo:build_joined_index}). The \mbox{\textit{map$_{buildJoinedIndex}$}} (\Algorithm~\ref{algo:map_build_joined_index}) algorithm splits a record into its key and its remaining attributes, and attaches the \emph{relation identifier}, $id_{rel}$, ($0$ for $\mathcal{R}$ and $1$ for $\mathcal{S}$). The \mbox{\textit{reduce$_{buildJoinedIndex}$}} algorithm (\Algorithm~\ref{algo:reduce_build_joined_index}) receives a key, and a list of tuples, each containing the relation identifier and the remaining attributes of a record, and outputs the key and its joined lists of remaining attributes. The keys have their joined lists of records joined independently, and the union of their join results constitutes the join results of $\mathcal{R}$ and $\mathcal{S}$. This \emph{key-independence} observation simplifies the analysis in \Section~\ref{sec:chunking} and onwards.

\begin{minipage}[t]{.5\textwidth}

\begin{minipage}[t]{\smallfigwidth}
    \centering

\begin{algorithm}[H]
\scriptsize
\caption{\small \mbox{\textit{buildJoinedIndex}}($\mathcal{R}$, $\mathcal{S}$)}
\label{algo:build_joined_index}
\begin{algorithmic}[1]
\INPUT Two relations to be joined.
\OUTPUT A joined index mapping 
\INDENTOUTPUT each key to the pair of lists of records.

 \STATE $keyed_r$ = 
  \CONTINUELINE $\mathcal{R}$.map($rec$ $\rightarrow{}$ map$_{buildJoinedIndex}$($rec$, $id_{rel_i}$ = $0$))
 \STATE $keyed_s$ = 
  \CONTINUELINE $\mathcal{S}$.map($rec$ $\rightarrow{}$ map$_{buildJoinedIndex}$($rec$, $id_{rel_i}$ = $1$))
 \STATE $joined\_index$ = $keyed_r$
 \STATE \quad .union($keyed_s$)
 \STATE \quad .groupByKey
 \STATE \quad .reduce(reduce$_{buildJoinedIndex}$)
 \RETURN $joined\_index$
\end{algorithmic}
\end{algorithm}

\begin{algorithm}[H]
\scriptsize
\caption{\small \mbox{\textit{map$_{buildJoinedIndex}$}}($rec_i$, $id_{rel_i}$)}
\label{algo:map_build_joined_index}
\begin{algorithmic}[1]
\INPUT A record from a relation,
\INDENTINPUT and the relation-identifier of the record.
\OUTPUT A tuple of the key, the relation-identifier and
\INDENTOUTPUT the remaining attributes of $rec_i$.
\INDENTOUTPUT The output tuple is of the form
\INDENTOUTPUT $\langle key_{rec_i}, \langle id_{rel_i}, attrib_{rec_i} \rangle \rangle$.

 \STATE $key_{rec_i}$ = $getKey$($rec_i$)
 \STATE $attrib_{rec_i}$ = $getAttrib$($rec_i$)
 \RETURN $\langle key_{rec_i}, \langle id_{rel_i}, attrib_{rec_i} \rangle \rangle$
\end{algorithmic}
\end{algorithm}

\end{minipage}

\end{minipage}
\begin{minipage}[t]{.5\textwidth}

\begin{minipage}[t]{\smallfigwidth}
    \centering
\begin{algorithm}[H]
\scriptsize
\caption{\small \mbox{\textit{reduce$_{buildJoinedIndex}$}}( \newline $\langle key_{rec_i}, \langle id_{rel_i}, attrib_{rec_i} \rangle^* \rangle$ )}
\label{algo:reduce_build_joined_index}
\begin{algorithmic}[1]
\INPUT A list of records keyed by the same key. 
\INDENTINPUT Each record in the list contains the
\INDENTINPUT  relation-identifier of a record, and its
\INDENTINPUT remaining attributes.
\OUTPUT A tuple of the key, and its joined lists of records.
\INDENTOUTPUT The remaining attributes of the $\mathcal{R}$ records, and
\INDENTOUTPUT the remaining attributes of the $\mathcal{S}$ records.
\INDENTOUTPUT The output tuple is of the form
\INDENTOUTPUT $\langle key_{rec_i}, \langle attrib_{\mathcal{R}_{rec_i}}^*, attrib_{\mathcal{S}_{rec_i}}^* \rangle \rangle$.

 \STATE u$_\mathcal{R}$ = empty Buffer
 \STATE u$_\mathcal{S}$ = empty Buffer
 \FORALL{$\langle id_{rel_i}, attrib_{rec_i} \rangle \in \langle id_{rel_i}, attrib_{rec_i} \rangle^*$}
   \IF{$id_{rel_i} = 0$}
     \STATE u$_\mathcal{R}$.append($attrib_{rec_i}$)
   \ELSE
     \STATE u$_\mathcal{S}$.append($attrib_{rec_i}$)
   \ENDIF
 \ENDFOR
 \RETURN $\langle key_{rec_i}, u_\mathcal{R}, u_\mathcal{S} \rangle$
\end{algorithmic}
\end{algorithm}
\end{minipage}

\end{minipage}

\vspace{20pt} 
The keys are joined by \mbox{\textit{\MultistageJoinCodeName{}Basic}} in stages, i.e., iteratively. In each iteration, \mbox{\textit{\MultistageJoinCodeName{}Iteration}} (\Algorithm~\ref{algo:tree_join_iteration}), each partition of the joined index is processed by an executor. The executor reads the partition and distributes its records among two partitions local to the executor (Line $1$ in \Algorithm~\ref{algo:tree_join_iteration}). This process is referred to as \emph{local splitting} of the partitions. One of the two partitions have the records with cold keys, and is hence called the \emph{cold partition}, and other has the records with hot keys, the \emph{hot partition}. The \mbox{\textit{isHotKey}} (\Algorithm~\ref{algo:is_hot_key}) decides if the joined lists of a key are short enough (based on the analysis in \Section~\ref{sec:what_is_hot}) to be assigned to the cold partition. Otherwise, it is assigned to the hot partition. If the key is cold, the join results of this key are obtained by outputting the key with all pairs of records from both relations (Line $2$ in \Algorithm~\ref{algo:tree_join_iteration}) using the \mbox{\textit{map$_{getAllValuePairs}$}} algorithm (\Algorithm~\ref{algo:map_get_all_value_pairs}). However, if the record lists are long, the key belongs to the hot partition. The executor processing this partition acts as a \emph{splitter} for the key. The executor chunks each of the two joined lists into sub-lists (Line $4$ in \Algorithm~\ref{algo:tree_join_iteration}) using \mbox{\textit{map$_{chunkPairOfLists}$}} (\Algorithm~\ref{algo:map_chunk_pair_of_lists}) that calls \mbox{\textit{chunkList}} (\Algorithm~\ref{algo:chunk_list}) on each list independently. The \mbox{\textit{chunkList}} algorithm simply chunks a list of length $\ell$ into a number of sub-lists, each of length $\ell'$. Then, \mbox{\textit{\MultistageJoinCodeName{}Iteration}} outputs all pairs of sub-lists (Line $5$ in \Algorithm~\ref{algo:tree_join_iteration}) again using the \mbox{\textit{map$_{getAllValuePairs}$}} algorithm, but producing pairs of sub-lists this time, instead of pairs of records. This dataset of pairs of sub-lists produced from the hot partitions constitutes the joined index to be processed in the next iteration. In the next iteration, the keys and their pairs of joined lists are assigned to random partitions to distribute the load among the executors (Line $6$ in \Algorithm~\ref{algo:tree_join_basic}).

\begin{minipage}[t]{.5\textwidth}

\begin{minipage}[t]{\smallfigwidth}
    \centering

\begin{algorithm}[H]
\scriptsize
\caption{\small \mbox{\textit{\MultistageJoinCodeName{}Iteration}}($joined\_index$)}
\label{algo:tree_join_iteration}
\begin{algorithmic}[1]
\INPUT A joined index of two relations to be joined.
\OUTPUT Join results of some keys,
\INDENTOUTPUT and the joined index (of the remaining keys)
\INDENTOUTPUT for next iteration.

 \STATE $\langle cold\_index, hot\_index \rangle$ = 
 \CONTINUELINE $splitPartitionsLocally$(
 \CONTINUELINEINDENTED $joined\_index$, $isHotKey$)
 \STATE $partial\_results$ =
 \CONTINUELINE $cold\_index$.map(map$_{getAllValuePairs}$)
 \STATE $new\_index$ = $hot\_index$
 \STATE \quad .map(map$_{chunkPairOfLists}$)
 \STATE \quad .map(map$_{getAllValuePairs}$)
 \RETURN 
   $\langle partial\_results, new\_index \rangle$
\end{algorithmic}
\end{algorithm}

\begin{algorithm}[H]
\scriptsize
\caption{\small \mbox{\textit{map$_{getAllValuePairs}$}}( \newline $\langle key, \mathcal{L}_1, \mathcal{L}_2 \rangle$)}
\label{algo:map_get_all_value_pairs}
\begin{algorithmic}[1]
\INPUT A tuple of a key, and two generic lists, $\mathcal{L}_1$, and $\mathcal{L}_2$.
\OUTPUT A list of tuples,
\INDENTOUTPUT each tuples has $key$ and a pair from $\mathcal{L}_1$ and $\mathcal{L}_2$.
\INDENTOUTPUT All pairs from $\mathcal{L}_1$ and $\mathcal{L}_2$ are output.

 \STATE u = empty Buffer
 \FORALL{$v_i \in \mathcal{L}_1$, $v_j \in \mathcal{L}_2$}
  \STATE u.append($\langle key, v_i, v_j \rangle$)
 \ENDFOR
\RETURN u
\end{algorithmic}
\end{algorithm}
\end{minipage}

\end{minipage}
\begin{minipage}[t]{.5\textwidth}

\begin{minipage}[t]{\smallfigwidth}
    \centering

\begin{algorithm}[H]
\scriptsize
\caption{\small \mbox{\textit{isHotKey}}($\langle key, \mathcal{L}_1, \mathcal{L}_2 \rangle$)}
\label{algo:is_hot_key}
\begin{algorithmic}[1]
\INPUT A tuple of the key, and two joined lists of records.
\OUTPUT Whether the joined lists should be chunked.
\CONSTANT $\lambda$ the ratio of network to disk costs.

 \STATE $\ell_{\mathcal{R}} = |\mathcal{L}|$ // The length of $\mathcal{L}_1$.
 \STATE $\ell_{\mathcal{S}} = |\mathcal{L}|$ // The length of $\mathcal{L}_2$.
 \STATE $\ell = \sqrt{\ell_{\mathcal{R}} \times \ell_{\mathcal{S}}}$ // The effective length if $\ell_{\mathcal{R}} = \ell_{\mathcal{S}}$.
 \RETURN $\ell > \left(1 + \lambda \right)^{\frac{3}{2}}$
\end{algorithmic}
\end{algorithm}

\begin{algorithm}[H]
\scriptsize
\caption{\small \mbox{\textit{map$_{chunkPairOfLists}$}}( \newline $\langle key, \mathcal{L}_1, \mathcal{L}_2 \rangle$)}
\label{algo:map_chunk_pair_of_lists}
\begin{algorithmic}[1]
\INPUT A tuple of the key, and two lists, $\mathcal{L}_1$ and $\mathcal{L}_2$, of records.
\OUTPUT A tuple of the key, and two lists of lists.
\INDENTOUTPUT The first set of lists are the chunks of $\mathcal{L}_1$.
\INDENTOUTPUT The second set of lists are the chunks of $\mathcal{L}_2$.

 \RETURN $\langle key, \mbox{\textit{chunkList}}(\mathcal{L}_1), \mbox{\textit{chunkList}}(\mathcal{L}_2) \rangle$
\end{algorithmic}
\end{algorithm}

\begin{algorithm}[H]
\scriptsize
\caption{\small \mbox{\textit{chunkList}}($\mathcal{L}$)}
\label{algo:chunk_list}
\begin{algorithmic}[1]
\INPUT A  lists, $\mathcal{L}$, of records.
\OUTPUT A list of sub-lists of records.

 \STATE $\ell = |\mathcal{L}|$ // The length of $\mathcal{L}$.
 \STATE $\delta(\ell) = \ceilX{\sqrt[3]{\ell}}$ // The number of sub-lists.
 \STATE $\ell' = \ceilX{\ell^{\frac{2}{3}}}$ // The length of a sub-list.
 \RETURN 
   $[\mathcal{L}_0 \dots \mathcal{L}_{\ell' - 1}], 
   [\mathcal{L}_{\ell'} \dots \mathcal{L}_{2 \times \ell' - 1}],
   \dots,
   [\mathcal{L}_{(\delta(\ell) - 1) \times \ell'} \dots \mathcal{L}_{C}]$
\end{algorithmic}
\end{algorithm}
\end{minipage}

\end{minipage}

\begin{figure*}[!ht]
\centering
\includegraphics[width = \bigfigwidth]{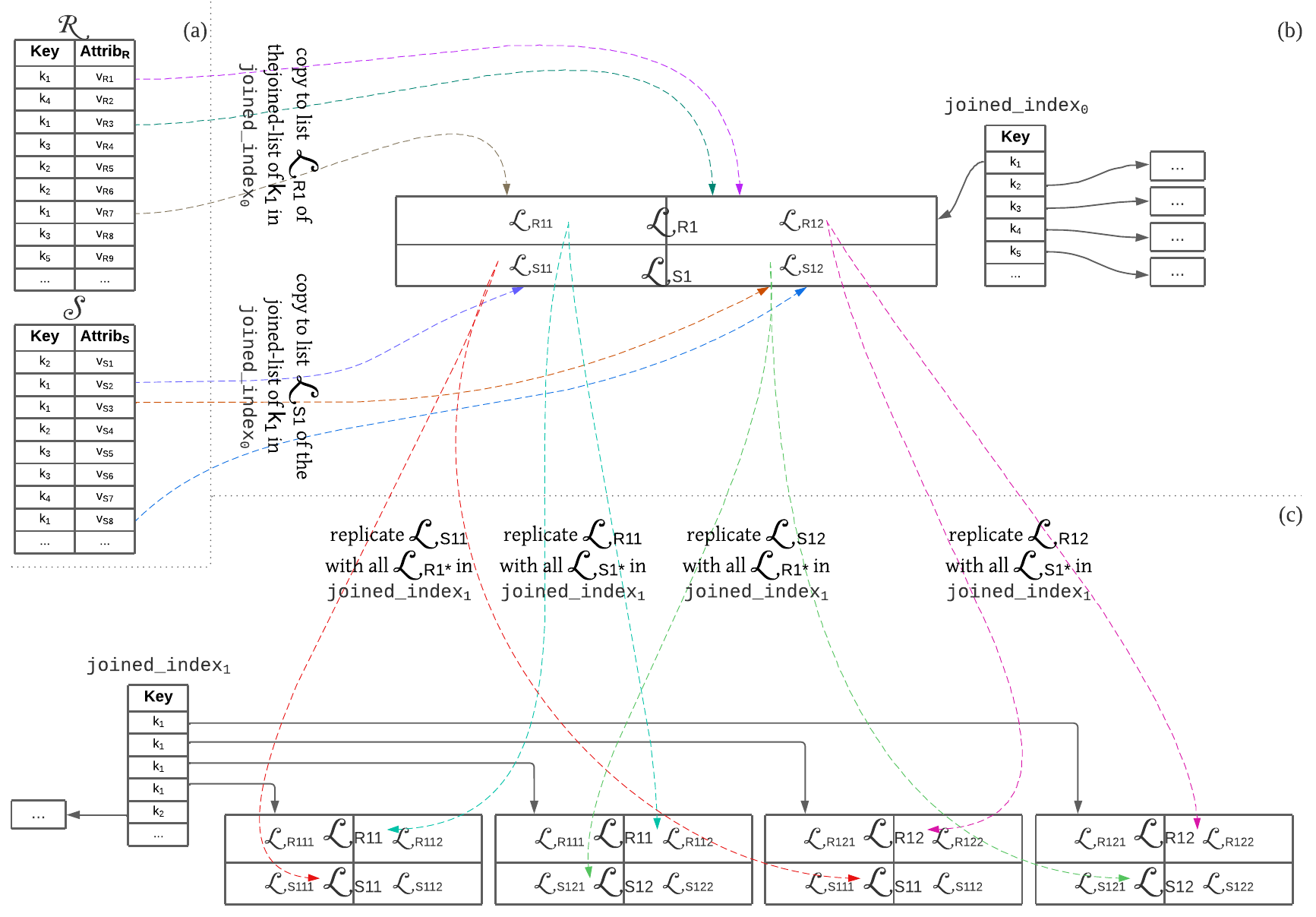}
\caption{A schematic example for joining a specific key, $k_1$, using the basic \MultistageJoinFullName{} algorithm. (a) The $\mathcal{R}$ and $\mathcal{S}$ relations to be joined. (b) The $k_1$ joined lists in the initial joined index formed using \mbox{\textit{groupByKey}}. Each of the joined lists are chunked into two sub-lists (to simplify the figure). (c) Each sub-list from $\mathcal{R}$ is produced with all the sub-lists from $\mathcal{S}$. These pairs of sublists can be processed by different executors in the next iteration. (b) and (c) together show the top two levels of the tree formed by the basic \MultistageJoinFullName{} algorithm.
}
\label{fig:basic_multistage_join_tree}
\end{figure*}

\vspace{20pt}
This tree of the execution iterations is illustrated in \Figure~\ref{fig:basic_multistage_join_tree} for an example key, $k_1$. \Figure~\ref{fig:basic_multistage_join_tree}(a) shows the $\mathcal{R}$ and $\mathcal{S}$ relations to be joined. \Figure~\ref{fig:basic_multistage_join_tree}(b) shows the basic \MultistageJoinFullName{} algorithm forming the joined lists of $k_1$ in the initial joined index, $joined\_index_0$, which is the top level of the tree. The joined lists are chunked into sub-lists. To simplify \Figure~\ref{fig:basic_multistage_join_tree}, each list is chunked into two sub-lists. That is, $\mathcal{L}_{\mathcal{R}1}$ is chunked into $\mathcal{L}_{\mathcal{R}11}$ and $\mathcal{L}_{\mathcal{R}12}$, and $\mathcal{L}_{\mathcal{S}1}$ is chunked into $\mathcal{L}_{\mathcal{S}11}$ and $\mathcal{L}_{\mathcal{S}12}$. \Figure~\ref{fig:basic_multistage_join_tree}(c) shows the basic \MultistageJoinFullName{} algorithm forming the second joined index, $joined\_index_1$, which is the next level of the tree. This is done by outputting each sub-list from $\mathcal{R}$, $\mathcal{L}_{\mathcal{R}11}$ and $\mathcal{L}_{\mathcal{R}12}$, with all the sub-lists from $\mathcal{S}$, $\mathcal{L}_{\mathcal{S}11}$ and $\mathcal{L}_{\mathcal{S}12}$. Notice these joined sub-lists have a repeated key in $joined\_index_1$ Hence, these keys can be distributed on different data partitions and hence processed by different executors for load balancing. To simplify \Figure~\ref{fig:basic_multistage_join_tree}, only the first two levels of the execution iteration tree are shown in \Figure~\ref{fig:basic_multistage_join_tree}(b) and \Figure~\ref{fig:basic_multistage_join_tree}(c), respectively.

\begin{figure*}[!ht]
\centering
\includegraphics[width = \bigfigwidth]{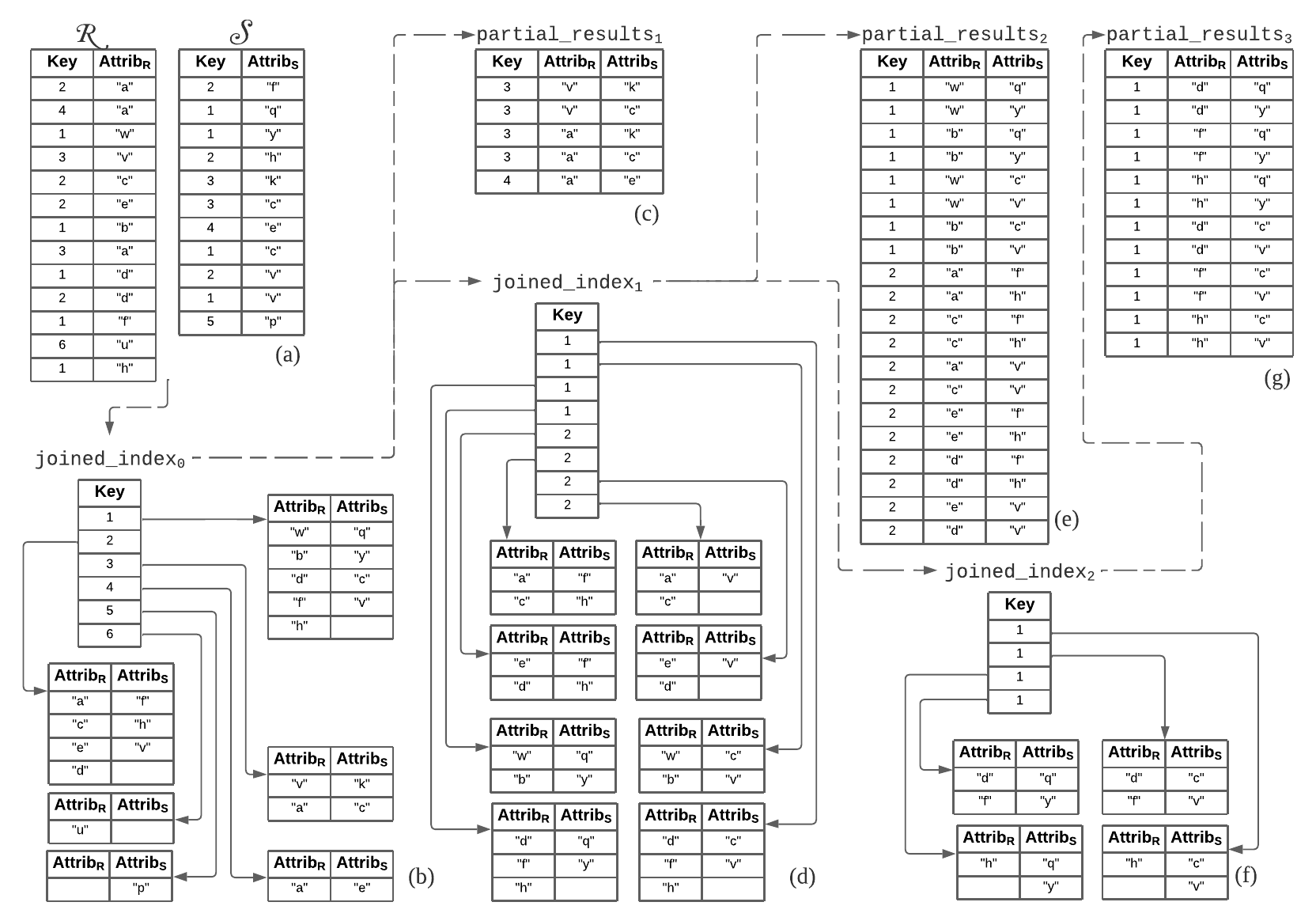}
\caption{An example of inner-joining two relations using \mbox{\textit{\MultistageJoinCodeName{}Basic}}. (a) The input relations. (b) The initial index built from the input relations in (a). The initial index is used to produce (c) the first partial results, and (d) the first joined index. The first joined index is used to produce (e) the second partial results, and (f) the second joined index. The second joined index is used to produce (g) the third partial results. The union of the partial results in (c), (e), and (g) constitutes the inner-join results.}
\label{fig:multistage_join}
\end{figure*}

\Figure~\ref{fig:multistage_join} shows an example of \mbox{\textit{\MultistageJoinCodeName{}Basic}} (\Algorithm~\ref{algo:tree_join_basic}). The \emph{buildJoinedIndex} algorithm (\Algorithm~\ref{algo:build_joined_index}) is invoked to build the initial joined index, $joined\_index_0$ (\Figure~\ref{fig:multistage_join}(b)), on the input relations in \Figure~\ref{fig:multistage_join}(a). Notice that all the keys in this initial index are distinct. Assuming that any key whose joined lists have $2$ or fewer records in both joined relations is considered cold, the joined lists of keys $3$ and $4$ are used to produce the first partial results, $partial\_results_1$ (\Figure~\ref{fig:multistage_join}(c)) and the joined lists of keys $1$ and $2$ are chunked into the first joined index, $joined\_index_1$ (\Figure~\ref{fig:multistage_join}(d)), and the joined lists of keys $5$ and $6$ are discarded, since they have records in only one relation. 

The \mbox{\textit{\MultistageJoinCodeName{}Iteration}} algorithm (\Algorithm~\ref{algo:tree_join_iteration}) uses the keys that have $2$ or fewer records in both joined relations in $joined\_index_1$ to produce the second partial results, $partial\_results_2$ (\Figure~\ref{fig:multistage_join}(e)), and uses the remaining keys to produce the second joined index, $joined\_index_2$ (\Figure~\ref{fig:multistage_join}(f)). While all the keys in $joined\_index_2$ have the same value, they are distributed among the executors for load balancing. All the keys in $joined\_index_2$ have $2$ or fewer records in both joined relations. Hence, no more iterations are executed, and $joined\_index_2$ is used to produce the third and final partial result, $partial\_results_3$ (\Figure~\ref{fig:multistage_join}(g)). The union of the three partial results constitutes the inner-join results of the input relations in \Figure~\ref{fig:multistage_join}(a).

We next analyze \mbox{\textit{\MultistageJoinCodeName{}Basic}}. We assume that $\lambda$ is the relative cost of sending data over the network vs. its IO from/to a local disk, and $\lambda > 0$.

\subsection{Chunking Lists of Hot Keys}
\label{sec:chunking}

As explained in \Section~\ref{sec:tree_join_basic}, The \mbox{\textit{\MultistageJoinCodeName{}Basic}} algorithm executes multiple iterations. During each iteration, for each hot key, its joined lists are chunked by the splitter, the executor assigned the key and its joined lists, into sub-lists to be processed by the executors of the subsequent iteration. The subsequent executors can (a) output the pairs of records in the joined sub-lists, or (b) act as further splitters of the sub-lists if they are still too long. The chunking of joined sub-lists continues until they are short enough.

Computing the optimal number of sub-lists for a list of records is crucial for load balancing between the executors and across iterations. Chunking a pair of joined lists into numerous pairs of sub-lists adds overhead to the splitter. Conversely, chunking the pair of joined lists into a few pairs of sub-lists adds overhead to the executors that handle these pairs in the subsequent iteration.

We choose to balance the load between any splitter at any iteration and the subsequent executors that process its key. For simplicity, this analysis ignores the key size, $m_{Key}$, and assumes that the record sizes, $m_{\mathcal{R}}$ and $m_{\mathcal{S}}$, for the records from ${\mathcal{R}}$ and ${\mathcal{S}}$, respectively, are roughly equal. For each key, the splitter chunks its joined lists, of lengths $\ell_{\mathcal{R}}$ and $\ell_{\mathcal{S}}$, respectively. The splitter chunks the first list into $\delta(\ell_{\mathcal{R}}) = \frac{\ell_{\mathcal{R}}}{\ell_{\mathcal{R}}^p} = \ell_{\mathcal{R}}^{1-p}$ sub-lists, each is of size $\ell_{\mathcal{R}}' = \ell_{\mathcal{R}}^p$ records, for some $p$, where $0 \leq p \leq 1$. Similarly, it chunks the second list into $\ell_{\mathcal{S}}^{1-p}$ sub-lists, each of size $\ell_{\mathcal{S}}' = \ell_{\mathcal{S}}^p$ records. The splitter outputs $(\ell_{\mathcal{R}} \times \ell_{\mathcal{S}})^{1-p}$ pairs of sub-lists, each of size $\ell_{\mathcal{R}}^p + \ell_{\mathcal{S}}^p$ records. Therefore, the splitter work is $\Theta$($(\ell_{\mathcal{R}} \times \ell_{\mathcal{S}})^{1-p} \times \max(\ell_{\mathcal{R}}^p, \ell_{\mathcal{S}}^p)$). Each subsequent executor receives a pair of sub-lists and outputs $(\ell_{\mathcal{R}} \times \ell_{\mathcal{S}})^p$ pairs of records, i.e., $2 \times (\ell_{\mathcal{R}} \times \ell_{\mathcal{S}})^p$ records. Hence, the subsequent executor work is $\Theta$($(\ell_{\mathcal{R}} \times \ell_{\mathcal{S}})^p$). To achieve decent load balancing between the splitter and each subsequent executor, the loads should be within a constant factor of each other, entailing $(\ell_{\mathcal{R}} \times \ell_{\mathcal{S}})^{1-p} \times \max(\ell_{\mathcal{R}}^p, \ell_{\mathcal{S}}^p) \approx (\ell_{\mathcal{R}} \times \ell_{\mathcal{S}})^p$. To simplify the expression, equality is assumed, yielding $\max(\ell_{\mathcal{R}}, \ell_{\mathcal{S}}) \times \min(\ell_{\mathcal{R}}, \ell_{\mathcal{S}})^{1-p} = (\ell_{\mathcal{R}} \times \ell_{\mathcal{S}})^p$. This can be simplified to $\max(\ell_{\mathcal{R}}, \ell_{\mathcal{S}})^{1-p} = \min(\ell_{\mathcal{R}}, \ell_{\mathcal{S}})^{2p-1}$. Taking the $\log$ and simplifying, yields $\frac{1-p}{2p-1} = \frac{\log(\max(\ell_{\mathcal{R}}, \ell_{\mathcal{S}}))}{\log(\min(\ell_{\mathcal{R}}, \ell_{\mathcal{S}}))}$. Therefore, $p$ can be expressed as shown in \Equation~\ref{eqn:general_p}.

\begin{equation}
\label{eqn:general_p}
p = \frac{1 + \log_{\min(\ell_{\mathcal{R}}, \ell_{\mathcal{S}}))}(\max(\ell_{\mathcal{R}}, \ell_{\mathcal{S}}))}{1 + 2 \times \log_{\min(\ell_{\mathcal{R}}, \ell_{\mathcal{S}}))}(\max(\ell_{\mathcal{R}}, \ell_{\mathcal{S}}))}
\end{equation}

The value $\log_{\min(\ell_{\mathcal{R}}, \ell_{\mathcal{S}}))}(\max(\ell_{\mathcal{R}}, \ell_{\mathcal{S}}))$ grows very slowly as the difference between $\ell_{\mathcal{R}}$ and $\ell_{\mathcal{S}}$ grows. Given that the final algorithm, \mbox{\textit{\FrameworkName{}}} (\Algorithm~\ref{algo:main_join}), employs \mbox{\textit{\MultistageJoinFullName{}}} for joining keys that are hot on both sides of the join, and  employs other joining algorithms if the key is hot on at most one side of the join, i.e., $\ell_{\mathcal{R}} \gg \ell_{\mathcal{S}}$ or $\ell_{\mathcal{R}} \ll \ell_{\mathcal{S}}$,  we simplify $p$ in \Equation~\ref{eqn:chunking_lists} below.

\begin{equation}
\label{eqn:chunking_lists}
p = \frac{2}{3}
\end{equation}

From \Equation~\ref{eqn:chunking_lists}, to achieve the load-balancing goal, $\delta(\ell)$, the number of sub-lists is $\ceilX{\sqrt[3]{\ell}}$, and each has $\ceilX{\ell^{\frac{2}{3}}}$ records, as shown in \Algorithm~\ref{algo:chunk_list}.

To illustrate using an example, assuming two joined lists, each has $10^5$ records. A {\naive} single-executor-per-key Shuffle-Join would output ${10^5}^2$ pairs of records, i.e., $2 \times 10^{10}$ IO, using one executor. Meanwhile, the \mbox{\textit{\MultistageJoinCodeName{}Basic}} splitter would chunk each list into $47$ sub-lists, and would output $47 \times 47 \times (2128 + 2128)$ records, which is $\approx 9.4 \times 10^6$ IO. These can be distributed on up to $2209$ subsequent executors. Each one of these executors would output $2128 \times 2128 \times 2 \approx 9.1 \times 10^6$ IO. Notice the only overhead for utilizing up to $2209$ executors to process the join of this key is producing $\approx 9.4 \times 10^6$ IO by the splitter, and sending this data over the network. Since $\frac{9.4 \times 10^6}{2 \times 10^{10}} < 0.05\%$, this overhead is $\approx (1 + \lambda) \times 0.05 \%$. If the lists instead had $10^4$ records, the load can be distributed on up to $484$ executors, and the overhead would only increase to $(1 + \lambda) \times 0.2 \%$. If the lists instead had $10^3$ records, the load can be distributed on up to $100$ executors, and the overhead would only increase to $(1 + \lambda) \times 1 \%$.

\subsection{The Fully-Balanced \MultistageJoinFullName{} Algorithm}
\label{sec:load_balancing_first_iteration}

The \mbox{\textit{\MultistageJoinCodeName{}Basic}} algorithm (\Section~\ref{sec:tree_join_basic}) and optimal list-chunking (\Section~\ref{sec:chunking}) achieve load balancing and good resource utilization. However, this is only true once the initial joined index has been built, and the first-iteration splitters are done splitting the joined lists of the hot keys (Lines $1$ and $4$ of \Algorithm~\ref{algo:tree_join_basic}). However, these early operations suffer from load imbalance, since early executors handle complete lists of hot keys. A first-iteration  executor that is assigned a hot key may (a) have higher splitting load than other splitters in the first-iteration, and (b) not be able to fit all the records of that hot key in its memory to perform the splitting, resulting in thrashing and hence throttling the join \cite{suri2011counting}. To achieve maximal load balancing among the executors building the initial joined index, and performing the first-iteration splitting, we identify hot keys, and we distribute the splitting of the initial joined index across a grid of partitions. The distribution of the splitting is similar to the \mbox{\textit{mapRec}} function in the multi-executor-per-key Shuffle-Join algorithms \cite{afrati2010optimizing,okcan2011processing,beame2014skew,li2018submodularity}, albeit working at the key level, instead of the executor level.

As a preprocessing step, hot keys are identified using the approximate distributed heavy-hitters algorithm in \cite{agarwal2013mergeable}. We only discuss $\mathcal{R}$, but the logic also applies equally to $\mathcal{S}$. The algorithm runs the local Space-Saving algorithms \cite{metwally2005efficient, metwally2006integrated} on individual $\mathcal{R}$ partitions, and then merges the local results. We can do the merging over the network in a tree-like manner using a priority queue of bounded size, $|\kappa_{\mathcal{R}}|_{max}$, where $|\kappa_{\mathcal{R}}|_{max}$ is the number of hot keys to be collected from $\mathcal{R}$\footnote{Quantifying $|\kappa_{\mathcal{R}}|_{max}$ and $|\kappa_{\mathcal{S}}|_{max}$ based on multiple parameters pertaining to the datasets and the hardware is discussed in \Section~\ref{sec:finding_hot_keys}.}. If $\mathcal{R}$ is already partitioned by the key, all records of any key reside on one partition, and its local frequency is equal to its global frequency. In that case, exact local counting and merging over the network with a bounded-size priority queue identifies the global hot keys exactly.

The load-balanced version of \MultistageJoinFullName{} is formalized in \Algorithm~\ref{algo:tree_join}. The hot keys in both $\mathcal{R}$ and $\mathcal{S}$ and their corresponding frequencies are identified in Lines $1$ and $2$, respectively. Their frequency maps, $\kappa_{\mathcal{R}}$ and $\kappa_{\mathcal{S}}$, are joined in Line $3$ yielding $\kappa_{\mathcal{R}\mathcal{S}}$, a map from each shared hot key to a tuple of two numbers: the key frequencies in $\mathcal{R}$ and $\mathcal{S}$, respectively. $\kappa_{\mathcal{R}\mathcal{S}}$ is then broadcasted to all executors to be cached and used later for local processing.

\begin{figure}[H]
\begin{center}
\begin{minipage}[t]{\smallfigwidth}
    \centering
\begin{algorithm}[H]
\scriptsize
\caption{\small \mbox{\textit{\MultistageJoinCodeName{}}}($\mathcal{R}$, $\mathcal{S}$)}
\label{algo:tree_join}
\begin{algorithmic}[1]
\INPUT Two relations to be joined.
\OUTPUT The join results.
\CONSTANT $|\kappa_{\mathcal{R}}|_{max}$ and $|\kappa_{\mathcal{S}}|_{max}$, 
\INDENTCONSTANT the number of hot keys to be collected 
\INDENTCONSTANT from $\mathcal{R}$ and $\mathcal{S}$, respectively.

 \STATE $\kappa_{\mathcal{R}}$ = $getHotKeys$($\mathcal{R}$, $|\kappa_{\mathcal{R}}|_{max}$)
 \STATE $\kappa_{\mathcal{S}}$ = $getHotKeys$($\mathcal{S}$, $|\kappa_{\mathcal{S}}|_{max}$)
 \STATE $\kappa_{\mathcal{R}\mathcal{S}}$ = $\kappa_{\mathcal{R}}$.join($\kappa_{\mathcal{S}}$)
 \STATE broadcast $\kappa_{\mathcal{R}\mathcal{S}}$ to all executors
 \STATE $\langle \mathcal{R}_{H}, \mathcal{R}_{C} \rangle$ = $splitPartitionsLocally$(
 \CONTINUELINE $\mathcal{R}$, $rec \rightarrow key_{rec} \in \kappa_{\mathcal{R}\mathcal{S}}.keys$)
 \STATE $\langle \mathcal{S}_{H}, \mathcal{S}_{C} \rangle$ = $splitPartitionsLocally$(
 \CONTINUELINE $\mathcal{S}$, $rec \rightarrow key_{rec} \in \kappa_{\mathcal{R}\mathcal{S}}.keys$)
 \STATE $joined\_index_{C}$ = $buildJoinedIndex$($\mathcal{R}_{C}$, $\mathcal{S}_{C}$)
 \STATE $unraveled_{\mathcal{R}_{H}}$ = $\mathcal{R}_{H}$.map(
 \CONTINUELINE $rec$ $\rightarrow$ map$_{unravel}$($rec$, $swap$ = false))
 \STATE $unraveled_{\mathcal{S}_{H}}$ = $\mathcal{S}_{H}$.map(
 \CONTINUELINE $rec$ $\rightarrow$ map$_{unravel}$($rec$, $swap$ = true))
 \STATE $joined\_index_{AK}$ = $buildJoinedIndex$(
 \CONTINUELINE $unraveled_{\mathcal{R}_{H}}$, $unraveled_{\mathcal{S}_{H}}$)
 \STATE $joined\_index_{H}$ = 
 \CONTINUELINE $joined\_index_{AK}$.map(map$_{stripKeyPadding}$)
 \STATE $joined\_index$ = 
 \CONTINUELINE $joined\_index_{H}$.union($joined\_index_{C}$)
 \STATE $\mathcal{Q}$ = empty Dataset
 \WHILE{$joined\_index$.nonEmpty}
  \STATE $\langle partial\_results, new\_index \rangle$ =
  \CONTINUELINEINDENTED $\MultistageJoinCodeName{}Iteration$($joined\_index$)
  \STATE $\mathcal{Q}$ = $\mathcal{Q} \cup partial\_results$
  \STATE $joined\_index$ = $new\_index$.randomShuffle
 \ENDWHILE
 \RETURN $\mathcal{Q}$
\end{algorithmic}
\end{algorithm}
\end{minipage}
\end{center}
\end{figure}

$\mathcal{R}$ is split into $\mathcal{R}_{H}$, a sub-relation that contains only the $\mathcal{R}$ records with keys in $\kappa_{\mathcal{R}\mathcal{S}}$, and $\mathcal{R}_{C}$, a sub-relation that contains the records whose keys are not in $\kappa_{\mathcal{R}\mathcal{S}}$ (Line $5$). $\mathcal{S}$ is split similarly into $\mathcal{S}_{H}$ and $\mathcal{S}_{C}$ (Line $6$). The cold sub-relations are used to build a cold joined index (Line $7$). The hot sub-relations, on the other hand, undergo the unravelling transformation in Lines $8$ and $9$. 

\begin{minipage}[t]{.5\textwidth}

\begin{minipage}[t]{\smallfigwidth}
    \centering
\begin{algorithm}[H]
\scriptsize
\caption{\small \mbox{\textit{map$_{unravel}$}}($rec_i$, $swap$)}
\label{algo:unravel}
\begin{algorithmic}[1]
\LOCAL $\kappa_{\mathcal{R}\mathcal{S}}$, a map from the hot keys to the frequencies of the
\INDENTINPUT keys in the relation and in the other relation,
\INPUT A record with a hot key from a relation,
\INDENTINPUT and a flag to swap the frequencies and the sub-list ids.
\OUTPUT The unraveled records of $rec_i$.
\ASSUMES The key of $rec_i$ exists in $\kappa_{\mathcal{R}\mathcal{S}}$.

 \STATE u = empty Buffer
 \STATE $key_{rec_i}$ = $getKey$($rec_i$)
 \IF{$swap$}
   \STATE  $\langle \ell_{\mathcal{S}}, \ell_{\mathcal{R}}  \rangle$ = $\kappa_{\mathcal{R}\mathcal{S}}$[$key_{rec_i}$]
 \ELSE
   \STATE  $\langle \ell_{\mathcal{R}}, \ell_{\mathcal{S}} \rangle$ = $\kappa_{\mathcal{R}\mathcal{S}}$[$key_{rec_i}$]
 \ENDIF
 \STATE $\delta_1 = \ceilX{\sqrt[3]{\ell_{\mathcal{R}}}}$
 \STATE $\delta_2 = \ceilX{\sqrt[3]{\ell_{\mathcal{S}}}}$
 \STATE $sub\_list\_id_1 = getRandom(\{0, \dots, \delta_1 - 1\})$ 
 \FORALL{$sub\_list\_id_2 \in \{0, \dots, \delta_2 - 1\}$}
   \IF{$swap$}
     \STATE $key' = \langle key_{rec_i}, sub\_list\_id_2, sub\_list\_id_1 \rangle$
   \ELSE
     \STATE $key' = \langle key_{rec_i}, sub\_list\_id_1, sub\_list\_id_2 \rangle$
   \ENDIF
   \STATE u.append($\langle key', rec_i \rangle$)
  \ENDFOR
 \RETURN u
\end{algorithmic}
\end{algorithm}
\end{minipage}

\end{minipage}
\begin{minipage}[t]{.5\textwidth}

\begin{minipage}[t]{\smallfigwidth}
    \centering

\begin{algorithm}[H]
\scriptsize
\caption{\small \mbox{\textit{map$_{stripKeyPadding}$}}($\langle \newline  \langle key, sub\_list\_id_1, sub\_list\_id_2 \rangle , \newline \mathcal{L}_1, \newline \mathcal{L}_2 \rangle$)}
\label{algo:map_strip_key_padding}
\begin{algorithmic}[1]
\INPUT A tuple of the augmented key, and two lists.
\OUTPUT A tuple of the stripped key, and the two lists.

 \RETURN $\langle key, \mathcal{L}_1, \mathcal{L}_2 \rangle$
\end{algorithmic}
\end{algorithm}

\end{minipage}
\end{minipage}

\vspace{20pt}
The unravelling transformation (formalized in \Algorithm~\ref{algo:unravel}) avoids collecting all the records of a hot key on one executor, and replaces the splitting of the joined lists of the hot key with local processing performed on the individual executors. Each partition of $\mathcal{R}$ and $\mathcal{S}$ is used to locally produce keyed records that when grouped by their keys, produce output that mimics the output of the first \mbox{\textit{\MultistageJoinCodeName{}Iteration}}. Hence, the load imbalance of the first \mbox{\textit{\MultistageJoinCodeName{}Iteration}} is evaded.

We only discuss $\mathcal{R}$, but the logic also applies equally to $\mathcal{S}$. For every record, $rec_i$, in $\mathcal{R}$, its key, $key_{rec_i}$, is extracted (Line $2$ in \Algorithm~\ref{algo:unravel}), and $\delta_1$ and $\delta_2$, the number of sub-lists from $\mathcal{R}$ and $\mathcal{S}$, respectively, are computed (Lines $3$ - $9$) based on \Equation~\ref{eqn:chunking_lists}. If the initial joined index was to be built as explained in \mbox{\textit{\MultistageJoinCodeName{}Basic}} (Line $1$ in \Algorithm~\ref{algo:tree_join_basic}), the two lists of that key would have been chunked into these many sub-lists during the splitting of the initial joined index (Line $4$ in \Algorithm~\ref{algo:tree_join_iteration}). That is, each of the $\delta_1$ sub-lists of the first list would have been produced with each of the $\delta_2$ sub-lists of the second list (Line $5$ in \Algorithm~\ref{algo:tree_join_iteration}). Knowing these numbers of sub-lists allows the executors to mimic this process without grouping the records of any hot key in a single partition. Each executor produces $rec_i$ $\delta_2$ times keyed by an \emph{augmented key}. The augmented key has the original key, and two sub-list ids: a sub-list id from the $\delta_1$ sub-lists and another from the $\delta_2$ sub-lists. The first sub-list id of the augmented key is randomly chosen from the $\delta_1$ sub-lists (Line $10$ in \Algorithm~\ref{algo:unravel}). The second sub-list id assumes all the possible values from the $\delta_2$ sub-lists (Lines $11$ - $17$ in \Algorithm~\ref{algo:unravel}). When processing a record from $\mathcal{S}$, the sub-list ids in its augmented keys are swapped, as per the different $swap$ values in Lines $8$ and $9$ in \Algorithm~\ref{algo:tree_join}.

The unraveled forms of $\mathcal{R}_{H}$ and $\mathcal{S}_{H}$ produced by \Algorithm~\ref{algo:unravel} (Lines $8$ and $9$ in \Algorithm~\ref{algo:tree_join}) have augmented keys, and are used to build a joined index, $joined\_index_{AK}$, based on these augmented keys. For each original hot key in $\kappa_{\mathcal{R}\mathcal{S}}$, each record in $\mathcal{R}$ is matched with every sub-list in $\mathcal{S}$. However, because this matching is done using $\delta_1 \times \delta_2$ augmented keys, the matching does not hot-spot on a single index partition, but is rather load-balanced between the $joined\_index_{AK}$ partitions. $joined\_index_{AK}$ mimics the hot-key entries of the index produced by the first \mbox{\textit{\MultistageJoinCodeName{}Iteration}} of \mbox{\textit{\MultistageJoinCodeName{}Basic}}. The only difference is that $joined\_index_{AK}$ contains augmented keys. These augmented keys are stripped back to the original keys (Line $11$ in \Algorithm~\ref{algo:tree_join}) using \mbox{\textit{map$_{stripKeyPadding}$}} (\Algorithm~\ref{algo:map_strip_key_padding}), a simple \mbox{\textit{mapRec}} function that when receiving an augmented key and a pair of records, only outputs the original key and the pair of records. Then, the resulting $joined\_index_{H}$ of the hot keys is unioned with the cold $joined\_index_{C}$ (Line $12$ in \Algorithm~\ref{algo:tree_join}), and the \mbox{\textit{\MultistageJoinCodeName{}}} algorithm proceeds exactly like \mbox{\textit{\MultistageJoinCodeName{}Basic}}.

Notice that load balancing based on augmented-keys assumes the records of the hot keys are randomly distributed among the input partitions of $\mathcal{R}$ and $\mathcal{S}$. If this is not the case, the records of the hot keys can be migrated and balanced among the input partitions. Detecting high variance in the hot key frequencies across partitions can be done while merging their frequencies in a tree-like manner. A similar approach has been employed in \cite{polychroniou2014track, polychroniou2018distributed} to minimize the communication cost in Track-Join, a PRPD-like algorithm.

\begin{figure*}[!ht]
\centering
\includegraphics[width = \bigfigwidth]{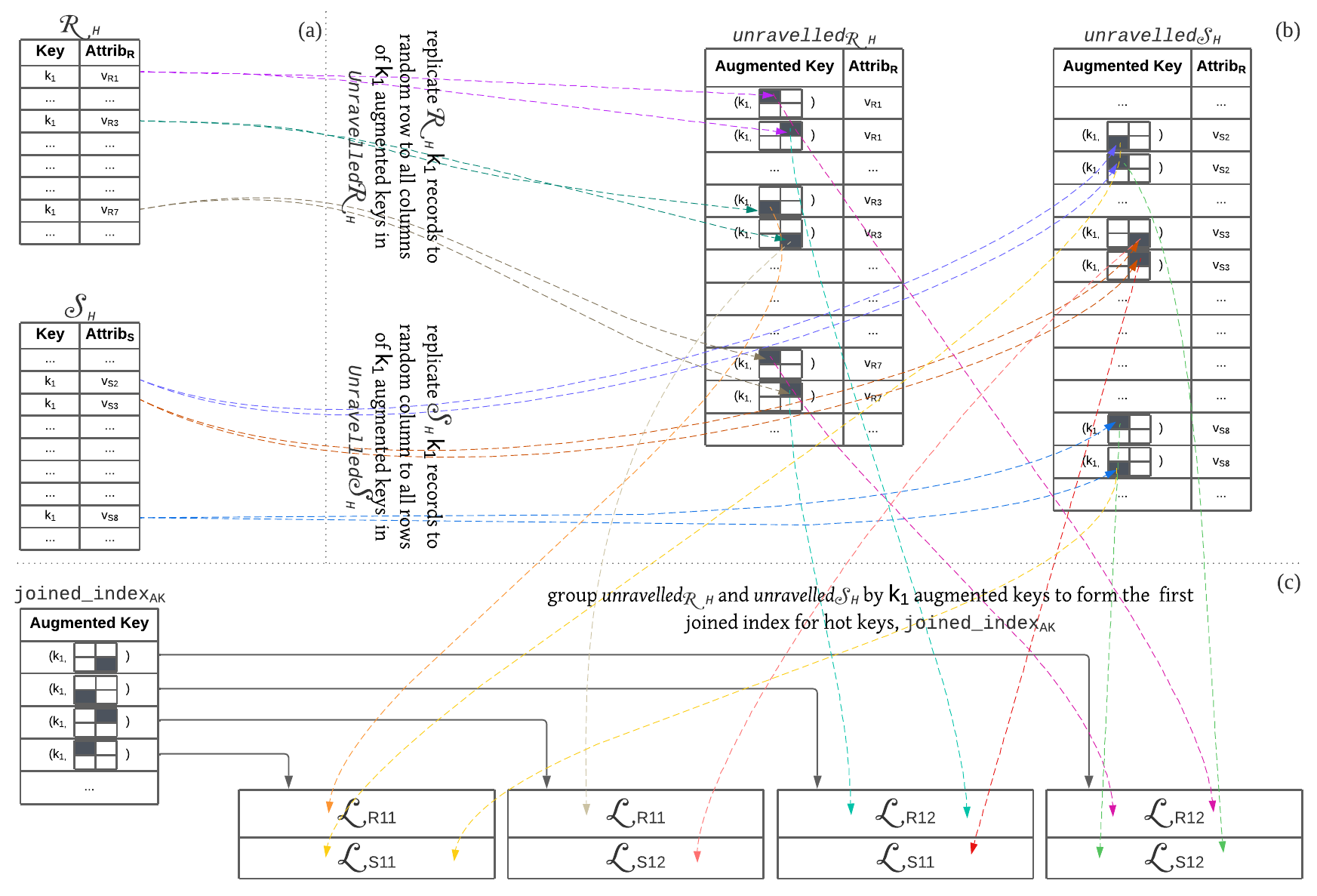}
\caption{A schematic example for joining a specific hot key, $k_1$, using the load-balanced \MultistageJoinFullName{} algorithm. (a) $\mathcal{R}_{H}$ and $\mathcal{S}_{H}$, the respective hot splits of the $\mathcal{R}$ and $\mathcal{S}$ relations to be joined. (b) The $k_1$ records replicated among multiple augmented $k_1$ keys in the unraveled forms of $\mathcal{R}_{H}$ and $\mathcal{S}_{H}$, $unraveled_{\mathcal{R}_{H}}$ and $unraveled_{\mathcal{S}_{H}}$, respectively. (c) The records of the unravelled relations grouped by the augmented keys using \mbox{\textit{groupByKey}} to form the first joined index, $joined\_index_{AK}$.
}
\label{fig:load_balanced_multistage_join_unravelling_transformation}
\end{figure*}

\Figure~\ref{fig:load_balanced_multistage_join_unravelling_transformation}  shows an example for joining the two relations using the load-balanced \MultistageJoinFullName{} algorithm. \Figure~\ref{fig:load_balanced_multistage_join_unravelling_transformation}(a) shows the splits of the joined $\mathcal{R}$ and $\mathcal{S}$ relations that contain hot keys, $\mathcal{R}_{H}$ and $\mathcal{S}_{H}$ respectively. \Figure~\ref{fig:load_balanced_multistage_join_unravelling_transformation}(b) shows the $k_1$ records replicated among multiple augmented $k_1$ keys in the unraveled forms of $\mathcal{R}_{H}$ and $\mathcal{S}_{H}$, $unraveled_{\mathcal{R}_{H}}$ and $unraveled_{\mathcal{S}_{H}}$, respectively. The $\mathcal{R}_{H}$ records are replicated among all columns of a random row of the $k_1$ augmented keys, while $\mathcal{S}_{H}$ records are replicated among all rows of a random columns of the $k_1$ augmented keys. This ensures each pair of $k_1$ records, one from $\mathcal{R_{H}}$ and another from $\mathcal{S}_{H}$, gets replicated to exactly one augmented key.  \Figure~\ref{fig:load_balanced_multistage_join_unravelling_transformation}(c) shows the records when grouped by the augmented keys using \mbox{\textit{groupByKey}} to form the first joined index, $joined\_index_{AK}$.

After the augmented keys of $k_1$ are stripped back to the original key, $k_1$, the $joined\_index_{AK}$ in \Figure~\ref{fig:load_balanced_multistage_join_unravelling_transformation}(c) resembles exactly the $joined\_index_1$ in \Figure~\ref{fig:basic_multistage_join_tree}(c). Hence, the load-balanced \MultistageJoinFullName{} algorithm achieves the same intermediary stage of the basic \MultistageJoinFullName{} without executing a \mbox{\textit{groupByKey}} on the hot keys, which enables it to evade the bottleneck of splitting the initial hot-key joined lists.

\begin{figure*}[!ht]
\centering
\includegraphics[width = \bigfigwidth]{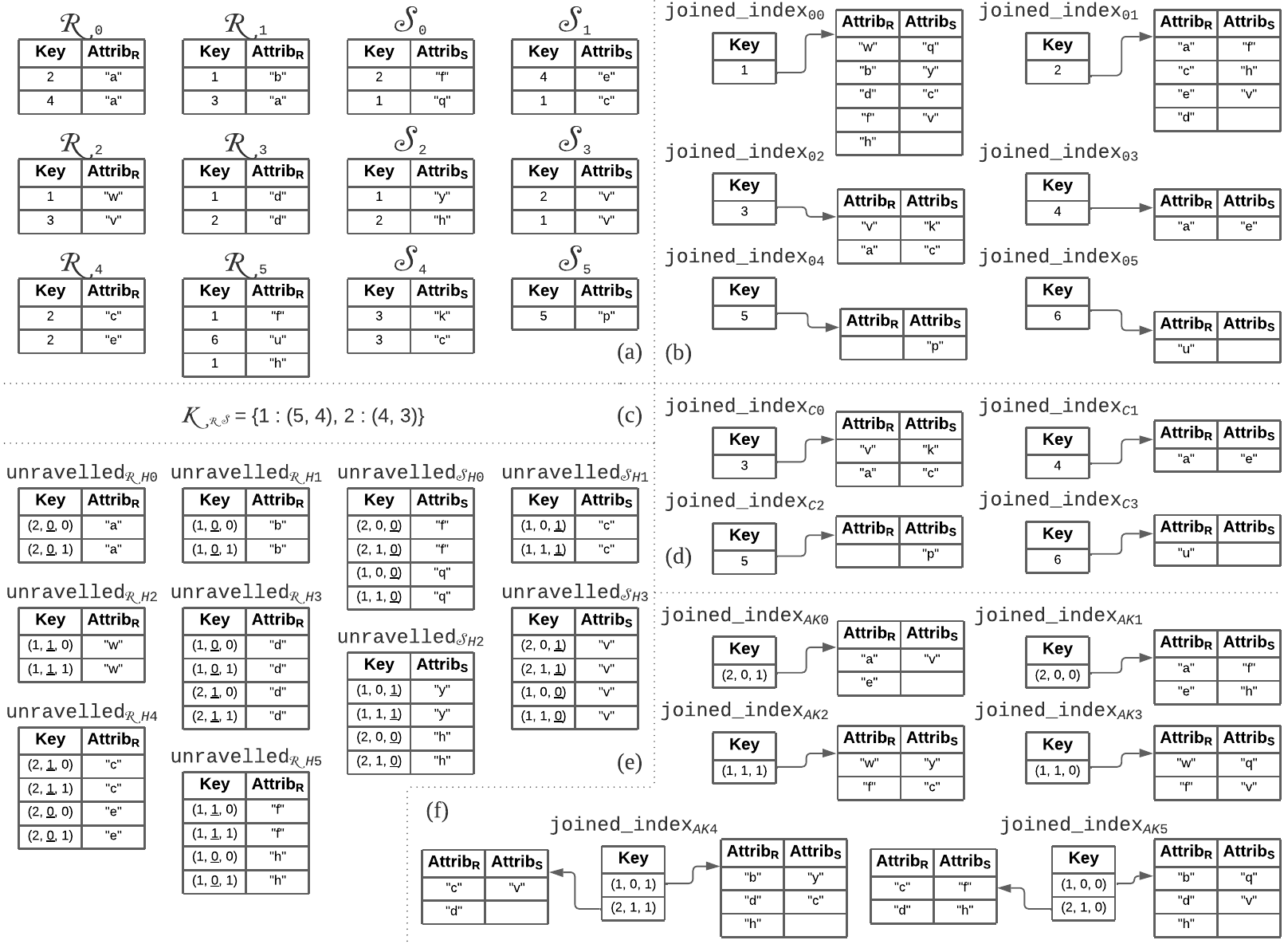}
\caption{An example showing the difference in load balancing in the first iteration between \mbox{\textit{\MultistageJoinCodeName{}Basic}} and \mbox{\textit{\MultistageJoinCodeName{}}} when joining the $\mathcal{R}$ and $\mathcal{S}$ given in \Figure~\ref{fig:multistage_join}. The example assumes each dataset has $6$ partitions. (a) The partitions of the input relations. (b) The partitions of the initial index, $joined\_index_0$, built from the input relations in (a) using \mbox{\textit{\MultistageJoinCodeName{}Basic}}. (c) through (f) show the processing steps of \mbox{\textit{\MultistageJoinCodeName{}}}. (c) The $\kappa_{\mathcal{R}\mathcal{S}}$ broadcasted to all executors containing the hot-keys and their frequencies in $\mathcal{R}$ and $\mathcal{S}$. (d) The partitions of the cold joined index. (e) The partitions of $unraveled_{\mathcal{R}_{H}}$ and $unraveled_{\mathcal{S}_{H}}$. (f) The partitions of $joined\_index_{AK}$ built using $unraveled_{\mathcal{R}_{H}}$ and $unraveled_{\mathcal{S}_{H}}$.}
\label{fig:multistage_join_first_iteration}
\end{figure*}

\Figure~\ref{fig:multistage_join_first_iteration} shows an example of building the initial joined index when joining the $\mathcal{R}$ and $\mathcal{S}$ given in \Figure~\ref{fig:multistage_join} using \mbox{\textit{\MultistageJoinCodeName{}Basic}} (\Algorithm~\ref{algo:tree_join_basic}) and of \mbox{\textit{\MultistageJoinCodeName{}}} (\Algorithm~\ref{algo:tree_join}). The example assumes each dataset has $6$ partitions, with the partition number shown as a subscript. \Figure~\ref{fig:multistage_join_first_iteration}(a) shows the partitions of the input relations. \Figure~\ref{fig:multistage_join_first_iteration}(b) shows the partitions of the initial index, $joined\_index_0$ produced by \mbox{\textit{\MultistageJoinCodeName{}Basic}}. \Figure~\ref{fig:multistage_join_first_iteration}(c) through \Figure~\ref{fig:multistage_join_first_iteration}(f) show the processing steps of \mbox{\textit{\MultistageJoinCodeName{}}}. \Figure~\ref{fig:multistage_join_first_iteration}(c) shows $\kappa_{\mathcal{R}\mathcal{S}}$, the hot-keys map broadcasted to all executors. Its keys are $1$ and $2$, and its values are tuples of their frequencies in $\mathcal{R}$ and $\mathcal{S}$, respectively. Based on \Equation~\ref{eqn:chunking_lists}, each list of records for keys $1$ and $2$ will be broken into two sub-lists. $\mathcal{R}_{C}$ and $\mathcal{S}_{C}$ are not shown for brevity, but the partitions of the initial cold joined index produced by them is shown in \Figure~\ref{fig:multistage_join_first_iteration}(d). \Figure~\ref{fig:multistage_join_first_iteration}(e) shows the partitions of $unraveled_{\mathcal{R}_{H}}$ and $unraveled_{\mathcal{S}_{H}}$, with the randomly generated sub-list ids underlined. $unraveled_{\mathcal{R}_{H}}$ and $unraveled_{\mathcal{S}_{H}}$ are used to produce $joined\_index_{AK}$, whose partitions are shown in \Figure~\ref{fig:multistage_join_first_iteration}(f).

\Figure~\ref{fig:multistage_join_first_iteration}(b) and \Figure~\ref{fig:multistage_join_first_iteration}(f) contrasts the load balancing of \mbox{\textit{\MultistageJoinCodeName{}Basic}} and \mbox{\textit{\MultistageJoinCodeName{}}}. In the case of \mbox{\textit{\MultistageJoinCodeName{}Basic}} (\Figure~\ref{fig:multistage_join_first_iteration}(b)), the partition with the heaviest load, $joined\_index_{00}$, produces $20$ pairs of records, while that with the lightest load, $joined\_index_{04}$ and $joined\_index_{05}$, produces $0$ pairs of records.  In the case of \mbox{\textit{\MultistageJoinCodeName{}}}, we assume each partition of the initial cold joined index is unioned with the corresponding partition of the initial hot joined index, produced by stripping the keys of $joined\_index_{AK}$ in \Figure~\ref{fig:multistage_join_first_iteration}(f). The partition with the heaviest load, $joined\_index_{05}$, the union of $joined\_index_{AK5}$ and $joined\_index_{H5}$, produces $10$ pairs of records. On the other hand, the partitions with the lightest load, $joined\_index_{02}$ and $joined\_index_{03}$, the union of $joined\_index_{AK2}$ and $joined\_index_{H2}$, and the union of $joined\_index_{AK3}$ and $joined\_index_{H3}$, respectively, produce $4$ pairs of records each.

\subsection{Executing Natural Self-Joins}
\label{sec:tree_self_join}

While \MultistageJoinFullName{} can execute self-joins in the same way it handles normal equi-joins, there are special optimizations that can be introduced in that case. We discuss an optimization that results in reducing the processing and IO by roughly half. 

Because we are processing a single relation, this relation is split into its cold and hot sub-relations. Hence, in \Algorithm~\ref{algo:tree_join}, Lines $2$, $6$, and $9$ are omitted, and $\kappa_{\mathcal{R}\mathcal{S}}$ = $\kappa_{\mathcal{R}}$. In Line $7$, instead of building a cold joined index, where each key in the index has two joined lists of records, a cold index is built, where each key has a single list of records. The join results of a cold key in the cold index are produced by outputting the key with all pairs of records where the first record is before the second in the record list. This corresponds to producing the upper triangle in a matrix whose rows and columns represent the records of the key, and each cell represents the pair of corresponding records.

For a hot key that has a long list of records, the unravelling (Line $8$) happens in a way simpler than \Algorithm~\ref{algo:unravel} since there is no $swap$ parameter. The sub-list ids are picked from the range $[0, \dots, \delta - 1]$, where $\delta = \ceilX{\sqrt[3]{\ell}}$, and $\ell$ is the length of the records in the list of this key. To realize the processing and IO savings, In Lines $12$ - $17$ in \Algorithm~\ref{algo:unravel}, a record with an augmented key $\langle key, sub\_list\_id_1, sub\_list\_id_2\rangle$ is only output if $sub\_list\_id_1  \geq sub\_list\_id_2$. Otherwise, it is output with the reversed augmented key, $\langle key, sub\_list\_id_2, sub\_list\_id_1\rangle$.

Notice that while the processing and the IO is reduced by roughly half, the overall runtime may not drop by the same ratio. The reason is this optimization does not alleviate other overheads, such as scheduling executors, as discussed later in \Section~\ref{sec:results}. These overheads may vary among different distributed-processing frameworks.

If the schematic figure of the basic \MultistageJoinFullName{} algorithm, \Figure~\ref{fig:basic_multistage_join_tree}, was to illustrate a natural self-join, both $\mathcal{R}$ and $\mathcal{S}$ in \Figure~\ref{fig:basic_multistage_join_tree}(a) would be one and the same relation. Hence, in \Figure~\ref{fig:basic_multistage_join_tree}(b), there would only be one list, $\mathcal{L}_{\mathcal{R}1}$ indexed by $k_1$. This list is chunked into $\mathcal{L}_{\mathcal{R}11}$ and $\mathcal{L}_{\mathcal{R}12}$. In \Figure~\ref{fig:basic_multistage_join_tree}(c), there would only be the upper triangle of the sub-lists matrix, i.e., the joined sub-lists $\mathcal{L}_{\mathcal{R}11}$ with $\mathcal{L}_{\mathcal{R}11}$, $\mathcal{L}_{\mathcal{R}11}$ with $\mathcal{L}_{\mathcal{R}12}$, and $\mathcal{L}_{\mathcal{R}12}$ with $\mathcal{L}_{\mathcal{R}12}$. \Figure~\ref{fig:basic_multistage_join_tree}(c) would not contain the lower matrix combination, i.e., the joined sub-lists $\mathcal{L}_{\mathcal{R}12}$ with $\mathcal{L}_{\mathcal{R}11}$.

If the schematic figure of the load balanced \MultistageJoinFullName{} algorithm, \Figure~\ref{fig:load_balanced_multistage_join_unravelling_transformation}, was to illustrate a natural self-join, both $\mathcal{R}$ and $\mathcal{S}$ in \Figure~\ref{fig:load_balanced_multistage_join_unravelling_transformation}(a) would be one and the same relation. Hence, in \Figure~\ref{fig:load_balanced_multistage_join_unravelling_transformation}(b), the augmentations of $k_1$ would have embodied an upper triangle matrix, instead of a full matrix. In \Figure~\ref{fig:load_balanced_multistage_join_unravelling_transformation}(c), these augmented $k_1$ keys would only produce the upper triangle of the sub-lists matrix, i.e., the joined sub-lists $\mathcal{L}_{\mathcal{R}11}$ with $\mathcal{L}_{\mathcal{R}11}$, $\mathcal{L}_{\mathcal{R}11}$ with $\mathcal{L}_{\mathcal{R}12}$, and $\mathcal{L}_{\mathcal{R}12}$ with $\mathcal{L}_{\mathcal{R}12}$. \Figure~\ref{fig:load_balanced_multistage_join_unravelling_transformation}(c) would not contain the lower matrix combination, i.e., the joined sub-lists $\mathcal{L}_{\mathcal{R}12}$ with $\mathcal{L}_{\mathcal{R}11}$.

\subsection{Defining Hot Keys}
\label{sec:what_is_hot}

From the key-independence observation, keys can be considered in isolation. If the joined lists are not chunked, the time to produce the join results on $1$ executor using single-executor-per-key Shuffle-Join is $\Delta_{ShuffleJoin} \approx \Theta(\ell^2)$. Meanwhile, the end-to-end \MultistageJoinFullName{} processing time, $\Delta_{\MultistageJoinCodeName{}}$, is the time taken to chunk the joined lists, the time to randomly shuffle the pairs of sub-lists over the network, and the time taken by the $\ceilX{\ell^{\frac{2}{3}}}$ subsequent executors in the next iteration. A key is considered hot and is worth chunking if $\Delta_{ShuffleJoin} > \Delta_{\MultistageJoinCodeName{}}$ using the worst-case scenario for \MultistageJoinFullName{}, i.e.,  a single stage and exactly one splitter. Given $n$ available executors, this condition is expressed in \Relation{~\ref{ineq:hot_keys}.

\begin{equation}
\label{ineq:hot_keys}
\ell^2 > 
\left((1 + \lambda) + \ceilX{\frac{\ell^{\frac{2}{3}}}{n}}\right) \times \ell^{\frac{4}{3}}
\end{equation}

The ceiling can be relaxed while preserving the correctness of \Relation~\ref{ineq:hot_keys}. One (non-tight) value for $\ell$ that provably satisfies \Relation~\ref{ineq:hot_keys} for any $\lambda > 0$ and any $n > 1$ is $\left(1 + \lambda \right)^{\frac{3}{2}}$. This value is used in the rest of the paper to define the minimum frequency for a key to be hot.

Given the current standards of data centers, we can estimate $\lambda$ and hence, quantify the minimum frequency for a key to be hot. Transferring $1$MB over a point on a $10$Gbps fiber channel from a server to a leaf switch takes $\approx 1$ms. This includes neither serialization, switching, and queuing delays, nor protocol overhead. Reading the same amount of data from a Solid-State Disk (SSD) takes as low as $0.14$ms on modern SSDs \footnote{\url{https://www.pcmag.com/news/ssd-vs-hdd-whats-the-difference}, Updated August 26, 2022.}. Therefore, given the current network and IO latencies, the minimum frequency for a key to be hot should be within the range of $[10, 100]$ records.

\subsection{The Number of Iterations}
\label{sec:num_iterations}

Since the overhead of allocating the executors for a stage was ignored, we make the case that the number of iterations in this multistage join is very small. 

From the key-independence observation, the join is concluded when the results for the last key are computed. Assuming all other factors are equal, the last key to compute has the longest pair of joined lists, $\ell_{max}$. After the first stage, each of the subsequent executors of the next iteration inherits a pair of lists whose lengths are $\ell_{max}^{\frac{2}{3}}$. This chunking continues for $t$ times as long as the lengths exceed $\left(1 + \lambda \right)^{\frac{3}{2}}$. Hence, the following relation holds for $t$.

\begin{equation*}
  \left(1 + \lambda \right)^{\frac{3}{2}} < \overbrace{
    {{{\ell_{max}^{\frac{2}{3}}}^{\frac{2}{3}}}^{\dots}}^{\frac{2}{3}}
   }^\text{$t$ times}
\end{equation*}
 
Hence, $\left(1 + \lambda \right)^{\frac{3}{2}} < \ell_{max}^{{(\frac{2}{3})}^t}$. Raising both sides to the ${{(\frac{3}{2})}^t}$ power yields ${\left(1 + \lambda \right)^{(\frac{3}{2})}}^{t+1} < \ell_{max}$. Taking the log with base $1 + \lambda$ on both sides yields the following.

\begin{equation*}
{\left(\frac{3}{2}\right)}^{t+1}
< \log_{1 + \lambda}(\ell_{max})
\end{equation*}

Taking the log with base $\frac{3}{2}$ yields \Relation~\ref{ineq:num_iterations}.

\begin{equation}
\label{ineq:num_iterations}
t <
\log_{\frac{3}{2}}\left(\log_{1 + \lambda}(\ell_{max})\right)
-1
\end{equation}

From \Relation~\ref{ineq:num_iterations}, the number of iterations is $O$($\log(\log(\ell_{max}))$), which grows very slowly with $\ell_{max}$.

\subsection{Comparison with the State-of-the-Art Algorithms}
\label{sec:amjoin_comparison}

The only join algorithms that handle keys that are hot on both sides are the multi-executor-per-key Shuffle-Join algorithms \cite{afrati2010optimizing,okcan2011processing,beame2014skew,li2018submodularity}. These algorithms distribute the work of one hot key on multiple executors. While these algorithms achieve excellent load balancing, they put immense pressure on the memory of the executors. This high executor memory requirements makes them less scalable when compared to \MultistageJoinFullName{} that requires less executor memory. On the other hand, \MultistageJoinFullName{} may take longer to compute the join in the case of very skewed datasets, since it executes potentially multiple iterations.

\subsubsection{Algorithmic Comparison}

\Figure~\ref{fig:grid_schematic}, \Figure~\ref{fig:basic_multistage_join_tree}, and \Figure~\ref{fig:load_balanced_multistage_join_unravelling_transformation} illustrate the difference between multi-executor-per-key Shuffle-Join algorithms \cite{afrati2010optimizing,okcan2011processing,beame2014skew,li2018submodularity}, the basic \MultistageJoinFullName{} algorithm and the load-balanced \MultistageJoinFullName{} algorithm. The three figures focus on the join of a specific key, $k_1$, in the input datasets.

\Figure~\ref{fig:grid_schematic} shows the multi-executor-per-key Shuffle-Join algorithms forming a grid of executors for $k_1$. These algorithms assign each $\mathcal{R}$ record to a specific random grid row, and replicate the record to all grid columns, and assign each $\mathcal{R}$ record to a specific random grid column, and replicate the record to all grid rows. Each grid executor outputs all possible pairs of records.

\Figure~\ref{fig:basic_multistage_join_tree} shows the basic \MultistageJoinFullName{} algorithm forming the joined lists of $k_1$ in the initial joined index (the top level of the tree). Each of these lists are chunked into sub-lists. At each iteration, the basic \MultistageJoinFullName{} algorithm forms the second joined index (the next level of the tree) by outputting each sub-list from $\mathcal{R}$, with all the sub-lists from $\mathcal{S}$. This chunking can continue over multiple iterations. At the tree leaves, each executor outputs all possible pairs of records from its joined sub-lists.

\Figure~\ref{fig:load_balanced_multistage_join_unravelling_transformation} shows the load-balanced \MultistageJoinFullName{} algorithm forming the augmented keys of $k_1$ and their joined lists, and skipping the top level of the tree. Starting from the second level of the tree, it continues the execution exactly like the basic \MultistageJoinFullName{} algorithm for the rest of the join.

The multi-executor-per-key Shuffle-Join algorithms produce the join results in one MapReduce job, while the \MultistageJoinFullName{} algorithms take multiple iterations. These iterations may take longer time to finish but consume less memory at each iteration, as shown next in \Section~\ref{sec:memory_requirements}. Overall this enhances the scalability of \MultistageJoinFullName{}.

\subsubsection{Memory Requirements Analysis}
\label{sec:memory_requirements}

For simplicity, we assume the input records are distributed randomly on the input dataset partitions. and we ignore the $\lceil . \rceil$ $\lfloor . \rfloor$ functions. For the sake of discussion, we focus on $\mathcal{R}$-$\mathcal{S}$ equi-joins, but the discussion also applies equally to natural self-joins \cite{okcan2011processing,li2018submodularity}. We also assume in-memory processing, where the input and output of each MapReduce function have to fit in memory during processing. That is, for equi-joins, the records input to and output by any executor for any given key have to fit in the memory of that executor. 

We compare the \MultistageJoinFullName{} algorithms with the most recent algorithm from the multi-executor-per-key Shuffle-Join family of algorithms. ExpVar-Join \cite{li2018submodularity} tries to find the optimal grid dimensions for each hot key that would balance the load on all executors.

\paragraph{Memory Requirements for ExpVar-Join}

Given the hottest key, with frequencies $\ell_{\mathcal{R}}$ and $\ell_{\mathcal{S}}$ in $\mathcal{R}$ and $\mathcal{S}$, respectively, ExpVar-Join requires the least memory on the reducers when it distributes the join of the records of this key across all $n$ executors. Assuming for simplicity a square grid, for each record from $\mathcal{R}$ or $\mathcal{S}$, each mapper produces $\sqrt{n}$ copies of it. Hence, its memory requirement is $\max(m_{\mathcal{R}}, m_{\mathcal{S}}) \times (1 + \sqrt{n})$, where $m_{\mathcal{R}}$ and $m_{\mathcal{S}}$ are the average size of the records in $\mathcal{R}$ and $\mathcal{S}$, respectively. Each reducer receives $\frac{\ell_{\mathcal{R}}}{\sqrt{n}}$ records from $\mathcal{R}$ and $\frac{\ell_{\mathcal{S}}}{\sqrt{n}}$ records from $\mathcal{S}$, and produces $\frac{\ell_{\mathcal{R}} \ell_{\mathcal{S}}}{n}$ pairs of records. Hence, the memory requirements on each reducer is $\frac{\ell_{\mathcal{R}} m_{\mathcal{R}} + \ell_{\mathcal{S}} m_{\mathcal{S}}}{\sqrt{n}} + \frac{\ell_{\mathcal{R}} \ell_{\mathcal{S}} (m_{\mathcal{R}} + m_{\mathcal{S}})}{n}$. The ExpVar-Join algorithm would typically be memory-bottlenecked by the reducers.

\paragraph{Memory Requirements for Basic \MultistageJoinFullName{}}

For the same scenario, in the first iteration of the basic \MultistageJoinFullName{} algorithm (i.e., the iteration that consumes the most memory), the splitter of the hottest key receives $\ell_{\mathcal{R}}$ records from $\mathcal{R}$ and $\ell_{\mathcal{S}}$ records from $\mathcal{S}$, and produces $\sqrt[3]{\ell_{\mathcal{R}} \ell_{\mathcal{S}}}$ pairs of sub-lists, each has $\ell_{\mathcal{R}}^{\frac{2}{3}}$ records from $\mathcal{R}$ and $\ell_{\mathcal{S}}^{\frac{2}{3}}$  records from $\mathcal{S}$. Hence, its total memory requirements are $\ell_{\mathcal{R}} m_{\mathcal{R}} + \ell_{\mathcal{S}} m_{\mathcal{S}} + \sqrt[3]{\ell_{\mathcal{R}} \ell_{\mathcal{S}}} \times (\ell_{\mathcal{R}}^{\frac{2}{3}} m_{\mathcal{R}} + \ell_{\mathcal{S}}^{\frac{2}{3}} m_{\mathcal{S}})$.

Each first-iteration subsequent executor receiving a pair of sub-lists, one of length $\ell_{\mathcal{R}}^{\frac{2}{3}}$ from $\mathcal{R}$ and one of length $\ell_{\mathcal{S}}^{\frac{2}{3}}$ from $\mathcal{S}$. If the sub-lists are not long enough, the first-iteration subsequent executor joins the sub-lists directly. In that case, each first-iteration subsequent executor produces $(\ell_{\mathcal{R}} \ell_{\mathcal{S}})^{\frac{2}{3}}$ pairs of records. However, since this is the hottest key in the dataset, each first-iteration subsequent executor receiving a pair of sub-lists almost surely chunks the pair of sub-lists further for subsequent iterations. In that case, each first-iteration subsequent executor produces $(\ell_{\mathcal{R}} \ell_{\mathcal{S}})^{\frac{2}{9}}$ pairs of sub-lists, each has $\ell_{\mathcal{R}}^{\frac{4}{9}}$ records from $\mathcal{R}$ and $\ell_{\mathcal{S}}^{\frac{4}{9}}$  records from $\mathcal{S}$.

Hence, if the first-iteration subsequent executors produce the join results directly, the memory requirements of each such subsequent executor of the first iteration is $\ell_{\mathcal{R}}^{\frac{2}{3}} m_{\mathcal{R}} + \ell_{\mathcal{S}}^{\frac{2}{3}} m_{\mathcal{S}} + (\ell_{\mathcal{R}} \ell_{\mathcal{S}})^{\frac{2}{3}} \times (m_{\mathcal{R}} + m_{\mathcal{S}})$. In that case, the basic \MultistageJoinFullName{} algorithm would balance the memory requirements of the splitter and the first-iteration subsequent executors. If the first-iteration subsequent executors chunk their pairs of sub-lists further for subsequent iterations, the memory requirements of each such subsequent executor of the first iteration is $\ell_{\mathcal{R}}^{\frac{2}{3}} m_{\mathcal{R}} + \ell_{\mathcal{S}}^{\frac{2}{3}} m_{\mathcal{S}} + (\ell_{\mathcal{R}} \ell_{\mathcal{S}})^{\frac{2}{9}} \times (\ell_{\mathcal{R}}^{\frac{4}{9}} m_{\mathcal{R}} + \ell_{\mathcal{S}}^{\frac{4}{9}} m_{\mathcal{S}})$. In that case, the basic \MultistageJoinFullName{} algorithm would typically have higher memory requirements for the first-iteration splitter.

\paragraph{Memory Requirements for Load-Balanced \MultistageJoinFullName{}}

For the same scenario, in the first iteration of the load-balanced \MultistageJoinFullName{} algorithm, the splitter work is distributed on all the executors, and splitting does not entail a \mbox{\textit{groupByKey}}. Hence the executors memory requirements for splitting is $\max(m_{\mathcal{R}} (1 + \sqrt[3]{\ell_{\mathcal{S}}}) , m_{\mathcal{S}} (1 + \sqrt[3]{\ell_{\mathcal{R}}}))$. The first-iteration subsequent executors memory requirements are the same as the case for the basic \MultistageJoinFullName{} algorithm. That is, if the first-iteration subsequent executors produce the join results, their memory requirements are $\ell_{\mathcal{R}}^{\frac{2}{3}} m_{\mathcal{R}} + \ell_{\mathcal{S}}^{\frac{2}{3}} m_{\mathcal{S}} + (\ell_{\mathcal{R}} \ell_{\mathcal{S}})^{\frac{2}{3}} \times (m_{\mathcal{R}} + m_{\mathcal{S}})$, and if they chunk their sub-lists for subsequent iterations, their memory requirements are $\ell_{\mathcal{R}}^{\frac{2}{3}} m_{\mathcal{R}} + \ell_{\mathcal{S}}^{\frac{2}{3}} m_{\mathcal{S}} + (\ell_{\mathcal{R}} \ell_{\mathcal{S}})^{\frac{2}{9}} \times (\ell_{\mathcal{R}}^{\frac{4}{9}} m_{\mathcal{R}} + \ell_{\mathcal{S}}^{\frac{4}{9}} m_{\mathcal{S}})$. The load-balanced \MultistageJoinFullName{} algorithm would typically have higher memory requirements for the first-iteration subsequent executors.

\paragraph{Illustrative Examples}

Considering some realistic examples would illustrate the difference in memory requirements of the algorithm. For all examples, let $m_{\mathcal{R}} = m_{\mathcal{S}} = 500$B. Since we are considering the hottest key, let the subsequent executors of \MultistageJoinFullName{} chunk their sub-lists for subsequent iterations. If ${\ell_{\mathcal{R}}} = {\ell_{\mathcal{S}}} = 10^4$, and $n = 100$ executors were used, ExpVar-Join would require each reducer to have $\approx 1$GB of memory, the basic \MultistageJoinFullName{} algorithm would require $\approx 225$MB of memory for its splitter, and $\approx 4$MB of memory for each of the first-iteration subsequent executors. The load-balanced \MultistageJoinFullName{} algorithm would require only $\approx11$KB for each first-iteration splitter, and $\approx 4$MB of memory for each first-iteration subsequent executor. If ${\ell_{\mathcal{R}}} = {\ell_{\mathcal{S}}} = 10^5$, and $n = 1000$ executors were used, ExpVar-Join would require each reducer to have $\approx 10$GB of memory, the basic \MultistageJoinFullName{} algorithm would require $\approx 4.6$GB of memory for its splitter, and $\approx 30$MB of memory for each of the first-iteration subsequent executors. The  load-balanced \MultistageJoinFullName{} algorithm would require only $\approx24$KB for each first-iteration splitter, and $\approx 30$MB of memory for each first-iteration subsequent executor.

\section{The Small-Large Outer-Joins}
\label{sec:broadcast}

\begin{minipage}[t]{.5\textwidth}

\begin{minipage}[t]{\smallfigwidth}
    \centering
\begin{algorithm}[H]
\scriptsize
\caption{\small \mbox{\textit{\SmallLargeJoinCodeName{}}}($\mathcal{R}$, $\mathcal{S}$)}
\label{algo:map_side_join}
\begin{algorithmic}[1]
\INPUT Two relations to be joined.
\OUTPUT The join results.
\ASSUMES $\mathcal{S}$ fits in memory. $\mathcal{R} \gg \mathcal{S}$.

 \STATE $index$ = $\mathcal{S}$.map(map$_{buildIndex}$).groupByKey
 \STATE broadcast $index$ to all executors
 \STATE $\mathcal{Q}$ = $\mathcal{R}$.map(map$_{\SmallLargeJoinCodeName{}}$)
 \RETURN $\mathcal{Q}$
\end{algorithmic}
\end{algorithm}
\end{minipage}

\end{minipage}
\begin{minipage}[t]{.5\textwidth}

\begin{minipage}[t]{\smallfigwidth}
    \centering
\begin{algorithm}[H]
\scriptsize
\caption{\small \mbox{\textit{map$_{buildIndex}$}}($rec_i$)}
\label{algo:map_build_index}
\begin{algorithmic}[1]
\INPUT A record from a relation.
\OUTPUT A tuple of the key, and
\INDENTOUTPUT the remaining attributes of $rec_i$.
\INDENTOUTPUT The output tuple is of the form
\INDENTOUTPUT $\langle key_{rec_i}, attrib_{rec_i} \rangle$.

 \STATE $key_{rec_i}$ = $getKey$($rec_i$)
 \STATE $attrib_{rec_i}$ = $getAttrib$($rec_i$)
 \RETURN $\langle key_{rec_i}, attrib_{rec_i} \rangle$
\end{algorithmic}
\end{algorithm}
\end{minipage}

\end{minipage}

\vspace{20pt}
The scenario of Small-Large joins arise when $\mathcal{R} \gg \mathcal{S}$ and $\mathcal{S}$ can be assumed to fit in the memory of each executor. An index can be built for the small relation, $\mathcal{S}$, the driver collects the index locally using the \mbox{\textit{treeAggregate}} operation that is implemented as a series of MapReduce jobs. The driver then broadcasts the index to all $n$ executors performing the join. Broadcasting $\mathcal{S}$ can be done in time that is logarithmic in $n$ \cite{bar1994designing,sanders2009two}\footnote{The tree-based broadcast algorithms stream the data from the driver to a layer of executors. The data is then streamed from these executors to a wider layer of executors, and so on. So, each executor acts as a receiver and a sender. Multiple concurrent such trees could be used to optimally utilize the entire bandwidth of the network, and hence finish the broadcast in $O(n)$ rounds of communication.}, and can hence be faster than shuffling the large relation, $\mathcal{R}$, across the network.

\begin{figure*}[!ht]
\centering
\includegraphics[width = \bigfigwidth]{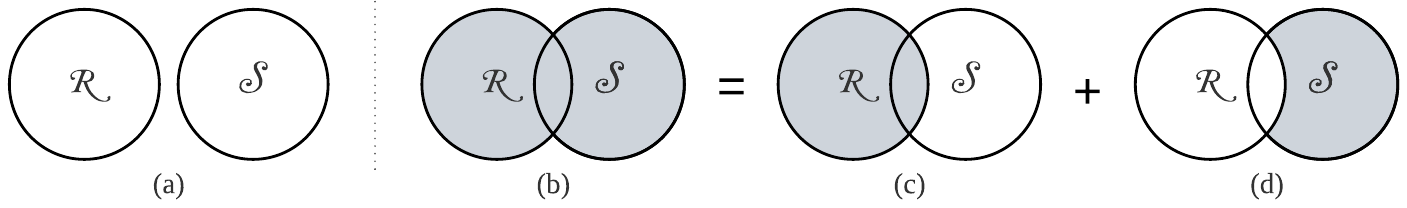}
\caption{Synthesizing the results of the full-outer-join from other joins. (a) shows the input relations whose full-outer-join results, shown in (b), can be expressed as a union of the left-outer-join results in (c) and the right-anti-join results in (d). The shaded areas represent keys whose records are included in the join results.}
\label{fig:full_outer_join}
\end{figure*}

The \mbox{\textit{\SmallLargeJoinCodeName{}} algorithm (\Algorithm~\ref{algo:map_side_join})} builds the index by invoking \mbox{\textit{map$_{buildIndex}$}} (\Algorithm~\ref{algo:map_build_index}) on each record, yielding its key and its remaining attributes, and then grouping the mapped records by their keys. Then \mbox{\textit{\SmallLargeJoinCodeName{}}} broadcasts the index instead of $\mathcal{S}$, and hence the algorithm family name, \SmallLargeJoinFullName{}. Assuming each executor, $e_i$, can accommodate the index of $\mathcal{S}$ in memory. For each record in its local partition, $\mathcal{R}_i$, the executor runs \mbox{\textit{map$_{\SmallLargeJoinCodeName{}}$}} (\Algorithm~\ref{algo:map_map_side_join}) on it. That is, it extracts the key, probes the local index of $\mathcal{S}$ with that key, and produces the join results. The local join is shown in \Algorithm~\ref{algo:map_map_side_join}.

\FrameworkName{} utilizes Broadcast-Joins for keys that are hot in only one relation. Extending \FrameworkName{} to outer-joins entails performing Small-Large outer-joins. We extend \SmallLargeJoinFullName{} (\SmallLargeJoinName{}) to outer-joins for the best performance of \FrameworkName{} outer-joins. The \SmallLargeFullOuterJoinFullName{} (\SmallLargeFullOuterJoinName{}) algorithm is in \Algorithm~\ref{algo:map_side_full_outer_join}. The left-outer-join and the right-outer-join are mere simplifications. 

The full-outer-join results can be computed in multiple ways. One viable way is unioning the results of the left-outer-join, and the \emph{right-anti-join}. The right-anti-join produces the records in the right relation that are not joinable with the left relation. This method is illustrated in \Figure~\ref{fig:full_outer_join}. These left-outer-join and right-anti-join can be computed in parallel on potentially-separate sets of executors. \SmallLargeFullOuterJoinName{} computes the full-outer-join results in that way.

\begin{minipage}[t]{.5\textwidth}

\begin{minipage}[t]{\smallfigwidth}
    \centering
\begin{algorithm}[H]
\scriptsize
\caption{\small \mbox{\textit{map$_{\SmallLargeJoinCodeName{}}$}}($rec_i$)}
\label{algo:map_map_side_join}
\begin{algorithmic}[1]
\LOCAL $index$, a map from each key in the replicated relation
\INDENTLOCAL to all the records with that key.
\INPUT A record from the non-replicated relation.
\OUTPUT A list of tuples representing the join of $rec_i$ 
\INDENTOUTPUT with $index$.

 \STATE u = empty Buffer
 \STATE $key_{rec_i}$ = $getKey$($rec_i$)
 \FORALL{$rec_j \in index[key_{rec_i}]$}
  \STATE u.append($\langle key_{rec_i}, rec_i, rec_j \rangle$)
 \ENDFOR
 \RETURN u
\end{algorithmic}
\end{algorithm}
\end{minipage}

\end{minipage}
\begin{minipage}[t]{.5\textwidth}

\begin{minipage}[t]{\smallfigwidth}
    \centering
\begin{algorithm}[H]
\scriptsize
\caption{\small \mbox{\textit{\SmallLargeFullOuterJoinCodeName{}}}( \newline $\mathcal{R}$, $\mathcal{S}$)}
\label{algo:map_side_full_outer_join}
\begin{algorithmic}[1]
\INPUT Two relations to be joined.
\OUTPUT The join results.
\ASSUMES $\mathcal{S}$ fits in memory. $\mathcal{R} \gg \mathcal{S}$.

 \STATE $index$ = $\mathcal{S}$.map(map$_{buildIndex}$).groupByKey
 \STATE broadcast $index$ to all executors
 \STATE $\mathcal{Q}_{leftOuter}$ = $\mathcal{R}$.map(map$_{broadcastLeftOuterJoin}$)
 \STATE $keys_{joined}$ = $\mathcal{R}$
 \STATE \quad .map(map$_{getRightJoinableKey}$)
 \STATE \quad .combine($unionSets$)
 \STATE \quad .treeAggregate($unionSets$)
 \STATE $keys_{unjoinable}$ = $index$.keys - $keys_{joined}$
 \STATE broadcast $keys_{unjoinable}$ to all executors
 \STATE $\mathcal{Q}_{rightAnti}$ = $\mathcal{S}$.map(map$_{rightAntiJoin}$)
 \RETURN $\mathcal{Q}_{leftOuter} \cup \mathcal{Q}_{rightAnti}$
\end{algorithmic}
\end{algorithm}
\end{minipage}

\end{minipage}

\begin{minipage}[t]{.5\textwidth}

\begin{minipage}[t]{\smallfigwidth}
    \centering
\begin{algorithm}[H]
\scriptsize
\caption{\small \mbox{\textit{map$_{broadcastLeftOuterJoin}$}}($rec_i$)}
\label{algo:map_map_side_left_outer_join}
\begin{algorithmic}[1]
\LOCAL $index$, a map from each key in the replicated relation
\INDENTLOCAL to all the records with that key.
\INPUT A record from the non-replicated relation.
\OUTPUT A list of truples representing the
\INDENTOUTPUT left-outer-join of $rec_i$ with $index$.

 \STATE u = empty Buffer
 \STATE $key_{rec_i}$ = $getKey$($rec_i$)
 \IF{$key_{rec_i}$ in $index$}
  \FORALL{$rec_j \in index[key_{rec_i}]$}
   \STATE u.append($\langle key_{rec_i}, rec_i, rec_j \rangle$)
  \ENDFOR
 \ELSE
  \STATE u.append($\langle key_{rec_i}, rec_i, null \rangle$)
 \ENDIF
 \RETURN u
\end{algorithmic}
\end{algorithm}
\end{minipage}

\end{minipage}
\begin{minipage}[t]{.5\textwidth}

\begin{minipage}[t]{\smallfigwidth}
    \centering
\begin{algorithm}[H]
\scriptsize
\caption{\small \mbox{\textit{map$_{getRightJoinableKey}$}}($rec_i$)}
\label{algo:get_right_joinable_keys}
\begin{algorithmic}[1]
\LOCAL $index$, a map from each key in the replicated relation
\INDENTLOCAL to all the records with that key.
\INPUT A record from the non-replicated relation.
\OUTPUT The key of $rec_i$ if it is joinable with $index$.

 \STATE u = empty Set
 \STATE $key_{rec_i}$ = $getKey$($rec_i$)
 \IF{$key_{rec_i}$ $\in$ $index$.keys}
  \STATE u.append($key_{rec_i}$)
 \ENDIF
 \RETURN u
\end{algorithmic}
\end{algorithm}

\begin{algorithm}[H]
\scriptsize
\caption{\small \mbox{\textit{map$_{rightAntiJoin}$}}($rec_j$)}
\label{algo:right_anti_join}
\begin{algorithmic}[1]
\LOCAL $keys_{unjoinable}$, a set of unjoinable keys.
\INPUT A record from the replicated relation.
\OUTPUT The join results of $rec_j$ if its key is in $keys_{unjoinable}$.

 \STATE $key_{rec_i}$ = $getKey$($rec_j$)
 \IF{$key_{rec_i}$ $\in$ $keys_{unjoinable}$}
  \RETURN $\langle key_{rec_i}, null, rec_j \rangle$
 \ENDIF
\end{algorithmic}
\end{algorithm}
\end{minipage}

\end{minipage}

\vspace{20pt}
Like the inner-join (\Algorithm~\ref{algo:map_side_join}), \SmallLargeFullOuterJoinName{} builds an index on the small relation, collects it at the driver, and broadcasts it to all the executors. The results of the left-outer-join are then computed. The \mbox{\textit{map$_{broadcastLeftOuterJoin}$}} function (\Algorithm~\ref{algo:map_map_side_left_outer_join}) is similar to \Algorithm~\ref{algo:map_map_side_join}, but also produces \emph{unjoined} records from the non-replicated relation.

To produce the full-outer-join, \SmallLargeFullOuterJoinName{} needs to also produce the right-anti-join results, i.e., the \emph{unjoinable} records in the small replicated relation. The driver finds the unjoinable keys in the replicated relation by utilizing the already replicated relation. The \mbox{\textit{map$_{getRightJoinableKey}$}} (\Algorithm~\ref{algo:get_right_joinable_keys}) produces a set for each record in the large non-replicated relation. The set either has a single key from the replicated relation if the key is joinable with the non-replicated relation, or is empty otherwise. The driver then unions these sets, first on each executor, and then across the network. The union constitutes the set of joined keys in the replicated relation. The index keys not in this union constitute the unjoinable keys.

The driver then broadcasts these unjoinable keys back to the executors, and maps the original small relation using \mbox{\textit{map$_{rightAntiJoin}$}} (\Algorithm~\ref{algo:right_anti_join}). Each executor scans its local partition of the small relation, and outputs only the records whose keys are unjoinable. The left-outer-join results are unioned with the right-anti-join results to produce the full-outer-join results.

\subsection{Optimizations}
\label{sec:broadcast_optimizations}

\SmallLargeFullOuterJoinName{} can be optimized as follows.
\begin{packed_enum}
  \item{The driver should send the joinable keys over the network if they are fewer than the unjoinable keys (Line $9$ of \Algorithm~\ref{algo:map_side_full_outer_join}). In that case, the condition in Line $2$ of \Algorithm~\ref{algo:right_anti_join} has to be reversed.}
  \item{Lines $3$ and $5$ in \Algorithm~\ref{algo:map_side_full_outer_join} can be combined into a single \mbox{\textit{mapRec}} function to reduce the number of scans on the large relation.}
  \item{In a shared-nothing architecture where an executor, $e_i$, can scan its data partition, Lines $5$ and $6$ in \Algorithm~\ref{algo:map_side_full_outer_join} can be combined. A single set can be allocated and populated as $\mathcal{R}_i$ is scanned.}
  \item{In a shared-nothing architecture that supports both multicasting and broadcasting, instead of broadcasting the index to all the executors (Line $2$ in \Algorithm~\ref{algo:map_side_join} and \Algorithm~\ref{algo:map_side_full_outer_join}), the driver may broadcast the unique keys only. Instead of the local joins (Line $3$ in \Algorithm~\ref{algo:map_side_join} and \Algorithm~\ref{algo:map_side_full_outer_join}), each executor, $e_i$, joins the $\mathcal{S}$ keys with its $\mathcal{R}_i$ and sends the joinable keys  to the driver. The driver multicasts each $\mathcal{S}$ record only to the executors that have its joinable partitions. Each $e_i$ joins the received records with its $\mathcal{R}_i$. The driver need not compute the unjoinable $\mathcal{S}$ keys (Lines $4$--$8$ in \Algorithm~\ref{algo:map_side_full_outer_join}), since they are the keys not multicasted at all. This optimization is a form of semi-join reduction, which was first introduced in \cite{bernstein1981query, apers1983optimization}, and later revived in other optimizations, e.g., \cite{stocker2001integrating,hill2009reducing}.}
\end{packed_enum}

\subsection{Comparison with the State-of-the-Art Algorithms}
\label{sec:broadcast_comparison}

A main difference between the proposed \SmallLargeFullOuterJoinName{} (\Algorithm~\ref{algo:map_side_full_outer_join}), DER \cite{xu2010new} and DDR \cite{cheng2017design} is subtle but effective. To flag the unjoined records in the $n$ executors, \SmallLargeFullOuterJoinName{} utilizes the unique keys in a semi-join fashion, instead of the DBMS-assigned record ids in case of DER\footnote{A minor advantage of the \SmallLargeJoinName{} family is abolishing the dependency on DBMS-specific functionality. This facilitates its adoption in any shared-nothing architecture.} or entire records in case of DDR. Assuming the key is of the same size as the record id, and is significantly smaller than the record, sending unique keys reduces the network load, and scales better with a skewed $\mathcal{S}$.

The other major difference is how the unjoinable records are identified after the first Broadcast-Join. DER hashes the unjoined ids from all executors over the network, and performs an inner Hash-Join. This entails hashing the records of $\mathcal{R}$, each of size $m_{\mathcal{R}}$, and the unjoinable ids of $\mathcal{S}$, each of size $m_{id}$ over the network. The communication cost of this step is $(n + 1) \times |\mathcal{S}| \times m_{id} + |\mathcal{R}| \times m_{\mathcal{R}}$. DDR hashes the $\mathcal{S}$ records, each of size $m_{\mathcal{S}}$, from all executors, incurring a communication cost of $n \times |\mathcal{S}| \times m_{\mathcal{S}}$, which is as costly as the first Broadcast-Join itself. \SmallLargeFullOuterJoinName{} collects and broadcasts the unique keys, each of size $m_{Key}$ (Lines $4-7$ and $9$ of \Algorithm~\ref{algo:map_side_full_outer_join}, respectively), with a communication cost of $2n \times |\mathcal{S}| \times m_{Key}$. This is much more efficient than DER and DDR, even without considering that broadcasting the unique keys is done in time logarithmic in $n$.

\section{The \FrameworkName{} Algorithm}
\label{sec:main_algorithm}

\FrameworkName{} employs the \MultistageJoinFullName{} and the Broadcast-Joins (we utilize \SmallLargeJoinFullName{} for its communication efficiency) to maximize the executors' utilization, and minimize the communication cost. \FrameworkName{} starts by collecting a map from hot keys to their frequencies for both relations, $\kappa_{\mathcal{R}}$ and $\kappa_{\mathcal{S}}$. Then, it splits each relation into four sub-relations containing the keys that are hot in both sides, the keys that are hot in its side but cold in the other side, the keys that are hot in the other side but cold in its side, and the keys that are cold in both sides, respectively (\Section~\ref{sec:splitting_relations}). Therefore, $\mathcal{R}$ is split into $\mathcal{R}_{HH}$, $\mathcal{R}_{HC}$, $\mathcal{R}_{CH}$ and $\mathcal{R}_{CC}$, and $\mathcal{S}$ is split similarly into $\mathcal{S}_{HH}$, $\mathcal{S}_{HC}$, $\mathcal{S}_{CH}$ and $\mathcal{S}_{CC}$. Irrespective of the keys in $\kappa_{\mathcal{R}}$ and $\kappa_{\mathcal{S}}$, it is provable that the join results are the union of the four joins in \Equation~\ref{eqn:main_join}.

\begin{figure}[H]
\begin{center}
\begin{minipage}[t]{\smallfigwidth}
    \centering
\begin{algorithm}[H]
\scriptsize
\caption{\small \mbox{\textit{\FrameworkCodeName{}}}($\mathcal{R}$, $\mathcal{S}$)}
\label{algo:main_join}
\begin{algorithmic}[1]
\INPUT Two relations to be joined.
\OUTPUT The join results.
\CONSTANT $|\kappa_{\mathcal{R}}|_{max}$ and $|\kappa_{\mathcal{S}}|_{max}$,
\INDENTCONSTANT the number of hot keys to be collected 
\INDENTCONSTANT from $\mathcal{R}$ and $\mathcal{S}$, respectively.

 \STATE $\kappa_{\mathcal{R}}$ = $getHotKeys$($\mathcal{R}$, $|\kappa_{\mathcal{R}}|_{max}$)
 \STATE $\kappa_{\mathcal{S}}$ = $getHotKeys$($\mathcal{S}$, $|\kappa_{\mathcal{S}}|_{max}$)
 \STATE $\langle \mathcal{R}_{HH}, \mathcal{R}_{HC},
 \mathcal{R}_{CH}, \mathcal{R}_{CC} \rangle$ = 
 $splitRelation$($\mathcal{R}$, $\kappa_{\mathcal{R}}$, $\kappa_{\mathcal{S}}$)
 \STATE $\langle \mathcal{S}_{HH}, \mathcal{S}_{HC},
 \mathcal{S}_{CH}, \mathcal{S}_{CC} \rangle$ = 
 $splitRelation$($\mathcal{S}$, $\kappa_{\mathcal{S}}$, $\kappa_{\mathcal{R}}$)
 \STATE $\mathcal{Q}$ = $\MultistageJoinCodeName{}$($\mathcal{R}_{HH}, \mathcal{S}_{HH}$)
 \STATE \quad $\cup \ \SmallLargeJoinCodeName{}$($\mathcal{R}_{HC}$, $\mathcal{S}_{CH}$)
 \STATE \quad $\cup \ \SmallLargeJoinCodeName{}$($\mathcal{S}_{HC}$, $\mathcal{R}_{CH}$)
 \CONTINUELINEINDENTED .map(map$_{swapJoinedRecords}$)
 \STATE \quad $\cup \ shuffleJoin$($\mathcal{R}_{CC}$, $\mathcal{S}_{CC}$)
 \RETURN $\mathcal{Q}$
\end{algorithmic}
\end{algorithm}
\end{minipage}
\end{center}
\end{figure}

\begin{figure*}[!ht]
\centering
\includegraphics[width = \bigfigwidth]{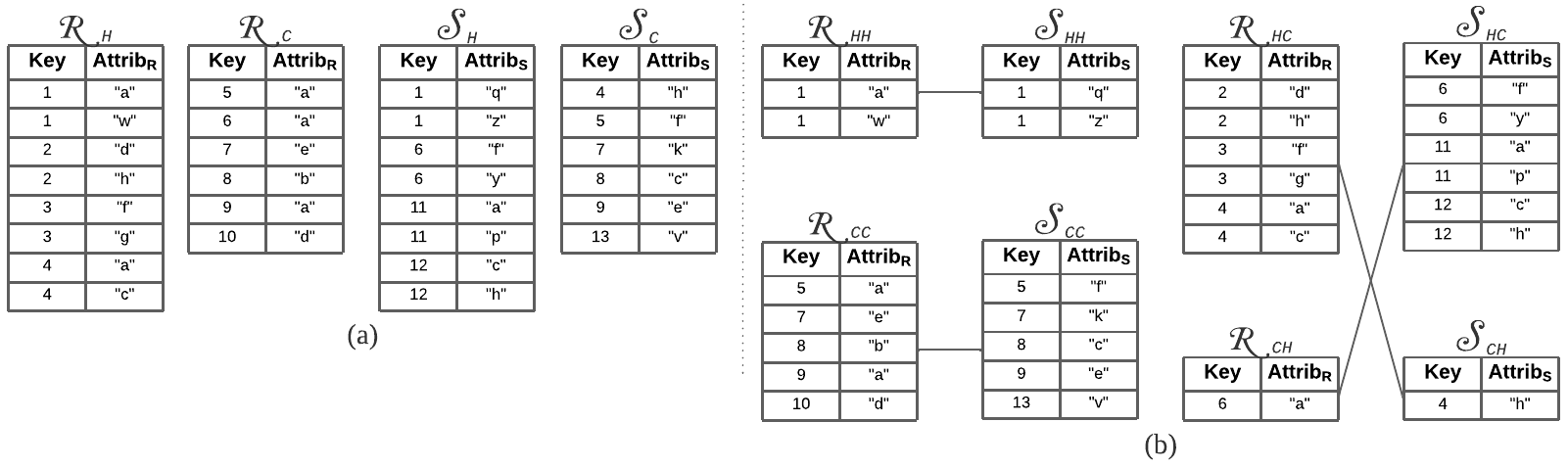}
\caption{
An example for joining the two relations in \Figure~\ref{fig:main_join_example_inner_and_outer} using \FrameworkName{}. (a) The hot-key and cold-key sub-relations of $\mathcal{R}$ and $\mathcal{S}$ after the first split. (b) The four sub-relations for each of the two input relations after the second split, and their joins based on \Equation~\ref{eqn:main_join}.
}
\label{fig:main_join_example_intermediate}
\end{figure*}

For the example of the two relations in \Figure~\ref{fig:main_join_example_inner_and_outer}(a), assuming the hot keys on either $\mathcal{R}$ and $\mathcal{S}$ are those occurring multiple times, then $\kappa_{\mathcal{R}}$ and $\kappa_{\mathcal{S}}$ are the maps $\{ 1: 2, 2 : 2, 3: 2, 4: 2\}$ and $\{ 1: 2, 6: 2, 11: 2, 12: 2\}$, respectively. \Figure~\ref{fig:main_join_example_intermediate}(b) shows the sub-relations of $\mathcal{R}$ and $\mathcal{S}$ paired together to be joined, according to \Equation~\ref{eqn:main_join}. If all the joins in \Equation~\ref{eqn:main_join} are inner-joins, then the inner-join results of $\mathcal{R}$ and $\mathcal{S}$ are given in \Figure~\ref{fig:main_join_example_inner_and_outer}(b).

\begin{equation}
\label{eqn:main_join}
  \begin{alignedat}{3}
\mathcal{Q} = & \ \mathcal{R}_{HH} & \ \Join & \ \mathcal{S}_{HH} \\
 \cup & \ \mathcal{R}_{HC} & \ \Join & \ \mathcal{S}_{CH} \\
 \cup & \ \mathcal{R}_{CH} & \ \Join & \ \mathcal{S}_{HC} \\
 \cup & \ \mathcal{R}_{CC} & \ \Join & \ \mathcal{S}_{CC}
  \end{alignedat}
\end{equation}

Since the first join involves keys that are hot in both the joined relations, \MultistageJoinFullName{} (\Algorithm~\ref{algo:tree_join}) is used. The second and the third joins in \Equation~\ref{eqn:main_join} lend themselves naturally to Small-Large joins. We only discuss the second join, but the logic also applies equally to the third join. $\mathcal{R}_{HC}$ and $\mathcal{S}_{CH}$ contain the $\mathcal{R}$ and $\mathcal{S}$ records, respectively, with keys that are hot in $\mathcal{R}$ but are cold in $\mathcal{S}$. Hot keys are rare by definition, since a hot key requires a minimum number of records to be hot, i.e., a minimum percentage of the records in the relation. Hence, the number of keys in $\mathcal{R}_{HC}$ and $\mathcal{S}_{CH}$ is limited. Moreover, the number of records with these keys in $\mathcal{S}$ is limited, since these keys are cold in $\mathcal{S}$. Therefore, $\mathcal{S}_{CH}$ is bounded in size, and should fit in the memory of the executors. This is discussed in more detail in \Section~\ref{sec:number_of_hot_keys}, where tight bounds on the number of hot keys are established to accommodate the smaller relation in the memory of the executors.

Each record resulting from the third join is swapped using \mbox{\textit{map$_{swapJoinedRecords}$}} (\Algorithm~\ref{algo:map_swap_joined_records}) so $Attrib_{\mathcal{R}}$ precedes $Attrib_{\mathcal{S}}$. Finally, the fourth join involves keys that are hot on neither side, and is hence performed using a single-executor-per-key Shuffle-Join. The algorithm is shown in \Algorithm~\ref{algo:main_join}. One optimization is passing the hot key maps, $\kappa_{\mathcal{R}}$ and $\kappa_{\mathcal{S}}$, to \MultistageJoinFullName{} instead of double-computing them.

Notice that in the case of processing natural self-joins, the hot keys are identical on both sides of the join. Hence, the second and the third Small-Large joins produce no results. In that case, \FrameworkName{} reduces to \MultistageJoinFullName{}.

While the \FrameworkName{} algorithm bears some resemblance to the PRPD family, it deviates in a core aspect. PRPD assigns a key as hot to the relation that has more records for that key. \FrameworkName{} independently collects the hot keys for $\mathcal{R}$ and $\mathcal{S}$, resulting in the simple and elegant \Equation~\ref{eqn:main_join}. \Equation~\ref{eqn:main_join} leads to (a) \FrameworkName{} utilizing the scalable \MultistageJoinFullName{} to join the keys that are hot in both $\mathcal{R}$ and $\mathcal{S}$, and (b) extending \FrameworkName{} smoothly to all outer-join variants (\Section~\ref{sec:outer_joins}).

\subsection{Splitting the Relations}
\label{sec:splitting_relations}

Once the hot keys for both $\mathcal{R}$ and $\mathcal{S}$ are collected, each relation is split into its four sub-relations. We only discuss $\mathcal{R}$, but the logic also applies equally to $\mathcal{S}$. This \emph{splitRelation} module (\Algorithm~\ref{algo:split_relation}) proceeds in two rounds. It first splits $\mathcal{R}$ into $\mathcal{R}_{H}$, the sub-relation whose keys are in $\kappa_{\mathcal{R}}$, and $\mathcal{R}_{C}$, the cold-key records in $\mathcal{R}$. In the example in \Figure~\ref{fig:main_join_example_inner_and_outer}, the results of the first-round of splitting is in \Figure~\ref{fig:main_join_example_intermediate}(a). In the second round, $\mathcal{R}_{H}$ is split into $\mathcal{R}_{HH}$, the sub-relation whose keys are in $\kappa_{\mathcal{S}}$, and $\mathcal{R}_{HC}$, the remaining records in $\mathcal{R}_{H}$. $\mathcal{R}_{C}$ is similarly split using $\kappa_{\mathcal{S}}$ into $\mathcal{R}_{CH}$ and $\mathcal{R}_{CC}$. In the example in \Figure~\ref{fig:main_join_example_inner_and_outer}, the second-round of splitting is in \Figure~\ref{fig:main_join_example_intermediate}(b).
 
\begin{minipage}[t]{.5\textwidth}

\begin{minipage}[t]{\smallfigwidth}
    \centering
\begin{algorithm}[H]
\scriptsize
\caption{\small \mbox{\textit{map$_{swapJoinedRecords}$}}( \newline $\langle key, attrib_{\mathcal{S}_s}, attrib_{\mathcal{R}_r} \rangle$)}
\label{algo:map_swap_joined_records}
\begin{algorithmic}[1]
\INPUT A tuple of the key, 
\INDENTINPUT and remaining attributes in reverse order.
\OUTPUT A tuple of the key, 
\INDENTOUTPUT and remaining attributes in the correct order.

 \RETURN $\langle key, attrib_{\mathcal{R}_r}, attrib_{\mathcal{S}_s} \rangle$
\end{algorithmic}
\end{algorithm}
\end{minipage}

\end{minipage}
\begin{minipage}[t]{.5\textwidth}

\begin{minipage}[t]{\smallfigwidth}
    \centering
\begin{algorithm}[H]
\scriptsize
\caption{\small \mbox{\textit{splitRelation}}($\mathcal{R}$, $\kappa_{\mathcal{R}}$, $\kappa_{\mathcal{S}}$)}
\label{algo:split_relation}
\begin{algorithmic}[1]
\INPUT A relation to be split,
\INDENTINPUT its hot keys map,
\INDENTINPUT and the hot keys map of the other relation.
\OUTPUT The four splits of $\mathcal{R}$.

 \STATE $\langle \mathcal{R}_{H}, \mathcal{R}_{C} \rangle$ = $splitPartitionsLocally$(
 \CONTINUELINE $\mathcal{R}$, $rec \rightarrow key_{rec} \in \kappa_{\mathcal{R}}.keys$)
 \STATE $\langle \mathcal{R}_{HH}, \mathcal{R}_{HC} \rangle$ = $splitPartitionsLocally$(
 \CONTINUELINE $\mathcal{R}_{H}$, $rec \rightarrow key_{rec} \in \kappa_{\mathcal{S}}.keys$)
 \STATE $\langle \mathcal{R}_{CH}, \mathcal{R}_{CC} \rangle$ = $splitPartitionsLocally$(
 \CONTINUELINE $\mathcal{R}_{C}$, $rec \rightarrow key_{rec} \in \kappa_{\mathcal{S}}.keys$)
 \RETURN $\langle \mathcal{R}_{HH}, \mathcal{R}_{HC}, \mathcal{R}_{CH}, \mathcal{R}_{CC} \rangle $
\end{algorithmic}
\end{algorithm}
\end{minipage}

\end{minipage}

\vspace{20pt}
The splitting is done by each executor reading its local partition sequentially, checking if the record key exists in the hot-key set, and writing it to either the hot-key sub-relation or the cold-key sub-relation. The splitting involves no communication between the executors. The two rounds can be optimized into one.

\subsection{When not to Perform Broadcast-Joins?}

The second and third joins in \Equation~\ref{eqn:main_join} are executed using Broadcast-Joins. We only discuss the second join, but the logic also applies equally to the third. For the Broadcast-Join assumption to hold, broadcasting the records of the small relation, $\mathcal{S}_{CH}$, over the network has to be faster than splitting the large relation, $\mathcal{R}_{HC}$, among the $n$ executors over the network. From \cite{bar1994designing}, the time to read $\mathcal{S}_{CH}$ and broadcast it follows the Big-$\Theta$ below.

\begin{equation*}
\Delta_{broadcast_{\mathcal{S}_{CH}}} \approx \Theta\left(
|\mathcal{S}_{CH}| \times m_{\mathcal{S}} \times \left( 1 + \lambda \times \log_{\lambda + 1}(n)
\right)
\right)
\end{equation*}

 Let $m_{\mathcal{R}}$ be the average size of a record in $\mathcal{R}$. The time to read $\mathcal{R}_{HC}$ and split it among the $n$ executors follows the Big-$\Theta$ below.

\begin{equation*}
\Delta_{split_{\mathcal{R}_{HC}}} \approx \Theta \left(
|\mathcal{R}_{HC}| \times m_{\mathcal{R}} \times (1 + \lambda)
\right)
\end{equation*}

At the time of executing the second join in \Equation~\ref{eqn:main_join}, $\mathcal{R}_{HC}$ and $\mathcal{S}_{CH}$ have already been computed. A Broadcast-Join is performed if $\Delta_{split_{\mathcal{R}_{HC}}} \geq \Delta_{broadcast_{\mathcal{S}_{CH}}}$. Otherwise, single-executor-per-key Shuffle-Join is performed.

\subsection{The Outer Variants of \FrameworkName{}}
\label{sec:outer_joins}

\begin{table*}[ht]
  \caption{The algorithms for the four sub-joins of \Equation~\ref{eqn:main_join} of the outer variants of \FrameworkName{}.}
  \label{table:main_join}
  \scriptsize
  \centering
  \begin{tabular}{ c | l l l }
 
  \toprule

                                                                       & \textbf{left-outer \FrameworkName{}} & \textbf{right-outer \FrameworkName{}} & \textbf{full-outer \FrameworkName{}} \\
  \midrule
                        $\mathcal{R}_{HH} \Join \mathcal{S}_{HH}$   & \MultistageJoinFullName{}                        & \MultistageJoinFullName{}                      & \MultistageJoinFullName{} \\
                        $\mathcal{R}_{HC} \Join \mathcal{S}_{CH}$   & \SmallLargeLeftOuterJoinFullName{}       & \SmallLargeJoinFullName{}                    & \SmallLargeLeftOuterJoinFullName{} \\
                        $\mathcal{S}_{HC} \Join \mathcal{R}_{CH}$     & \SmallLargeJoinFullName{}                      & \SmallLargeLeftOuterJoinFullName{}     & \SmallLargeLeftOuterJoinFullName{} \\
                        $\mathcal{R}_{CC} \Join \mathcal{S}_{CC}$   & Shuffle left-outer-join                                 & Shuffle right-outer-join                            & Shuffle full-outer-join \\
  \bottomrule

\end{tabular}
\end{table*}

Due to the \FrameworkName{} elegant design, the outer-join variants are minor modifications of the inner-join in \Algorithm~\ref{algo:main_join}. The only difference is that some of the four joins of \Equation~\ref{eqn:main_join} are executed using outer-join algorithms. \Table~\ref{table:main_join} shows the algorithm used for each of the four joins to achieve the outer-join variants of \FrameworkName{}. The first and the fourth joins are performed using \MultistageJoinFullName{} and a Shuffle outer-join, respectively. The second and third joins are performed using a Broadcast-Join or its left-outer variant (we utilize \SmallLargeJoinFullName{} or \SmallLargeLeftOuterJoinFullName{} for their communication efficiency). Applying the outer-joins in \Table~\ref{table:main_join} to the joins in \Figure~\ref{fig:main_join_example_intermediate}(b) yields the left-outer, right-outer, and full-outer-join results in \Figure~\ref{fig:main_join_example_inner_and_outer}(c), \Figure~\ref{fig:main_join_example_inner_and_outer}(d),  and \Figure~\ref{fig:main_join_example_inner_and_outer}(e), respectively.

Unlike the state-of-the-art distributed industry-scale algorithms, the different flavors of SkewJoin \cite{bruno2014advanced} built on top of Microsoft SCOPE, \FrameworkName{} supports all variants of outer-joins without record deduplication or custom partitioning of the relations. Moreover, \FrameworkName{} does not require introducing a new outer-join operator variant that understands the semantics of \emph{witness} tuples, a form of tuples introduced in \cite{bruno2014advanced} to extend SkewJoin to outer-joins.

\section{Collecting Hot Keys}
\label{sec:finding_hot_keys}

Identifying hot keys in $\mathcal{R}$ and $\mathcal{S}$ is an integral part of \MultistageJoinFullName{} and \FrameworkName{}, since it allows for load-balancing the  executors (\Section~\ref{sec:load_balancing_first_iteration}), and for splitting the joined relations into the sub-relations going into \Equation~\ref{eqn:main_join}. The \mbox{\textit{getHotKeys}} algorithm (Lines $1$ and $2$ in \Algorithm~\ref{algo:tree_join} and Lines $1$ and $2$ in \Algorithm~\ref{algo:main_join}) collects the hot keys for both $\mathcal{R}$ and $\mathcal{S}$. The maximum numbers of hot keys collected for $\mathcal{R}$ and $\mathcal{S}$ are $|\kappa_{\mathcal{R}}|_{max}$ and $|\kappa_{\mathcal{S}}|_{max}$, respectively, are discussed in \Section~\ref{sec:number_of_hot_keys}. We only discuss $|\kappa_{\mathcal{R}}|_{max}$, but the logic also applies equally to $|\kappa_{\mathcal{S}}|_{max}$. In \Section~\ref{sec:hot_keys_cost} we discuss the cost of collecting hot keys. We assume $\mathcal{R}$ and $\mathcal{S}$ are evenly distributed on the $n$ executors and no record migration among the input partitions is necessary.

\subsection{How many Hot Keys to Collect?}
\label{sec:number_of_hot_keys}

We only discuss $|\kappa_{\mathcal{R}}|_{max}$, but the logic also applies equally to $|\kappa_{\mathcal{S}}|_{max}$. Any key with $\left(1 + \lambda \right)^{\frac{3}{2}}$ or more records is hot (\Relation~\ref{ineq:hot_keys}). Three upper bounds apply to $|\kappa_{\mathcal{R}}|_{max}$. 

\begin{packed_enum}
  \item{No more than $\frac{M}{m_{Key}}$ hot keys can be collected, since no more keys of size $m_{Key}$ can fit in a memory of size $M$. The data structure overhead of the heavy-hitters algorithm is ignored for simplicity, but can be accounted for in the implementation.}
  \item{The number of hot keys in $\mathcal{R}$ cannot exceed $\frac{|\mathcal{R}|}{\left(1 + \lambda \right)^{\frac{3}{2}}}$, from the Pigeonhole principle.}
  \item{The second join in \Equation~\ref{eqn:main_join} is executed using Broadcast-Join (e.g., \SmallLargeJoinFullName{}). Let $m_{\mathcal{S}}$ be the average record size in $\mathcal{S}$. $\mathcal{S}_{CH}$ is computed after \mbox{\textit{getHotKeys}}, but is bounded to fit in memory by \Relation~\ref{rel:slh_memory}.

\begin{equation}
\label{rel:slh_memory}
|\mathcal{S}_{CH}| \leq \frac{M}{m_{\mathcal{S}}}
\end{equation}
  
  The keys in $\mathcal{S}_{CH}$ are a subset of the keys in $\kappa_{\mathcal{R}}$. Each of these keys has frequency below $\left(1 + \lambda \right)^{\frac{3}{2}}$ in $\mathcal{S}_{CH}$. To ensure $\mathcal{S}_{CH}$ fits in memory, we limit $|\kappa_{\mathcal{R}}|$ by \Relation~\ref{rel:slh_upper_bound}.

\begin{equation}
\label{rel:slh_upper_bound}
|\kappa_{\mathcal{R}}|_{max} \leq \frac{|\mathcal{S}_{CH}|}{\left(1 + \lambda \right)^{\frac{3}{2}}}
\end{equation}

Substituting \Relation~\ref{rel:slh_memory} in \Relation~\ref{rel:slh_upper_bound} yields an upper bound on $|\kappa_{\mathcal{R}}|_{max}$.

}
\end{packed_enum}

Given the three upper bounds above, the maximum numbers of hot keys collected for $\mathcal{R}$, $|\kappa_{\mathcal{R}}|_{max}$, is given by \Equation~\ref{rel:num_hot_keys}.

\begin{equation}
\label{rel:num_hot_keys}
|\kappa_{\mathcal{R}}|_{max} =
\min\left(
\frac{
\min\left(
|\mathcal{R}|,
\frac{M}{m_{\mathcal{S}}}
\right)
}{
\left(1 + \lambda \right)^{\frac{3}{2}}
}
,
\frac{M}{m_{Key}}
\right)
\end{equation}

\subsection{The Hot Keys Cost}
\label{sec:hot_keys_cost}

The cost of collecting $|\kappa_{\mathcal{R}}|_{max}$ hot keys using the algorithm in \cite{agarwal2013mergeable} is the cost of scanning the local $\mathcal{R}_i$ partitions in parallel, and the cost of merging the hot keys over the network in a tree-like manner. Given $m_{b}$ and $m_{\mathcal{R}}$, the average sizes of a key and a $\mathcal{R}$ record, respectively, the total cost is given by \Equation~\ref{rel:hot_keys_cost}.

\begin{equation}
\label{rel:hot_keys_cost}
\Delta_{getHotKeys} = \frac{|\mathcal{R}| \times m_{\mathcal{R}}}{n} + |\kappa_{\mathcal{R}}|_{max} \times m_b \times \lambda \times \log(n)
\end{equation}

We like to contrast collecting the hot keys using a distributed heavy-hitters algorithm to histogramming the relation using a distributed quantiles algorithm \cite{okcan2011processing,vitorovic2016load,gavagsaz2019load}. Assuming $\mathcal{R}$ is distributed evenly between the $n$ executors, and given an error rate, $\epsilon$, the local heavy-hitters algorithm uses $\Theta(\frac{1}{\epsilon})$ space \cite{metwally2005efficient}, while the local quantiles algorithm uses $O(\frac{1}{\epsilon} \log(\epsilon \frac{|\mathcal{R}|}{n}))$ space \cite{cormode2020tight}\footnote{This is the theoretical bound on any deterministic single-pass comparison-based quantiles algorithm, even if the algorithms in \cite{okcan2011processing,vitorovic2016load,gavagsaz2019load} do not necessarily use it.}. Collecting heavy hitters is more precise given the same memory, and incurs less communication when merging the local statistics over the network to compute the global statistics.

\section{Evaluation Results}
\label{sec:results}

We evaluated the scalability and the handling of skewed data of \MultistageJoinFullName{}-Basic, \MultistageJoinFullName{}, \FrameworkName{}-Basic (i.e., \FrameworkName{} using \MultistageJoinFullName{}-Basic for joining keys that are hot on both sides of the join), \FrameworkName{}, Hash-Join, Broadcast-Join, Full-SkewJoin \cite{bruno2014advanced} (the state-of-the-art industry-scale algorithm of the PRPD family), and Fine-Grained partitioning for Skew Data (FGSD-Join) \cite{gavagsaz2019load} (the state-of-the-art of the key-range-division family). We also evaluated the Spark v$3.$x joins\footnote{\url{https://databricks.com/blog/2020/05/29/adaptive-query-execution-speeding-up-spark-sql-at-runtime.html}, \url{https://docs.databricks.com/spark/latest/spark-sql/aqe.html}.}, the most advanced open-source shared-nothing algorithm. The two main optimizations of Spark3-Join are (a) dynamically changing the execution plan from Shuffle-Join to Broadcast-Join when one relation can fit in memory, and (b) dynamically combining and splitting data partitions to reach almost equi-size partitions for load balancing. Multicasting is not supported by Spark, the framework we used for evaluation. Evaluating multicasting algorithms (e.g., \cite{polychroniou2018distributed}) is left for future work.

The basic Broadcast-Join algorithm was implemented. This requires either of the input partitions to fit in memory. To ensure Broadcast-Join does not run out of memory, the implementation could have been enhanced by pre-splitting one of the to-be relations into multiple sub-relations, such that each sub-relation fits in memory, running a Broadcast-Join between the other relation and each sub-relation, and unioning the results of all the joins. However, the standard implementation of Broadcast-Join was used since these optimizations are neither known to be the default out-of-the-box implementation in major DBMSs, nor typically used by the average DBMS user. For the ExpVar-Join algorithm, the model training time was not accounted for when measuring the runtime of any join. For all adaptive algorithms (\MultistageJoinFullName{} and \FrameworkName{} and variants, and Full-SkewJoin), and ExpVar-Join the hot keys were collected exactly, which is more time consuming, and does not show the strengths of the algorithms for less skewed data. For \FrameworkName{} and variants, the computed hot keys were passed to the employed \MultistageJoinFullName{} instead of recomputing them. The sample for FGSD-Join was $10 \times$ the number of partitions, and the number of partitions was $3 \times$ to $5 \times$ the number of executors for all algorithms.

The performance of the algorithms on outer-joins was in line with the inner-joins, reported below. However, we could build the outer-join variants of neither Full-SkewJoin, since it depends on Microsoft SCOPE-specific deduplication, nor FGSD-Join, since it was missing in \cite{gavagsaz2019load}. We compared the outer-join variant of the proposed \SmallLargeJoinName{} against DER \cite{xu2010new} and DDR \cite{cheng2017design} on Small-Large outer-joins.

We conducted comprehensive experiments on reproducible synthetic realistic data, and on real proprietary data. All the experiments were executed on machines (driver and executors) with $8$GB of memory, $20$GB of disk space, and $1$ vCPU. Amazon EC2 m7g.large instances were used with $2$ vCPUs, $8$GB of memory, and up to $12.5$Gbps network bandwidth (Gbps) \footnote{The reader is referred to \url{https://aws.amazon.com/ec2/instance-types/} and \url{https://aws.amazon.com/ec2/instance-types/m7g/} for more specifications.}.

The algorithms code and the code for the dataset generation and experiment running was written in Scala v$2.11.12$ and executed using Spark v$2.4.3$, except for the Spark v$3.$x experiments that were run on Spark v$3.0.2$, and the code was compiled using Scala v$2.12.12$. Spark RDDs were used to store the distributed datasets. Kryo serialization was used to (de)serialize the objects, and all the classes needed were registered with Kryo to minimize the (de)serialization overhead. Custom serialization was used for natural self-joins to avoid (de)serializing the same list of records twice when the two joined lists or sub-lists were one and the same. All the code, including the data generation and the evaluation code, was peer reviewed, and tested for correctness.

Each evaluation run was repeated $3$ times, and the median value is reported. The runtime almost never varied significantly, since all the experiments were run on a dedicated resource queue.

\subsection{Estimating $\lambda$}
\label{sec:estimating_lambda}

To set the algorithm parameters, it was crucial to estimate a very critical parameter, $\lambda$, since it impacts the minimum number of records for a key to be hot, according to \Relation~\ref{ineq:hot_keys} in \Section~\ref{sec:what_is_hot}. The $\lambda$ was estimated by shuffling a randomly-generated dataset between the executors over the network, and storing the same dataset to the local disks of the executors, and estimating the ratio in the time for these two operations.

The shuffling and writing were repeated 100 times, and the times were recorded. The datasets used were $10^2$ and $10^3$ Bytes per record. The keys were uniformly selected from the positive range of the $32$-bit Integer space, i.e., in the range $[0, \mbox{2,147,483,647}]$. The number of records was $10^6$ records per executor, and the number of executors assumed the values $10^2$ and $10^3$. The size of the records and the number of records were chosen so that for all the measurements, all the Spark RDD partitions can fit in memory for a more accurate estimation. \Table~\ref{tab:lambda_estimation} is the $\lambda$ as estimated by comparing the median times of sending data over the network vs. its IO from/to local disks. 

\begin{table}[ht]
\caption{The measurement of $\lambda$, the relative runtime of sending data over the network vs. its IO from/to a local disk.}
\label{tab:lambda_estimation}
\centering
\begin{tabular}{c | c c } 
  \toprule
  & \makecell{RecordSize = $10^2$} & \makecell{RecordSize = $10^3$} \\ [0.5ex] 
  \midrule
  \makecell{NumExecutors = $10^2$} & $7.34$ & $7.28$ \\ 
  \makecell{NumExecutors = $10^3$} & $6.98$ & $8.05$ \\
  \bottomrule
\end{tabular}
\end{table}

The empirical estimation of $\lambda$ does not show a huge variation in $\lambda$ even when varying the data size, or the number of executors by an order of magnitude. The average value for $\lambda$ was $7.4125$. This suggests that for the environment used in the experiment, the minimum frequency for a key to be hot should be within the range of $[10, 100]$ records. Based on \Relation~\ref{ineq:hot_keys} and \Equation~\ref{rel:num_hot_keys}, among the most popular $1000$ keys, those that have $100$ records or more were considered hot. These values agree with the guidelines of Full-SkewJoin.

\subsection{Experiments on Synthetic Data}
\label{sec:synthetic}

The synthetic data is a combination of random data generated using two processes. The first process generates a dataset of $10^9$ records whose keys are uniformly selected from the positive range of the $32$-bit Integer space, i.e., in the range $[0, \mbox{2,147,483,647}]$. The second process generates a dataset with $10^7$ records whose keys are drawn from a Zipf distribution with a skew-parameter $\alpha$ and a domain of $10^5$ keys, i.e., in the range $[0, \mbox{99,999}]$. The records generated by both processes were merged together using the RDD \mbox{\textit{union}}() function. All the records have the same size, $m_{\mathcal{R}\mathcal{S}}$. 

Initially, the experiments were conducted using the Zipfian keys only. The Zipfian keys were generated using the Apache Commons math package, which uses the state-of-the-art inverse-CDF algorithm. Yet, generating skewed keys from a space of $10^5$ was pushing the limits of the math package. The uniformly-distributed keys were eventually incorporated to increase the key space. 

We refer to the combined dataset using the notation $D(\alpha, m_{\mathcal{R}\mathcal{S}})$. For instance, $D(0.5, 10^3)$ refers to a dataset whose records have a size of $10^3$ Bytes, and has $10^9$ records whose keys are uniformly selected from the positive Integer space, and $10^7$ records whose keys are drawn from a Zipf-$0.5$ distribution with a range of $10^5$.

Two sizes of records were experimented with, $S = 10^2$, and $S = 10^3$ Bytes, representing relatively small and large records. The Zipf-$\alpha$ was varied between $0.0$ and $1.0$, representing uniform to moderately skewed data. Higher values of $\alpha$ are common in many applications\footnote{One example is the word frequency in Wikipedia \cite{Wikipedia}.}. However, higher values of $\alpha$ are expected to favor algorithms that handle data skew better, namely \FrameworkName{}(-Basic) and \MultistageJoinFullName{}(-Basic). These four algorithms split the processing of each hot key among multiple executors. Finally, the number of executors used to execute the join was varied from $100$ to $1000$ executors to evaluate how the algorithms utilize more resources to scale to more skewed datasets. All algorithms were allowed $2$ hours to execute each join.

\subsubsection{Equi-Join Scalability with Data Skew}
\label{sec:scalability_data_skew}

\begin{figure*}
\centering
\captionsetup{width=\smallfigwidth}
\begin{minipage}[t]{.5\textwidth}

  \centering
  \includegraphics[height=100pt, width = \smallfigwidth]{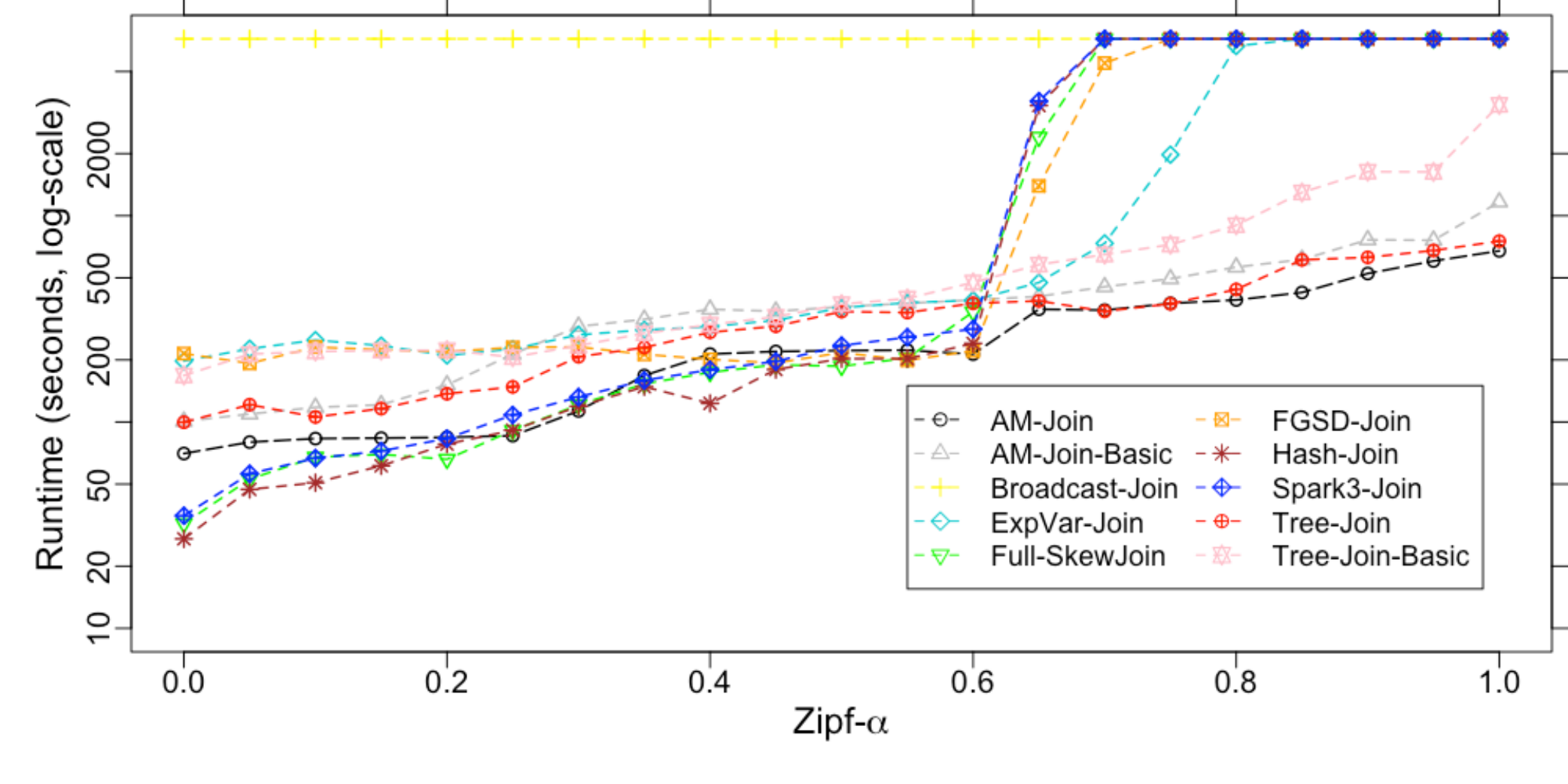}
  \caption{The runtime of the equi-join algorithms on $1000$ executors for $D(\alpha, 100)$ while varying the Zipf-$\alpha$. Notice the logarithmic scale on the vertical axes.}
  \label{fig:numRecords10000000_numDistinctKeys100000_valueSize100_VaryingAlpha}

\end{minipage}
\begin{minipage}[t]{.5\textwidth}

  \centering
  \includegraphics[height=100pt, width = \smallfigwidth]{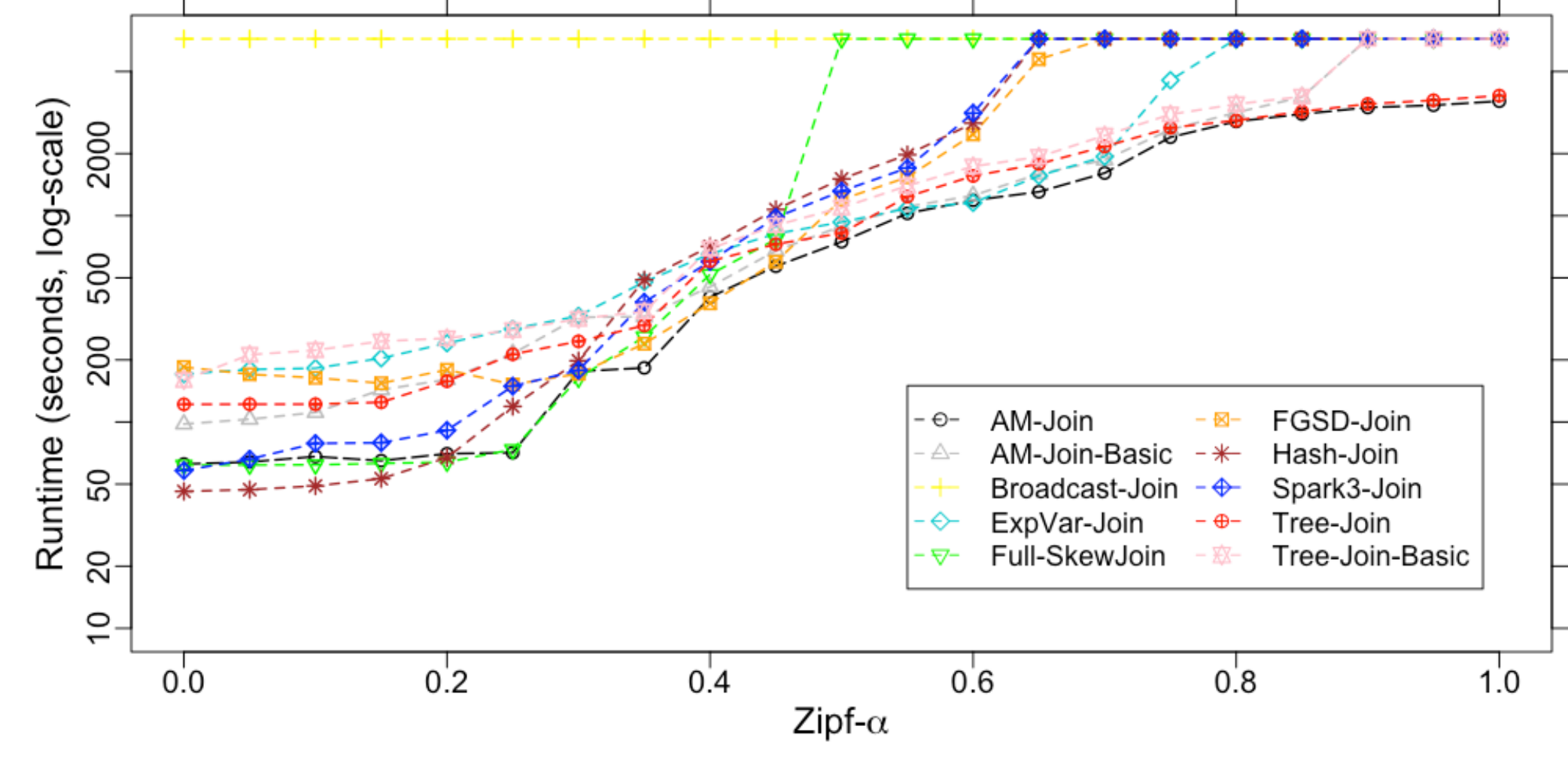}
  \caption{The runtime of the equi-join algorithms on $1000$ executors for $D(\alpha, 1000)$ while varying the Zipf-$\alpha$. Notice the logarithmic scale on the vertical axes}
  \label{fig:numRecords10000000_numDistinctKeys100000_valueSize1000_VaryingAlpha}

\end{minipage}
\end{figure*}

\Figure~\ref{fig:numRecords10000000_numDistinctKeys100000_valueSize100_VaryingAlpha} shows the runtimes of the algorithms when joining two relations, each has records with a size of $10^2$ Bytes, and has $10^9$ records whose keys are uniformly selected from the positive Integer space, and $10^7$ records whose Zipfian keys have a range of $10^5$ while varying the Zipf-$\alpha$.

For these relatively small records, the runtimes of all the algorithms were comparable until Zipf-$\alpha$ reached $0.6$ with a slight edge for Hash-Join until Zipf-$\alpha$ reached $0.15$. For Zipf-$\alpha$ in the range $[0.0, 0.25]$, the slowest algorithms were FGSD-Join, ExpVar-Join, and \MultistageJoinFullName{}-Basic. For Zipf-$\alpha$ in the range $[0.2, 0.55]$, none of the algorithms was a consistent winner. The \FrameworkName{} and \MultistageJoinFullName{} and their basic counterparts succeeded for all Zipf-$\alpha$ values. The other algorithms started running out of memory at different Zipf-$\alpha$ values. ExpVar-Join was able to succeed until Zipf-$\alpha$ was 0.8. The remaining algorithms did not finish the join for Zipf-$\alpha$ values larger than $0.7$. 

The experiment was repeated with records of size $10^3$ Bytes instead of $10^2$ Bytes, and the results are reported in \Figure~\ref{fig:numRecords10000000_numDistinctKeys100000_valueSize1000_VaryingAlpha}\footnote{This is a correction to the prior paper \cite{metwally2022scaling}. The size of the records, $m_{\mathcal{R}\mathcal{S}}$, was erroneously stated as $10^4$ Bytes in the \cite{metwally2022scaling}, but they were actually $10^3$ Bytes.}. For these relatively large records, the runtimes of all the algorithms were comparable until Zipf-$\alpha$ reached $0.3$ with a slight edge for Hash-Join in the Zipf-$\alpha$ $[0.0, 0.15]$ range, and FGSD-Join, ExpVar-Join, and \MultistageJoinFullName{}-Basic being the slowest algorithms for Zipf-$\alpha$ in the range $[0.0, 0.25]$. For the Zipf-$\alpha$ $[0.2, 0.3]$ range, none of the  algorithms was a consistent winner. \FrameworkName{} had the edge for Zipf-$\alpha$ above $0.35$. Full-SkewJoin, Hash-Join, Spark3-Join, FGSD-Join and ExpVar-Join could not finish within the deadline starting at Zipf-$\alpha$ of $0.5$, $0.65$, $0.65$, $0.7$ and $0.75$, respectively. Even \FrameworkName{}-Basic and \MultistageJoinFullName{}-Basic did not scale for Zipf-$\alpha$ starting $0.9$ since they could not fit all the relatively large records of the hottest key in the memory of one executor. Only the load-balanced \FrameworkName{} and \MultistageJoinFullName{} algorithms were able to succeed for all values of Zipf-$\alpha$.

The runtime of all the algorithms increased as the data skew increased. Not only do some executors become more loaded than others, but also the size of the results increases. The only exception was FGSD-Join, whose runtime was very stable through the low to mid range of Zipf-$\alpha$ ($[0.0, 0.6]$ and $[0.0, 0.3]$ in \Figure~\ref{fig:numRecords10000000_numDistinctKeys100000_valueSize100_VaryingAlpha} and \Figure~\ref{fig:numRecords10000000_numDistinctKeys100000_valueSize1000_VaryingAlpha}, respectively). The ExpVar-Join exhibited similar behavior but for smaller ranges of Zipf-$\alpha$. The pre-join steps were costly, but useless in load-balancing when the data was not skewed. In the mid Zipf-$\alpha$ range, FGSD-Join and ExpVar-Join allocated more executors to the hot keys than the simpler Hash-Join, and hence were faster. For higher Zipf-$\alpha$, FGSD-Join was bottlenecked by assigning each of the hottest keys to a single executor. ExpVar-Join was bottlenecked by the executors of the hottest key running out of memory, since they are producing the hottest-key join results in one MapReduce job, as explained in \Section~\ref{sec:amjoin_comparison}.

\FrameworkName{} and \MultistageJoinFullName{} scale almost linearly with the increased skewness, since they are able to distribute the load on the executors fairly evenly. Their basic counterparts still scaled almost linearly but were slower, due to the bottleneck of building the initial index and splitting the hot keys in the first iteration. From \Relation~\ref{ineq:num_iterations}, their runtimes are expected to increase with $\ell_{max}$, the frequency of the hottest key, which is impacted by Zipf-$\alpha$. ExpVar-Join scales as well as \FrameworkName{} and \MultistageJoinFullName{} and sometimes executes faster for all but the skewed relations. The bottleneck of ExpVar-Join is the high memory requirements discussed in \Section~\ref{sec:amjoin_comparison}.}Full-SkewJoin and Hash-Join perform relatively well for weakly skewed joins until both the joined relations become mildly skewed (at a Zipf-$\alpha$ of $0.6$ and $0.4$ in \Figure~\ref{fig:numRecords10000000_numDistinctKeys100000_valueSize100_VaryingAlpha} and \Figure~\ref{fig:numRecords10000000_numDistinctKeys100000_valueSize1000_VaryingAlpha}, respectively). For moderately skewed data, both algorithms failed to produce the results within the deadline, albeit for different reasons. Full-SkewJoin could not load the hot keys and all their records in the memory of all the executors, while the executors of Hash-Join were bottlenecked by the join results of the hottest keys.

The Hash-Join executes faster than the adaptive algorithms (\FrameworkName{} and Full-SkewJoin) for the majority of the weakly skewed range (Zipf-$\alpha$ $[0.0, 0.6]$ and $[0.0, 0.3]$ ranges in \Figure~\ref{fig:numRecords10000000_numDistinctKeys100000_valueSize100_VaryingAlpha} and \Figure~\ref{fig:numRecords10000000_numDistinctKeys100000_valueSize1000_VaryingAlpha}, respectively), since the adaptive algorithms are slowed down by computing the hot keys. However, the adaptive algorithms utilizing the Broadcast-Join to join the keys that are hot in only one of the joined relations pays off for more skewed data. For larger Zipf-$\alpha$ (Zipf-$\alpha$ of $0.65$ and $0.45$ in \Figure~\ref{fig:numRecords10000000_numDistinctKeys100000_valueSize100_VaryingAlpha} and \Figure~\ref{fig:numRecords10000000_numDistinctKeys100000_valueSize1000_VaryingAlpha}, respectively), Full-SkewJoin executed clearly faster than Hash-Join, and \FrameworkName{} executed significantly faster than both.

Since the data partitions are too big to fit in memory, and they are already of roughly equal sizes, Spark3-Join performed as a basic Sort-Merge-Join. This was comparable, but slightly slower than Hash-Join.

Broadcast-Join was never able to execute successfully, regardless of the data skew, since both relations were too large to load into the memory of the executors. Broadcast-Join is a fast algorithm, but is not a scalable one. This is clear from comparing the runtimes of Full-SkewJoin in \Figure~\ref{fig:numRecords10000000_numDistinctKeys100000_valueSize100_VaryingAlpha} and \Figure~\ref{fig:numRecords10000000_numDistinctKeys100000_valueSize1000_VaryingAlpha}, respectively. Since Full-SkewJoin employs Broadcast-Join for the hot keys, it was able to handle more skewed data when the records were smaller (until Zipf-$\alpha = 0.65$ and $0.45$ when the record sizes were $10^2$ and $10^3$, respectively).

\MultistageJoinFullName{} and \FrameworkName{} performed significantly faster than their basic counterparts. When the records have size $10^2$ Bytes, the average speedup of \FrameworkName{} over \FrameworkName{}-Basic is $1.55$, and 
the average speedup of \MultistageJoinFullName{} over \MultistageJoinFullName{}-Basic is $1.58$, where the speedup is calculated as the ratio of the runtime of \FrameworkName{}-Basic to that of \MultistageJoinFullName{}. When the records have size $10^2$ Bytes, the average speedups were $1.46$ and $1.36.$, respectively. For \MultistageJoinFullName{}(-Basic) and \FrameworkName{}(-Basic), there were regions of relative flattening of the runtime due to the stepwise increase in the number of iterations. This is clear in the $\alpha$ $[0.5, 0.7]$ and $[0.85, 1.0]$ ranges in \Figure~\ref{fig:numRecords10000000_numDistinctKeys100000_valueSize100_VaryingAlpha}, and in the $\alpha$ $[0.55, 0.7]$ and $[0.8, 1.0]$ ranges in \Figure~\ref{fig:numRecords10000000_numDistinctKeys100000_valueSize1000_VaryingAlpha}. This is because the number of iterations is a function of $\ell_{max}$ which is impacted by Zipf-$\alpha$.

\subsubsection{Exploring the Equi-Join Parameter Space}

Similar results were obtained for different parameters, e.g., scaling the keys by a multiplier to influence the join selectivity, using $10^8$ Zipfian records\footnote{Using $1000$ executors, each dataset took over $10$ hours to generate.}, using records of size $10^4$ Bytes, or allocating $4$ GB of memory per executor. For non-skewed data, Hash-Join was consistently the fastest and FGSD-Join was consistently the slowest. Full-SkewJoin was fast while it can fit the data in memory, and \FrameworkName{} and \MultistageJoinFullName{} were fast and able to scale for various $\alpha$ values. The runtime of Spark3-Join closely tracked that of Hash-Join but was slightly slower.

We also evaluated these algorithms on Small-Large joins. The Broadcast-Join performed best, and so did Spark3-Join since it also morphed into a Broadcast-Join. Full-SkewJoin performed slightly better than \FrameworkName{}, \MultistageJoinFullName{}, FGSD-Join, and Hash-Join with speedup in the range $[1.12, 1.18]$. As the size of the small relation increased and reached  the memory limit, the Spark3-Join did not execute a Broadcast-Join, and the Broadcast-Join could not accommodate the small relation in memory, and the results were very similar to those in \Section~\ref{sec:scalability_data_skew}.

\subsubsection{Equi-Join Scalability with Computational Resources}

Two $D(0.65, 100)$ relations were used to evaluate the scalability of the algorithms. While the Zipf-$\alpha$ of $0.65$ generates mildly skewed keys, this value was chosen, since this is the largest $\alpha$ where all the algorithms except Broadcast-Join were able to compute the join results. For all the scalability evaluation, the Broadcast-Join was never able to execute successfully, regardless of the number of executors used, since neither of the relations could be fit in the memory of the executors.

\begin{figure*}
\centering
\captionsetup{width=\smallfigwidth}
\begin{minipage}[t]{.5\textwidth}

  \centering
  \includegraphics[height=100pt, width = \smallfigwidth]{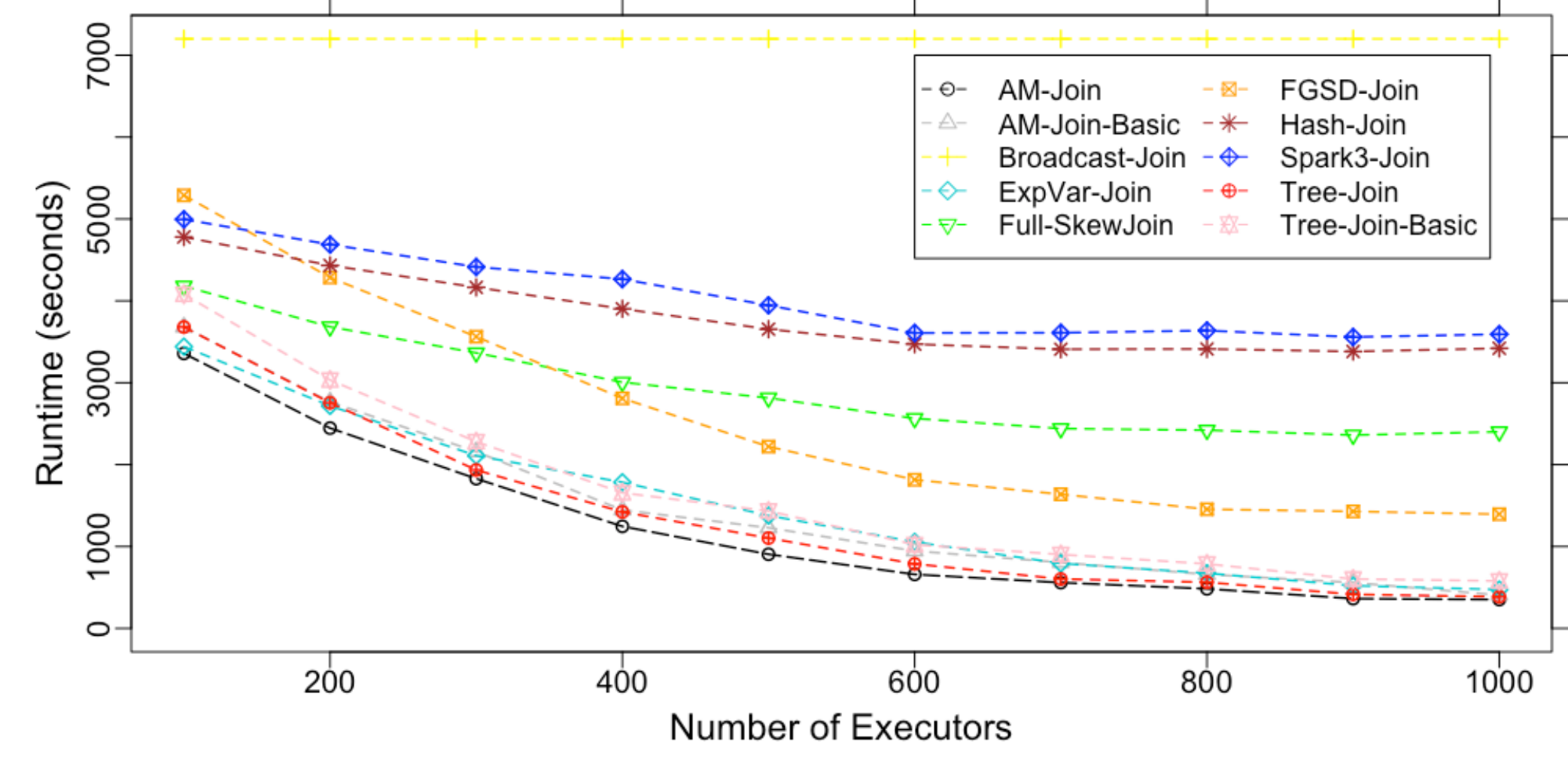}
  \caption{The runtime of the equi-join algorithms for $D(0.65, 100)$ while varying the number of executors.}
  \label{fig:numRecords10000000_numDistinctKeys100000_valueSize100_VaryingExecutors}

\end{minipage}
\begin{minipage}[t]{.5\textwidth}

  \centering
  \includegraphics[height=100pt, width = \smallfigwidth]{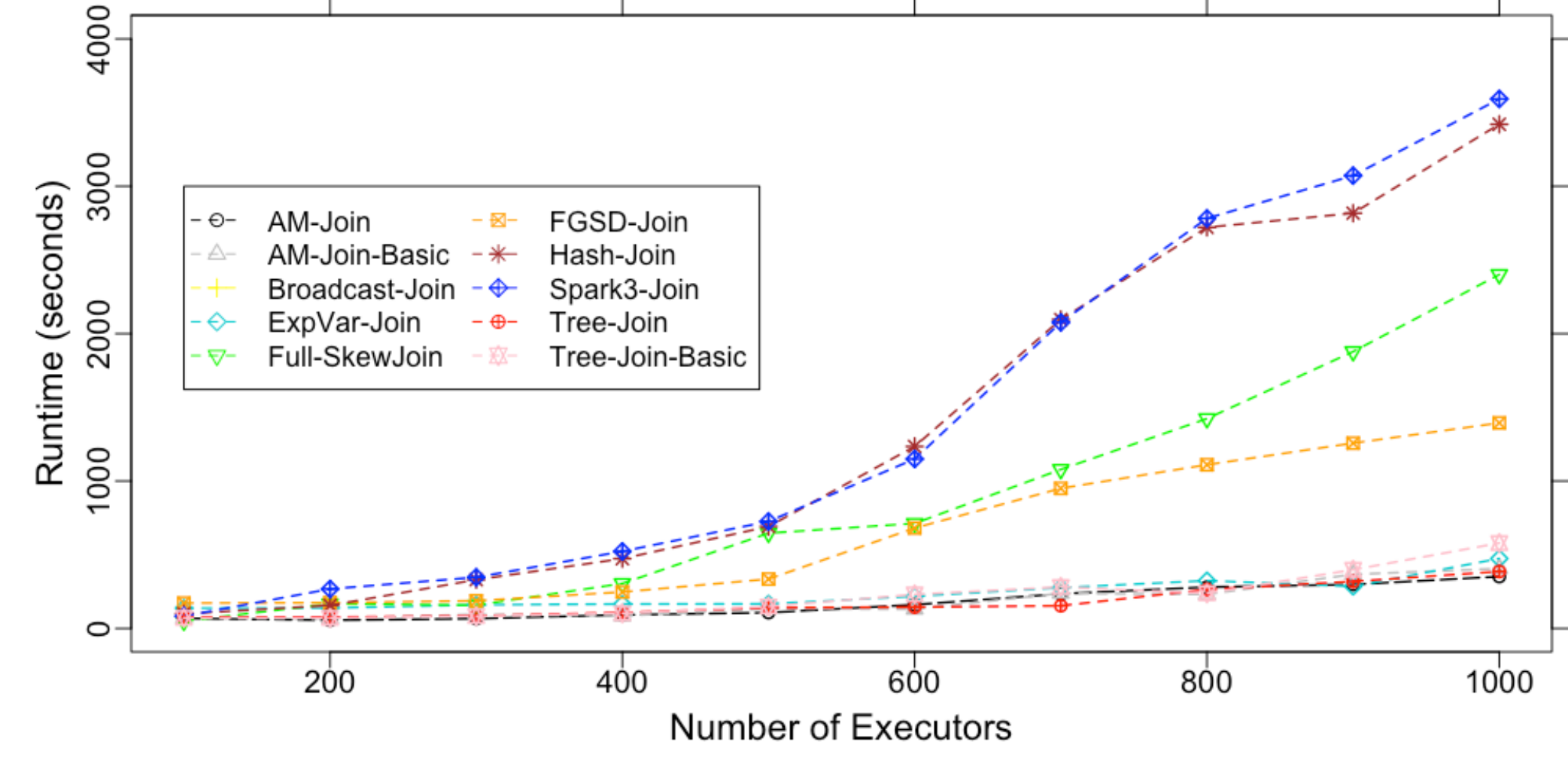}
  \caption{The runtime of the equi-join algorithms for a fraction of $D(0.65, 100)$ while varying the number of executors. The fraction is the number of executors $/1000$.}
  \label{fig:numRecords10000000_numDistinctKeys100000_valueSize100_VaryingExecutorsAndDataSize}

\end{minipage}
\end{figure*}

\paragraph{Strong Scalability Evaluation} 

\Figure~\ref{fig:numRecords10000000_numDistinctKeys100000_valueSize100_VaryingExecutors} shows the runtimes of the equi-join algorithms while varying the number of executors.

All the algorithms were able to scale with the increased number of resources. The algorithms showed a reduction in runtime as they were allowed more resources, albeit with various degrees. While FGSD-Join was the slowest at $100$ executors, its runtime showed the biggest absolute drop. As the number of executors increased, the sample used by FGSD-Join to determine the key-ranges grew, allowing FGSD-Join to better distribute the load. The improvements of Hash-Join and Spark3-Join were the fastest to saturate, and their run-times almost did not improve starting at $700$ executors, since the bottleneck was joining the hottest keys. Full-SkewJoin followed, and stopped having noticeable improvement at $800$ executors. \FrameworkName{}(-Basic), \MultistageJoinFullName{}(-Basic) and ExpVar-Join were able to improve their runtime as more executors were added, since they were able to split the load of joining the hottest keys among all the executors.

There is another phenomenon that is worth highlighting. The difference in the runtime between \FrameworkName{}, \MultistageJoinFullName{} and their basic counterparts diminished as more executors were used. The reason is the majority of the scalability comes from organizing the equi-join execution in multiple stages, which is shared by all the four algorithms.

\paragraph{Weak Scalability Evaluation} 

\Figure~\ref{fig:numRecords10000000_numDistinctKeys100000_valueSize100_VaryingExecutorsAndDataSize} shows the runtimes of the equi-join algorithms while varying both the number of executors and the sizes of the relations. For each number of executors on the horizontal axis in \Figure~\ref{fig:numRecords10000000_numDistinctKeys100000_valueSize100_VaryingExecutorsAndDataSize}, the joined relations were sampled from two $D(0.65, 100)$ relations. The sampling rate is the number of the executors $/1000$. Hence, only $\approx \frac{1}{10}$ of the two $D(0.65, 100)$ relations were joined when the number of executors was $100$, and the entire two $D(0.65, 100)$ were joined when the number of executors was $1000$.

All the algorithms were able to scale with the increased number of resources and as the joined relations grew in size. The algorithms showed an almost quadratic increase in runtime as they were allowed more resources and joined bigger relations. This can be attributed to the join complexity and output size increasing quadratically with the sampling rate. For instance, doubling the sizes of the input relations results in quadrupling the size of the join results, assuming the key frequency distributions, and the record sizes do not change.

When compared to ExpVar-Join and FGSD-Join, the other algorithms executed the smallest joins with $100$ executors in almost half the time. This can be attributed to the pre-join steps of ExpVar-Join and FGSD-Join. For ExpVar-Join, this pre-join overhead was still comparatively low, since the number of executors was small. As the number of machines increased, this overhead increased, but still improved the runtime of ExpVar-Join. Between $600$ and $1000$ executors, the fastest algorithms were \MultistageJoinFullName{}, \FrameworkName{}, \FrameworkName{}-Basic, and ExpVar-Join, in that order.

\subsubsection{Speedup of Natural Self-Joins}

\begin{figure*}
\centering
\captionsetup{width=\smallfigwidth}
\begin{minipage}[t]{.5\textwidth}

  \centering
  \includegraphics[height=100pt, width = \smallfigwidth]{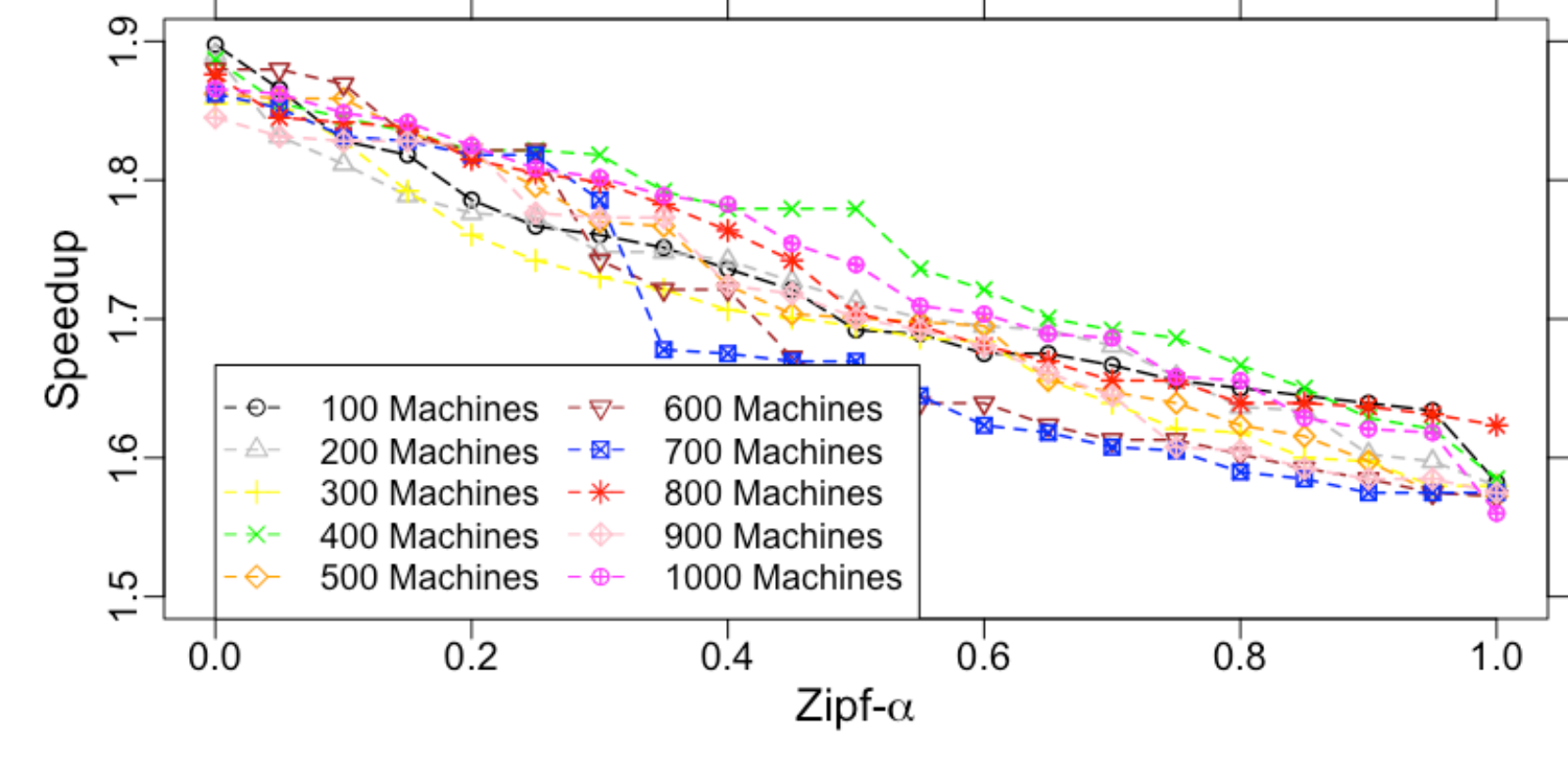}
  \caption{The natural equi-join speedup of \MultistageJoinFullName{} for $D(\alpha, 100)$ while varying the Zipf-$\alpha$ using various numbers of executors.}
  \label{fig:numRecords10000000_numDistinctKeys100000_valueSize100_VaryingExecutorsAndAlpha}

\end{minipage}
\begin{minipage}[t]{.5\textwidth}

  \centering
  \includegraphics[height=100pt, width = \smallfigwidth]{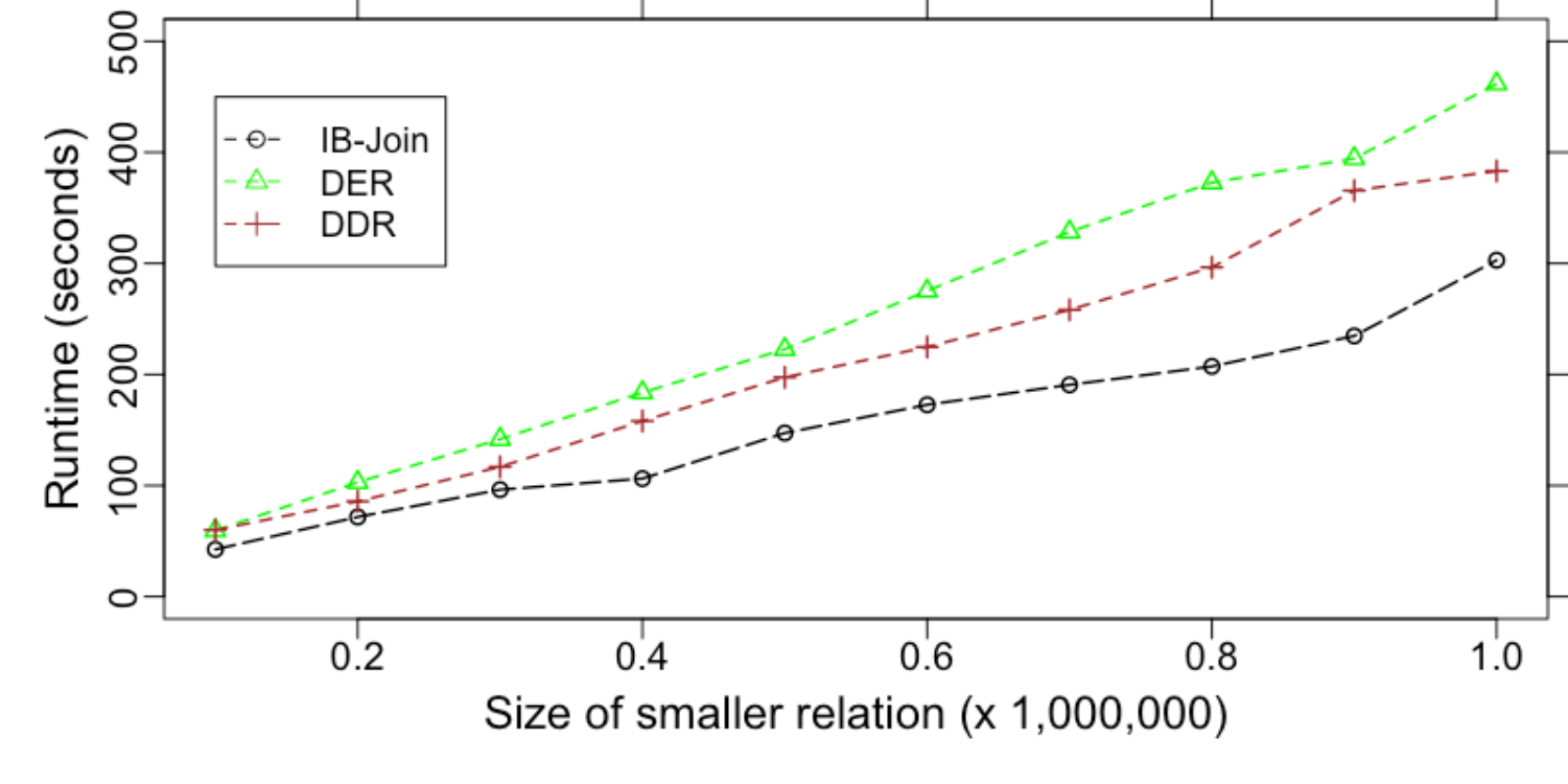}
  \caption{The runtime of the Small-Large outer-join algorithms while varying the size of the smaller relation.}
  \label{fig:numRecords100000000_numDistinctKeys200000_valueSize100_VaryingSmallerRelatoin}

\end{minipage}
\end{figure*}

We evaluate the speedup gained by the optimizations in \Section~\ref{sec:tree_self_join}. We report the \MultistageJoinFullName{} speedup between joining a relation with itself as a regular equi-join, and executing a natural self-join (i.e., eliminating the redundant join results). The speedup is reported only for \MultistageJoinFullName{}, since \FrameworkName{} reduces to \MultistageJoinFullName{} in the case of natural self-joins. The speedups for different Zipf-$\alpha$ values using various numbers of executors are reported in \Figure~\ref{fig:numRecords10000000_numDistinctKeys100000_valueSize100_VaryingExecutorsAndAlpha}. The experiments were run on a relation with records of size $10^2$, but records of size $10^3$ gave almost identical results.

From \Figure~\ref{fig:numRecords10000000_numDistinctKeys100000_valueSize100_VaryingExecutorsAndAlpha}, the speedup was roughly $1.67$ across the board regardless of the Zipf-$\alpha$, and the number of executors. There are two observations to be made. First, the speedup decreased as the Zipf-$\alpha$ increased. This can be attributed to the number of \MultistageJoinFullName{} iterations which are not impacted by the processing and IO savings discussed in \Section~\ref{sec:tree_self_join}. The larger the Zipf-$\alpha$, the more skewed the data, and the larger number of \MultistageJoinFullName{} iterations. Scheduling these iterations is a fixed overhead and does not enjoy the same savings discussed in \Section~\ref{sec:tree_self_join}. The other observation is as the number of executors increased, the savings tended to decrease. This can be attributed to the fixed overhead of communication among the executors as their number increases.

\subsubsection{Performance on Small-Large Outer-Joins}

\Figure~\ref{fig:numRecords100000000_numDistinctKeys200000_valueSize100_VaryingSmallerRelatoin} shows the runtimes of \SmallLargeJoinName{}, DER \cite{xu2010new} and DDR \cite{cheng2017design}\footnote{For DER and DDR, the Spark implementations provided by the authors of \cite{cheng2017design} at \url{https://github.com/longcheng11/small-large} were used in these experiments.} when performing a right-outer join of two relations with records of size $10^2$ Bytes, and keys uniformly selected in the range $[1, 2 \times10^5]$. The larger relation has $10^8$ records, while the size of the smaller relation was varied between $10^5$ and $10^6$ records. The keys in the smaller relation were all even to ensure a selectivity of $50\%$. This is the selectivity that least favors the \SmallLargeJoinName{} family optimizations in \Section~\ref{sec:broadcast_optimizations}.

\Figure~\ref{fig:numRecords100000000_numDistinctKeys200000_valueSize100_VaryingSmallerRelatoin} confirms the communication cost analysis done in \Section~\ref{sec:broadcast_comparison}. DDR was consistently faster than DER with one exception, while the right-outer variant of \SmallLargeJoinName{} was significantly faster than both. In reality, DER performs worse than shown in \Figure~\ref{fig:numRecords100000000_numDistinctKeys200000_valueSize100_VaryingSmallerRelatoin}, since the time for assigning unique ids to the records is not shown.

\subsection{Experiments on Real Data}
\label{sec:real}

\begin{figure}
\centering
\includegraphics[height=100pt, width = \smallfigwidth]{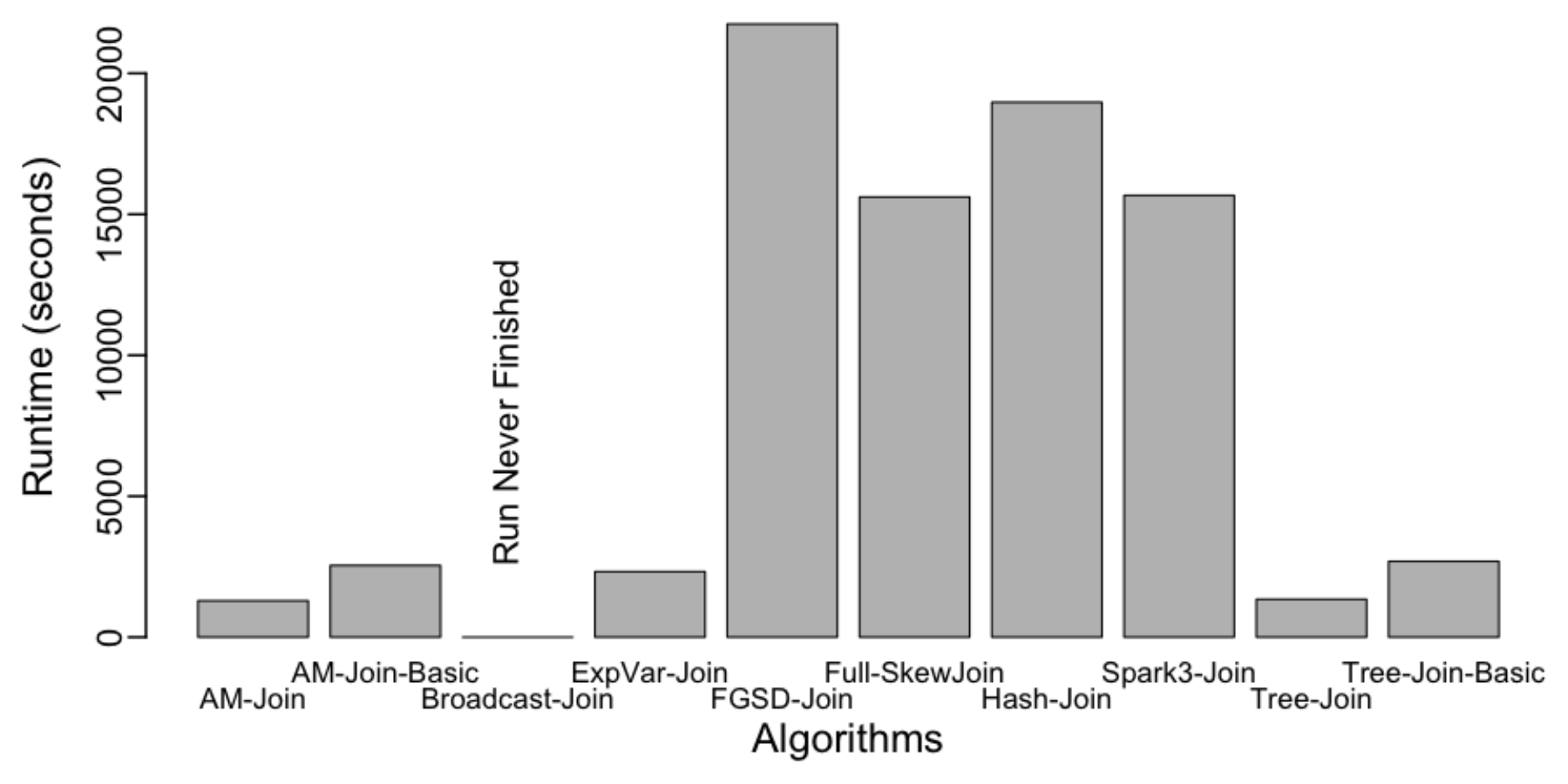}
\caption{The runtime of the equi-join algorithms on $1000$ executors on real data.}
\label{fig:real_data}
\end{figure}

We ran two experiments on real data. All the experiments were executed on machines (driver and executors) with $8$GB of memory, $20$GB of disk space, and $1$ vCPU. The first experiment was a natural self-join on the signup features of the riders of trips in a specific geo-location in October, 2020. These features are extracted from signup attributes, e.g., the account names. These signup attributes and the extracted features are not revealed due to their usage in fraud detection, a sensitive area for the majority of the companies. The self-joined relation had $5.1 \times 10^8$ distinct keys, and each record was of size $36$ Bytes. The total size of the data was $17.6$ TB. There were more than $1000$ keys with frequencies exceeding $100$. The frequencies of the hottest $10$ keys are reported in \Table~\ref{tab:real_data_frequency_distribution}. Only \FrameworkName{} and \MultistageJoinFullName{} were able to finish successfully. They both took $2.0$ hours to finish. ExpVar-Join, Full-SkewJoin, Hash-Join and Broadcast-Join did not finish in $48$ hours.

\begin{table}[ht]
\caption{The frequencies of the hottest $10$ keys used in real-data experiments.}
\label{tab:real_data_frequency_distribution}
\centering
\begin{tabular}{c | r r r r r r r r r r } 
  \toprule
  Rank & $1$ & $2$ & $3$ & $4$ & $5$ & $6$ & $7$ & $8$ & $9$ & $10$ \\
  \midrule
  Frequency & $21235$ & $21138$ & $21078$ & $20720$ & $20463$ & $20380$ & $20321$ & $20006$ & $19888$ & $19823$ \\
  \bottomrule
\end{tabular}
\end{table}

The second experiment was a join between two relations. The first relation is the one described above. The second relation represented data on a single day, October $14^{th}$, 2020. The size of the second relation was $0.57$ TB. The runtime of the equi-join algorithms on $1000$ executors is shown in \Figure~\ref{fig:real_data}. All the algorithms were able to finish successfully, except for Broadcast-Join since it could not fit the smaller relation in the memory of the executors. Thanks to organizing the equi-join execution in multiple stages, \FrameworkName{} and \MultistageJoinFullName{} were able to execute an order of magnitude faster than all the other algorithms. Their basic counterparts however, executed significantly slower due to the load imbalance of building the initial index and splitting the hot keys in the first iteration. ExpVar-Join executed faster than the basic counterparts but slower than \FrameworkName{} and \MultistageJoinFullName{}. The large number of hot keys in the real relations slowed ExpVar-Join since the complexity of its pre-join step grows with the number of hot keys.

\section{Conclusion}
\label{sec:conclusion}

This paper proposes \FrameworkFullName{} (\FrameworkName{}) a fast, efficient and scalable equi-join algorithm that is built using the basic MapReduce primitives, and is hence deployable in any distributed shared-nothing architecture. The paper started by proposing \MultistageJoinFullName{}, a novel algorithm that organizes the equi-join execution in multiple stages. \MultistageJoinFullName{} attains scalability by distributing the load of joining a key that is hot in both relations throughout the join execution. Such keys are the scalability bottleneck of most of the state-of-the-art distributed algorithms. \FrameworkName{} utilizes \MultistageJoinFullName{} for load balancing, high resource utilization, and scalability. Moreover, \FrameworkName{} utilizes Broadcast-Joins that reduce the network load when joining keys that are hot in only one relation. By utilizing \MultistageJoinFullName{} and Broadcast-Join, \FrameworkName{} achieves speed, efficiency, and scalability. \FrameworkName{} extends to all the outer-joins elegantly without record deduplication or custom table partitioning, unlike the state-of-the-art industry-scale algorithms \cite{bruno2014advanced}. For the fastest execution of \FrameworkName{} outer-joins, the paper proposed \SmallLargeJoinFullName{} (\SmallLargeJoinName{}) family that improves on the state-of-the-art Small-Large outer-join algorithms \cite{xu2010new,cheng2017design}. All the proposed algorithms use the basic MapReduce primitives only, and hence can be adopted on any shared-nothing architecture. 

This paper also tackles natural self-joins, which is at the intersection of equi-joins, inner-joins, and self-joins. The paper shows how to optimize \MultistageJoinFullName{} to save roughly half the processing and IO when executing natural self-joins.

All the theoretical claims have been verified using extensive evaluation. Our evaluation highlights the improved performance and scalability of \FrameworkName{} when applied to the general equi-joins. When compared to the sate-of-the-art algorithms \cite{bruno2014advanced,gavagsaz2019load}, \FrameworkName{} executed comparably fast on weakly-skewed synthetic tables and can join more-skewed or orders-of-magnitude bigger tables, including our real-data tables. These advantages are even more pronounced when applying the join algorithms to natural self-joins. The proposed \SmallLargeJoinName{} outer-join algorithm executed much faster than the state-of-the-art algorithms in \cite{xu2010new,cheng2017design}.

Our future directions focus on optimizing \FrameworkName{} for the general shared-nothing architecture that supports multicasting data, for NUMA machines connected by a high-bandwidth network, and learning from the RDMA and the work-stealing enhancements of \cite{rodiger2016flow}. Moreover, we plan to explore using Radix join \cite{manegold2002optimizing,barthels2015rack} that it is only bound by the memory bandwidth as the Shuffle-Join.
\begin{acks}
We like to express our appreciation to Sriram Padmanabhan, Vijayasaradhi Uppaluri, Gaurav Bansal, and Ryan Stentz from Uber for revising the manuscript and improving the presentation of the algorithms, and to Nicolas Bruno from Microsoft for discussing the SkewJoin algorithms, and to Shixuan Fan from Snowflake for revising the theoretical foundations.
\end{acks}

\balance
\bibliographystyle{ACM-Reference-Format}
\bibliography{equijoins_mr}

\end{document}